\documentclass[12pt,a4paper]{article}
\usepackage[T1]{fontenc}
\usepackage{amsfonts}
\usepackage{amsmath}
\usepackage{array}
\usepackage{bm}
\usepackage{booktabs}
\usepackage{tabularx}
\usepackage{framed}
\usepackage{changepage}
\usepackage{subcaption}
\usepackage[most]{tcolorbox}

\usepackage{longtable}
\usepackage{pdflscape}
\usepackage{lmodern}
\usepackage{caption}
\usepackage{fancyhdr}
\usepackage[a4paper, total={6.5in, 9.75in}]{geometry}
\usepackage{amssymb}
\usepackage{float}
\usepackage{titlesec}
\usepackage{caption}
\usepackage[backend=biber,style=apa,labelnumber]{biblatex}
\usepackage{graphicx}
\usepackage{hyperref}
\usepackage{booktabs}
\usepackage{chngcntr}
\usepackage{placeins}
\usepackage{authblk}
\usepackage{appendix}
\usepackage{lscape}
\usepackage{pdflscape}
\usepackage{lipsum}
\usepackage{rotating}
\usepackage{amsmath}
\usepackage{lscape}
\usepackage{afterpage}
\usepackage{subcaption}
\makeatletter
\let\sf@counterlist\@empty
\makeatother

\DeclareFieldFormat{url}{\href{#1}{Available here}}
\renewbibmacro*{url}{\iffieldundef{url}{}{\setunit{\addspace}\printfield{url}}}

\usepackage{tikz}
\usetikzlibrary{positioning}
\usetikzlibrary{decorations.pathreplacing, arrows.meta}

\providecommand{\U}[1]{\protect\rule{.1in}{.1in}}

\hypersetup{
  colorlinks   = true,
  urlcolor     = blue,
  linkcolor    = blue,
  anchorcolor = blue,
  citecolor   = blue
}

\title{Earthquakes, floods, and fiscal resilience: Evidence from Croatia and Slovenia\thanks{The opinions expressed in this publication are those of the authors. We are grateful to Leonardo Martinez for his careful reading and suggestions for improvement. A special thanks goes to Daniele Tavani for insightful correspondence on earlier drafts of the article. The usual caveats apply.}}

\bigskip

\author[a]{Luka Draganić}
\author[a\thanks{\emph{Corresponding author}: \ttfamily leonarda.srdelic@ijf.hr}]{Leonarda Srdelić}
\author[b]{Marwil J. Davila-Fernandez}
\affil[a]{\small \emph{Institute of Public Finance, Croatia}\smallskip}
\affil[b]{\small \emph{Colorado State University, United States}\smallskip}

\date{May 2026}

\begin{document}

\maketitle

\begin{abstract}


This paper studies how severe natural disasters affect public debt sustainability in two small open euro-area economies, Croatia and Slovenia, using the IMF's Natural Disaster Debt Dynamics Tool (ND-DDT). Croatia is calibrated to the 2020 Zagreb earthquake, and Slovenia to the 2023 floods. We compare baseline debt projections with four complementary disaster configurations: two empirical-distribution specifications (one-off and recurrent), one local projection, and one 95th-percentile quantile regression. We further conduct a four-channel robustness exercise allowing the disaster shock to affect inflation and the effective interest rate, in addition to real GDP growth and the primary balance. In the absence of an extreme natural event, we estimate that Croatian public debt declines from 63 per cent of GDP in 2024 to 52.7 per cent by 2034, while Slovenian debt rises from 66.6 to 74.7 per cent over the same horizon. However, a severe disaster shifts both trajectories upward and increases their persistence, raising the Croatian debt-to-GDP ratio to between 64.5 and 70.4 per cent of GDP, while the Slovenian ratio reaches between 83 and 95 per cent across scenarios. A counterfactual exercise confirms that pre-disaster fiscal space, adaptive capacity, and risk-transfer arrangements attenuate but do not eliminate the post-disaster debt burden. Our findings highlight that fiscal resilience must be built before the shock occurs.

\bigskip

\textbf{Keywords:} Public debt sustainability, Natural disasters, Fiscal risk, Stochastic simulations, Risk management.

\bigskip

\textbf{JEL:} J11; O41; O52.

\end{abstract}

\clearpage

\section{Introduction}

Public debt sustainability is a central concern for small open economies, where fiscal space is often limited and exposure to external shocks is high. Natural disasters such as earthquakes and floods pose substantial fiscal risks. They can abruptly widen deficits through emergency spending and revenue losses, leading to rapid increases in public debt. Climate change is projected to raise the frequency and severity of weather- and climate-related hazards, further amplifying these pressures \parencite{IPCC2021SPM}. The challenge reflects the difficulty of managing economic systems subject to infrequent but highly disruptive shocks \parencite{ArrowEtAl95, LevinEtAl98, CarpenterEtAl12, Maler08}. A sustainable debt trajectory requires fiscal resilience to large adverse events. Quantifying it matters for effective fiscal risk management and proactive policy design in economies vulnerable to natural hazards.

Empirical evidence on the macroeconomic consequences of natural disasters remains mixed \parencite{EvgenidisEtAl21, Taniguchi22, BatesEtAL24, GuoEtAL24}. Many studies document adverse output effects, but findings vary considerably across disaster types, magnitudes, and country contexts \parencite{Rahaman2025, TaghizadehHesary2021, ChaudhuriMenezes2025, ManagiGuan2017}. \textcite{Fomby2013} show that GDP growth responses differ markedly across droughts, floods, earthquakes, and storms, with developing economies typically suffering larger and more persistent losses than advanced ones. Severe disasters also have disproportionately larger effects than moderate events, and sectoral exposure is uneven. \textcite{Noy2009} broadly corroborates these findings, emphasising that country-specific characteristics matter. Stronger institutions, higher human capital, greater fiscal capacity, and more robust financial buffers all improve resilience to disaster shocks. \textcite{LeEtAl2025} link disaster risk explicitly to fiscal sustainability outcomes in a panel of 184 economies, showing that climate vulnerability shocks increase debt-to-GDP ratios and deteriorate fiscal balances (see also, \citeauthor{Saadaoui2026ClimateFiscalRisk}, \citeyear{Saadaoui2026ClimateFiscalRisk}).

Controlling for hazard intensity, asset exposure, and vulnerability, counterfactual GDP estimates suggest that average macroeconomic effects appear modest over the medium-term but grow substantially larger for severe events \parencite{Hochrainer2009}. A structural quantile VAR approach tells a similar story, with climate-related disasters shifting the full predictive distribution of output growth and inflation \parencite{ChavleishviliMoench2025}. Such extreme events push downside risks to growth and upside risks to inflation, with persistent effects running through higher conditional volatility and skewness. The picture is less clear-cut for long-run growth, however. Using synthetic controls, \textcite{Cavallo2013} find that even extremely large disasters do not systematically reduce long-run output once political shocks are disentangled from the natural event itself, suggesting that persistent losses often reflect post-disaster political instability rather than direct economic damage.

We contribute to this literature by examining how natural disasters affect public debt sustainability in two small open euro-area economies, Croatia and Slovenia.\footnote{These two countries share a common history as republics of the former Yugoslavia, both transitioning from centrally planned to market economies during the 1990s and early 2000s. Following independence, Croatia and Slovenia undertook substantial reforms to develop their private sectors and attract foreign investment, joining the WTO in 2000 and 1995, CEFTA in 2003 and 1996, and the EU in 2013 and 2004, respectively. As small open economies operating within a monetary union, monetary policy tools are unavailable at the national level, leaving fiscal policy as the primary instrument for absorbing large shocks. This makes fiscal resilience to natural disasters particularly consequential for both countries.} We study how large disaster events affect their debt trajectories through the IMF's Natural Disaster Debt Dynamics Tool \parencite{Acosta2025}, hereafter ND-DDT. The empirical backbone of the analysis draws on the cross-country panel local-projection and quantile-regression estimates of \textcite{NguyenFengGarciaEscribano2025}, evaluated at country-specific values for the two countries of interest. Understanding how natural disasters affect debt sustainability in this context is directly relevant to the design of EU fiscal rules and disaster risk frameworks. Croatia and Slovenia offer a rare opportunity to compare two different disaster types, seismic and hydrological, within otherwise similar institutional and macroeconomic settings.

Croatia accumulated public debt rapidly during the global financial crisis and the prolonged recession that followed, only stabilising and reducing it in the late 2010s. The twin shocks of the COVID-19 pandemic and the 2020 earthquakes then exposed how vulnerable public finances remained to large exogenous events. Slovenia faced similar pressures from the August 2023 floods, which affected a substantial share of its territory and population. We construct a baseline scenario reflecting expected macroeconomic conditions and fiscal policies, and a stress scenario in which a major earthquake hits Croatia in 2026 and a major flood hits Slovenia in the same year. The Croatian shock is calibrated to the 2020 Zagreb earthquake as documented in EM-DAT, while the Slovenian shock is anchored to the August 2023 floods. Results are stress-tested using four complementary simulation methods to assess robustness.

The first simulation applies shocks only in the initial period and is calibrated to the 5th percentile of the empirical distribution of historical disaster impacts. It captures an extreme one-off event. The second imposes repeated shocks over the entire projection horizon at the same percentile. This avoids the implausible assumption of an extreme disaster hitting every year while still testing fiscal vulnerability under sustained pressure; it is best read as a tail-risk envelope rather than a plausible expected path. The third method computes the dynamic response of macro-fiscal variables through the local-projection estimator of \textcite{Jorda2005}. We evaluate it at country-specific values of disaster severity, pre-disaster fiscal balance, adaptive capacity, income group, and disaster type. For real GDP growth, we report 95 per cent confidence intervals; for the primary balance, we report point estimates, consistent with the active confidence-interval selector in the ND-DDT scenario files. Finally, the fourth method consists of applying quantile regression at the upper conditional quantile $\tau = 0.95$. It captures high-risk states that are more vulnerable to debt accumulation and complements the average response estimated by the local projections.

Our contribution to the literature is threefold. First, we provide a comparative case study of public debt dynamics under two different disaster shocks in two small open euro-area economies, adding to recent efforts to incorporate natural disaster risk into debt sustainability assessments. Second, we show that the same broad class of disaster shock generates substantially different post-disaster debt trajectories depending on the baseline debt path. Croatia's declining baseline attenuates persistence, with debt rising to between 64.5 and 70.4 per cent of GDP against a no-disaster baseline of 52.7 per cent. Slovenia's rising baseline amplifies the effect, pushing public debt to between 83 and 95 per cent of GDP compared with 75 per cent without the shock. Third, we quantify the role of pre-disaster fiscal buffers and adaptive capacity in mitigating disaster-induced debt accumulation, including the risk of breaching the EU's 60 per cent of GDP Maastricht threshold. We report a four-channel robustness exercise that accounts for inflation and effective interest-rate channels.

The remainder of the paper proceeds as follows. Section 2 covers public debt trends in Croatia and Slovenia and their exposure to natural disasters. Section 3 outlines the methodology, including the debt dynamics model and data sources. Section 4 presents results under the baseline and disaster scenarios, the stochastic risk analysis, and the robustness exercise on the inflation and interest rate channels. Section 5 concludes with key findings, policy implications, and directions for future research.

\section{Background}

\subsection{Trends and structure of Croatian public debt}

\begin{figure}[tbp]
    \centering
    \captionsetup{position=above}
    \caption{Croatia's general government debt and its composition, 2010--2024.}
    \bigskip
    \begin{minipage}{0.8\textwidth} \centering
        {\footnotesize\textbf{a)} Trajectories of public debt in billion euros and as \% of GDP}\par\vspace{4pt}
        \includegraphics[width=\textwidth]{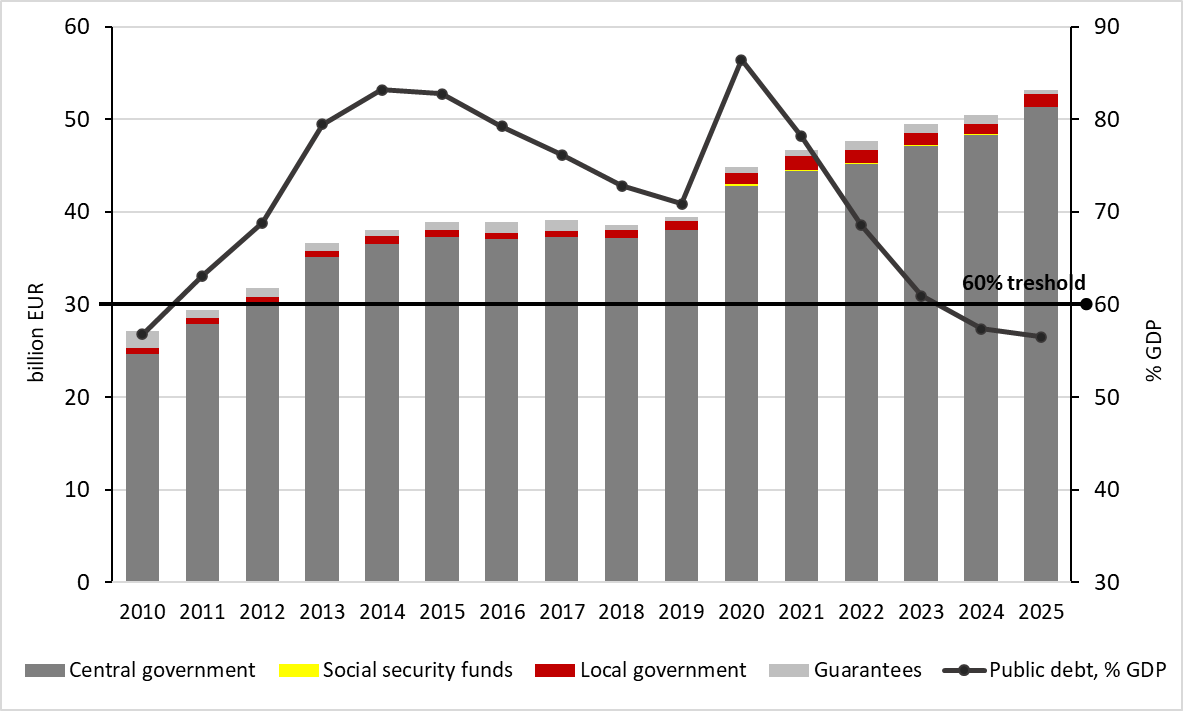}
    \end{minipage}

    \bigskip \bigskip

    \begin{minipage}{0.8\textwidth} \centering
        {\footnotesize\textbf{b)} General government deficit, debt change, and SFA}\par\vspace{4pt}
        \includegraphics[width=\textwidth]{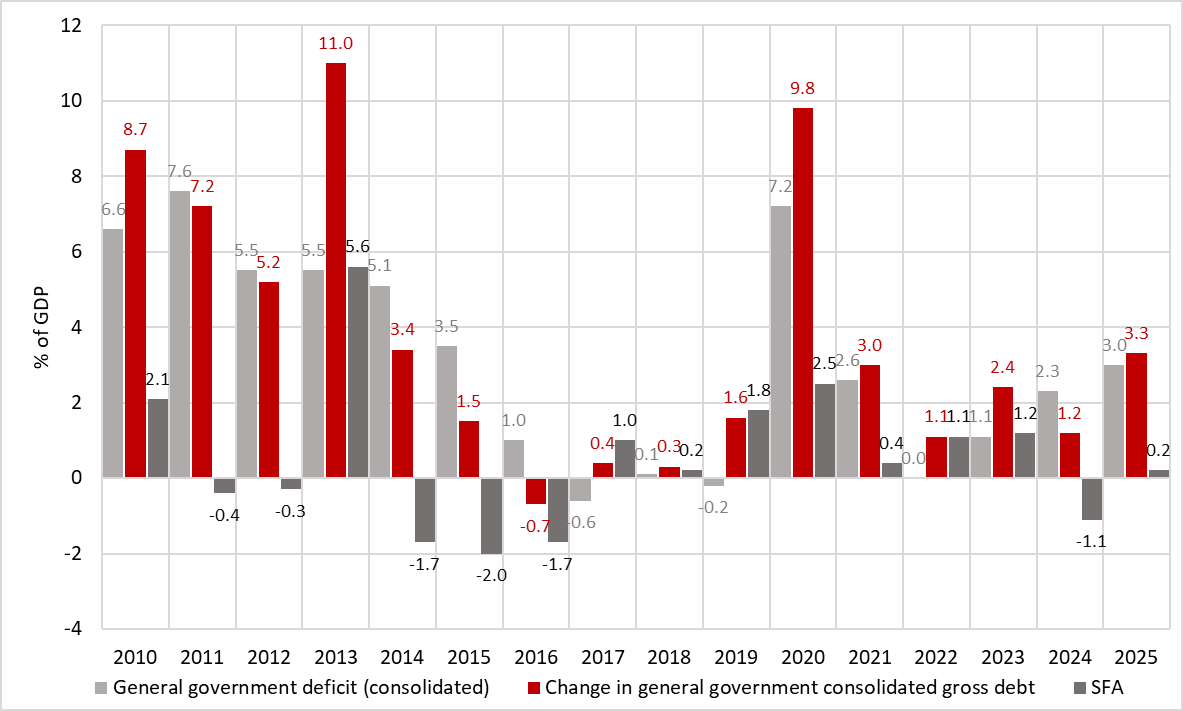}
    \end{minipage}
\caption*{\footnotesize \textit{Sources: Eurostat, Croatian National Bank.}}

    \label{fig:debt-dynamics-CRO}
\end{figure}

Croatia's general government debt between 2010 and 2025 passed through several distinct phases, reflecting macroeconomic conditions, fiscal policy measures, and institutional changes (for an overview of growth performance and its relationship with international trade, see \citeauthor{SrdelicDavila24}, \citeyear{SrdelicDavila24}). In nominal terms, debt increased from \texteuro 25.2 billion in 2010 to \texteuro 52.4 billion in 2025. The central government accounted for the predominant share throughout, rising from \texteuro 24.7 billion to \texteuro 51.3 billion. Local government debt grew gradually from \texteuro 0.6 billion to \texteuro 1.3 billion, while social security fund debt remained negligible, peaking at \texteuro 0.14 billion in 2020. Fig.~\ref{fig:debt-dynamics-CRO}, panel (a), reports the main trends.

\begin{figure}[tbp]
    \captionsetup{position=above}
    \caption{Structure and risk profile of the Croatian public debt portfolio, 2010--2024.}
    \centering

    \begin{minipage}{0.47\textwidth}
        \centering
        {\footnotesize\textbf{a)} Structure by financial instrument}\par\vspace{2pt}
        \includegraphics[width=\textwidth]{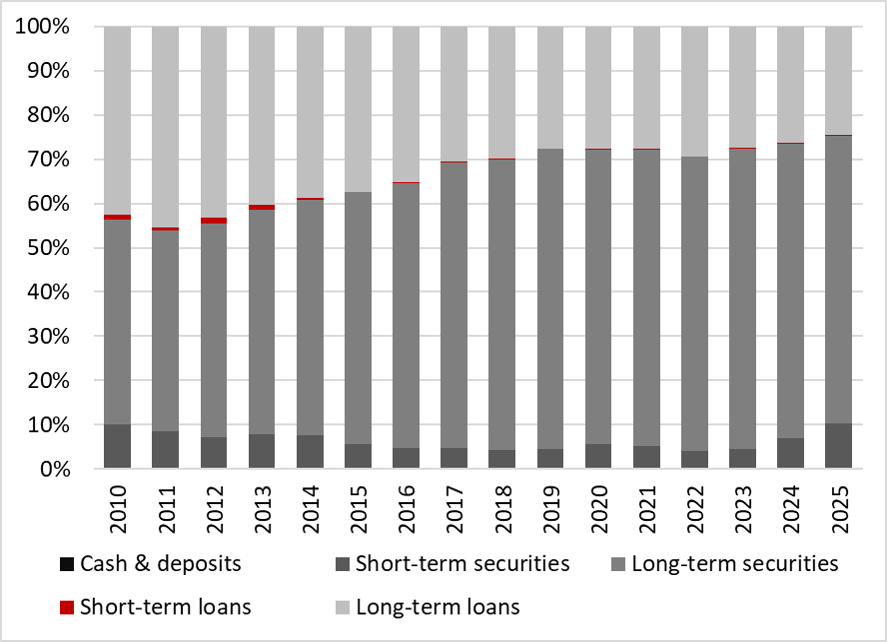}
    \end{minipage}
    \hfill
    \begin{minipage}{0.47\textwidth}
        \centering
        {\footnotesize\textbf{b)} Structure of debt by creditor residency}\par\vspace{2pt}
        \includegraphics[width=\textwidth]{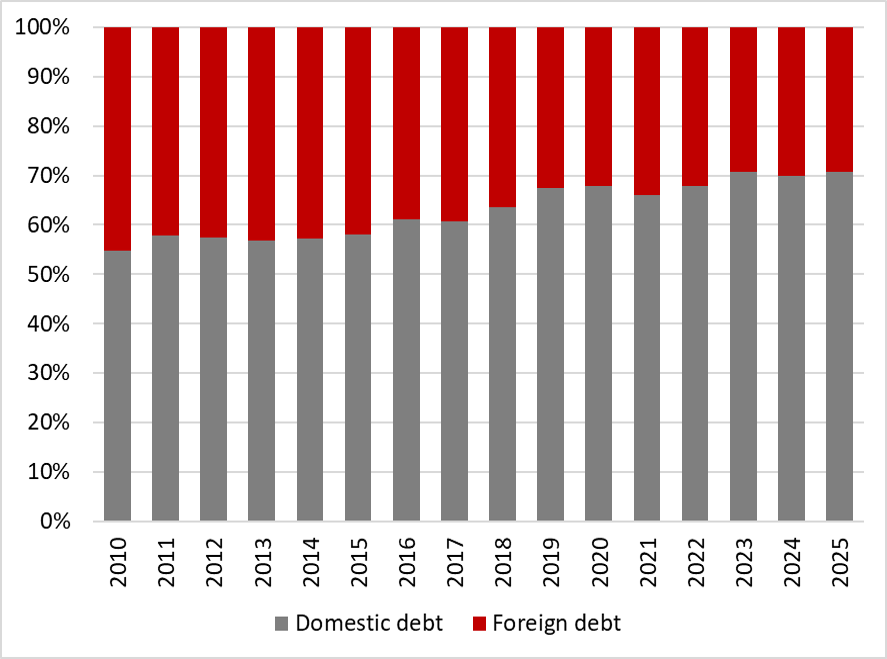}
    \end{minipage}

    \vspace{20pt}

    \begin{minipage}{0.47\textwidth}
        \centering
        {\footnotesize\textbf{d)} Domestic debt by instrument}\par\vspace{2pt}
        \includegraphics[width=\textwidth]{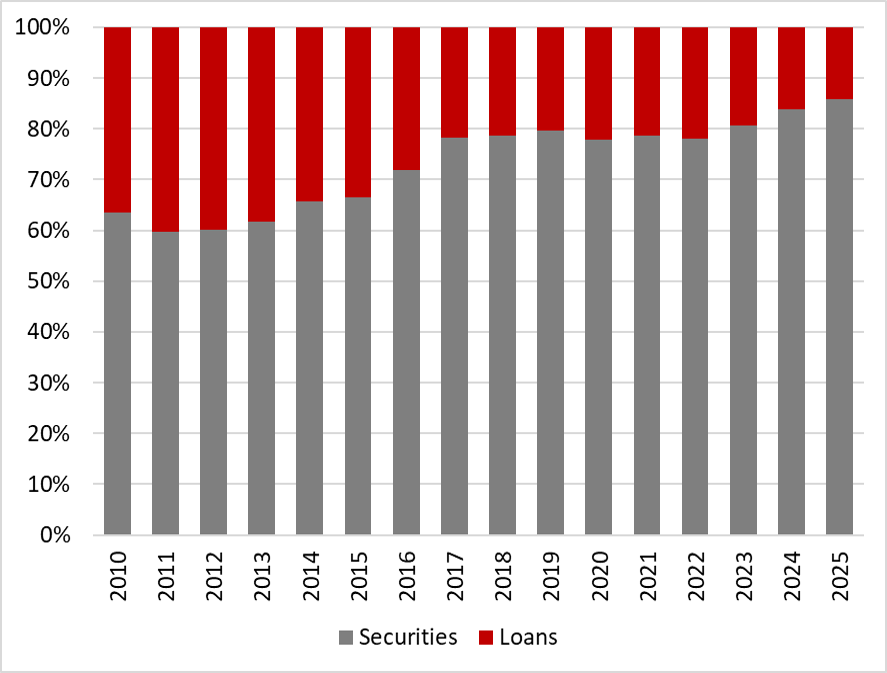}
    \end{minipage}
    \hfill
    \begin{minipage}{0.47\textwidth}
        \centering
        {\footnotesize\textbf{e)} Foreign debt by instrument}\par\vspace{2pt}
        \includegraphics[width=\textwidth]{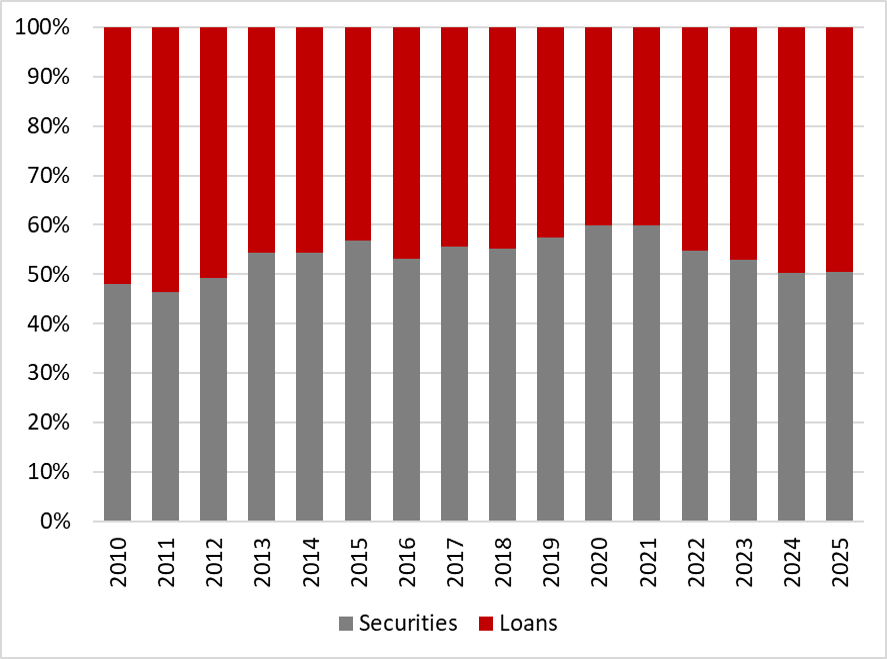}
    \end{minipage}

    \caption*{\footnotesize \textit{Sources: Eurostat, Ministry of Finance, Croatian Bureau of Statistics, Croatian National Bank.}}
    \label{fig:debt-structure1}
\end{figure}

Following the global financial crisis and the 2009 recession, a sharp fall in GDP, coupled with declining revenues and countercyclical spending, pushed the debt-to-GDP ratio from 57 per cent at the start of the period to a peak of 83 per cent by 2014. The introduction of the Excessive Deficit Procedure (EDP) in 2014 marked a turning point. Primary balances turned positive from 2015 onwards, expenditure was brought under tighter control, and revenues strengthened. Between 2016 and 2019, fiscal adjustment drew on both spending restraint and robust tax receipts, supported by cyclical recovery and policy efforts to improve revenue performance. The debt ratio fell to 71 per cent of GDP by 2019. Fiscal consolidation under the EDP delivered measurable improvements in fiscal outcomes, though the measures were assessed as less favourable to growth than alternative policy scenarios \parencite{DeskarDesign}.

The COVID-19 pandemic hit in 2020 and undid much of the fiscal progress achieved in the preceding years. Emergency spending and a sharp contraction in economic activity widened the deficit substantially. The blow was particularly hard for Croatia, where value-added tax accounts for more than half of total tax revenues and lockdowns compressed consumption sharply. Two major earthquakes that year, one in Zagreb in March and another in Petrinja in December, compounded the pressure, requiring large-scale spending on emergency relief and reconstruction. Precautionary borrowing added to this, pushing the debt ratio to 86 per cent of GDP. From 2021 onwards, economic recovery, stronger fiscal revenues, and the effect of inflation on nominal GDP pulled the ratio down to 78 per cent in 2021, 69 per cent in 2022, 62 per cent in 2023, and 58 per cent in 2024, falling back below the Maastricht reference value of 60 per cent.

An important technical aspect of public debt dynamics is the divergence between the annual deficit and the change in debt stock. In principle, a government deficit should correspond to an equivalent increase in debt, and a surplus to a reduction. In practice, debt developments also depend on operations outside the budget balance, a difference captured by the Stock-Flow Adjustment (SFA), which records transactions and accounting operations that affect the debt stock without passing through the reported deficit \parencite{Galinec2018}. Typical examples include the assumption of state-owned enterprise debt, loans extended to public entities, and the government's net acquisition of financial assets. Valuation changes and debt reclassifications also contribute. Exchange rate movements were a factor in the past, given the share of foreign-currency-denominated debt, but this ceased to be relevant when Croatia joined the euro area in 2023.

Croatia's experience illustrates the relevance of these adjustments. In 2010 and especially in 2013, the debt ratio rose by significantly more than the deficit, reflecting SFA of 2.1 per cent and 5.5 per cent of GDP, respectively. During the consolidation years 2014--2016, a negative SFA contributed to faster debt reduction than the headline deficit alone would imply. The pandemic year of 2020 offers an additional illustration. The deficit stood at 7.2 per cent of GDP, yet the debt ratio rose by 9.8 per cent, with SFA contributing 2.6 per cent of GDP through precautionary borrowing and other debt-increasing operations. In 2024, the pattern ran the other way. Despite a deficit of 2.4 per cent of GDP, the debt ratio rose by only 1.2 per cent, as the negative SFA of -1.2 per cent offset part of the accumulation. Fig.~\ref{fig:debt-dynamics-CRO}, panel (b), reports these trajectories.


Long-term securities dominate the structure of central government debt, growing from \texteuro 11.4 billion in 2010 to more than \texteuro 33 billion in 2025. Fig.~\ref{fig:debt-structure1}, panel (a) shows this trend. It reflects a policy orientation towards extending maturities and reducing refinancing risk. Long-term loans remain an important component, though their relative weight has declined as reliance on securities expanded. Short-term instruments, both securities and loans, play a minor and variable role, serving primarily for liquidity management. Cash and deposits are negligible throughout. The overall picture is one of a gradual shift towards long-term market financing. The domestic and external composition of debt has also shifted notably. By 2025, domestic investors had become predominant, holding above 70 per cent of the total, up from an almost equal split with foreign holders in the early 2010s, as shown in Fig.~\ref{fig:debt-structure1}, panel (b). Greater issuance on the local market and a reduction in external exposure drove this change, lowering vulnerability to rollover pressures and anchoring debt sustainability more firmly in domestic financial markets.

On the domestic side, central government debt is increasingly concentrated in securities, rising from around 64 per cent in 2010 to above 85 per cent in 2025, as reported in Fig.~\ref{fig:debt-structure1}, panel (c). External debt fluctuates around an equal split between securities and loans, see Fig.~\ref{fig:debt-structure1}, panel (d). Securities edged above 50 per cent in the mid-2010s but drifted back towards parity after the pandemic, reflecting a balance between market issuance and borrowing from international lenders.

\subsection{Slovenian public debt trends and structure}

\begin{figure}[tbp]
    \centering
    \captionsetup{position=above}
    \caption{Slovenia’s general government debt and its composition, 2010–2024.}
    \bigskip

    \begin{minipage}{0.8\textwidth} \centering
        {\footnotesize\textbf{a)} Trajectories of public debt in billion euros and as \% of GDP}\par\vspace{4pt}
        \includegraphics[width=\textwidth]{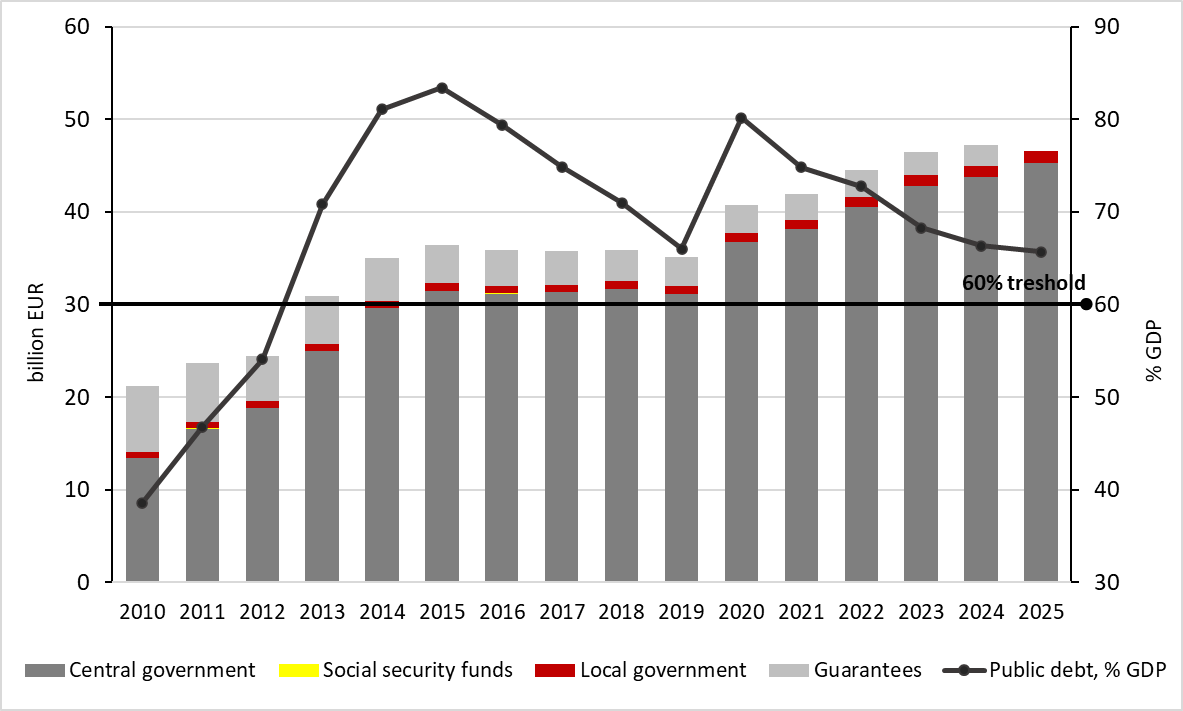}
    \end{minipage}

    \bigskip \bigskip

    \begin{minipage}{0.8\textwidth} \centering
        {\footnotesize\textbf{b)} General government deficit, debt change, and SFA}\par\vspace{4pt}
        \includegraphics[width=\textwidth]{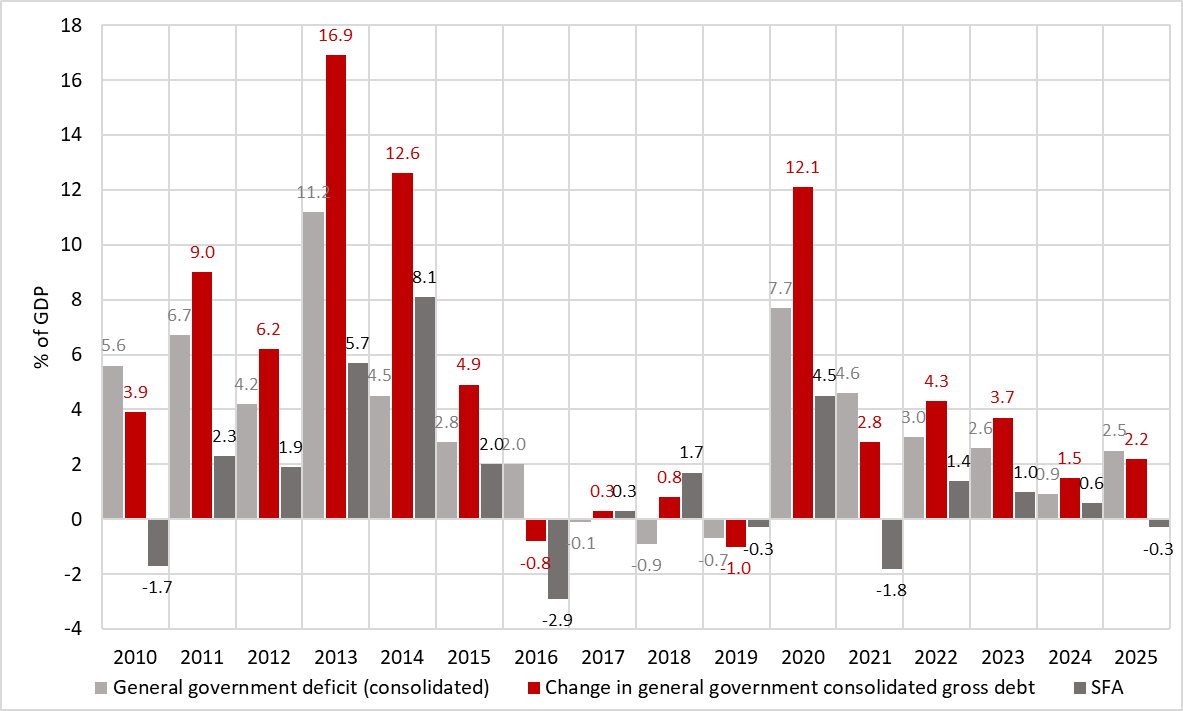}
    \end{minipage}
\caption*{\footnotesize \textit{Sources: Eurostat, Slovenian Statistical Office and Slovenian Ministry of Finance. Data on Slovenian guarantees for 2025 were unavailable at the time of writing.}}

    \label{fig:debt-dynamics-SLO}
\end{figure}

Slovenia's debt-to-GDP ratio followed a broadly similar trajectory over the period, as shown in Fig.~\ref{fig:debt-dynamics-SLO}, panel (a). The ratio rose sharply from 39 per cent in 2010 to 83 per cent by 2015, driven by two compounding factors. Between 2008 and 2013 Slovenia lost more than 9 per cent of GDP, one of the largest contractions among euro area countries \parencite{ECCPSLO}. At the same time, bank recapitalisation required an estimated \texteuro 4.8 billion, equivalent to around 13 per cent of GDP \parencite{SB13}. A period of recovery and fiscal consolidation followed, gradually bringing the ratio down before the pandemic reversed those gains. The COVID-19 shock produced a sharp spike, after which the ratio started to decline slowly until the August 2023 floods interrupted this process. Fiscal pressures slowed the pace of consolidation. Unlike Croatia, Slovenia's debt ratio has remained above the 60 per cent Maastricht threshold since 2012.

In nominal terms, general government debt rose from \texteuro 13.9 billion in 2010 to \texteuro 46.3 billion by 2025. As in Croatia, central government debt is the dominant component, corresponding to \texteuro 45.3 billion in 2025. Local government debt grew from \texteuro 0.6 billion to \texteuro 1.3 billion, nearly identical to Croatia over the same period. Social security funds contribute negligibly, with official data recording \texteuro 0 since 2021. Government guarantees occupy a somewhat more meaningful position than in Croatia, though they fell substantially from \texteuro 7.4 billion in 2010 to \texteuro 2.3 billion in 2024. 

\begin{figure}[tbp]
    \captionsetup{position=above}
    \caption{Structure and risk profile of the Slovenian public debt portfolio, 2010--2024.}
    \centering

    \begin{minipage}{0.47\textwidth}
        \centering
        {\footnotesize\textbf{a)} Structure by financial instrument}\par\vspace{2pt}
        \includegraphics[width=\textwidth]{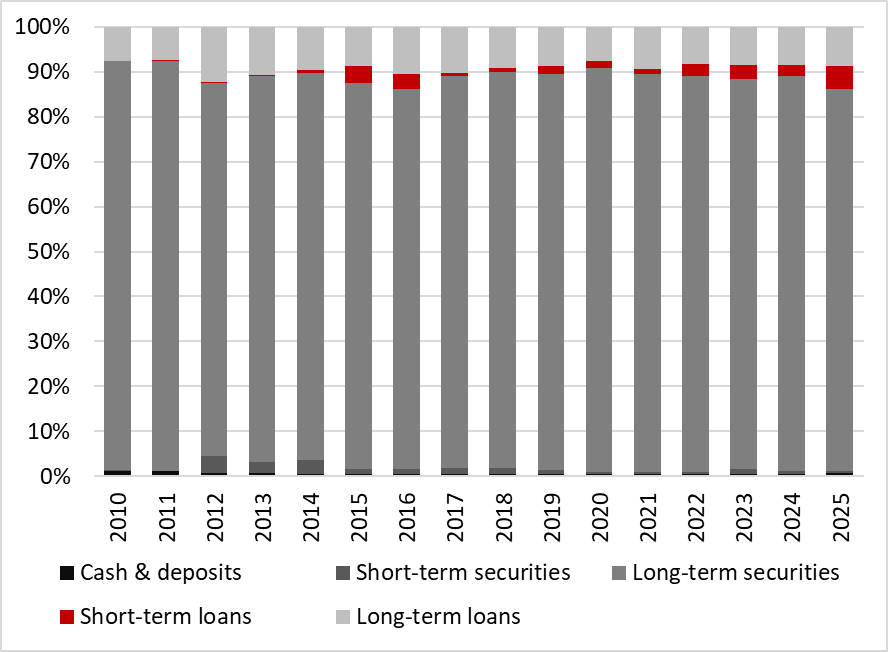}
    \end{minipage}
    \hfill
    \begin{minipage}{0.47\textwidth}
        \centering
        {\footnotesize\textbf{b)} Structure of debt by creditor residency}\par\vspace{2pt}
        \includegraphics[width=\textwidth]{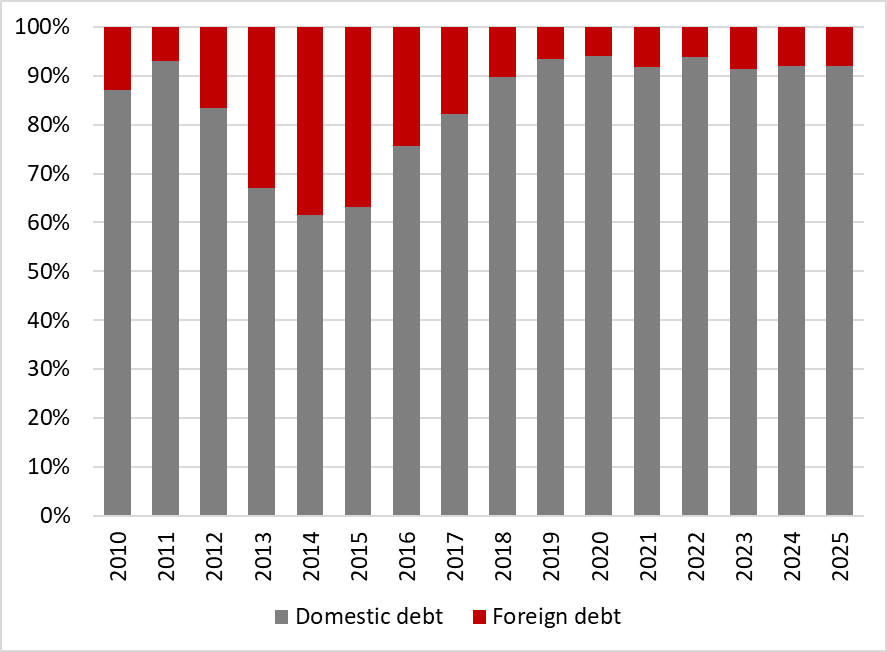}
    \end{minipage}

    \vspace{20pt}

    \begin{minipage}{0.47\textwidth}
        \centering
        {\footnotesize\textbf{d)} Domestic debt by instrument}\par\vspace{2pt}
        \includegraphics[width=\textwidth]{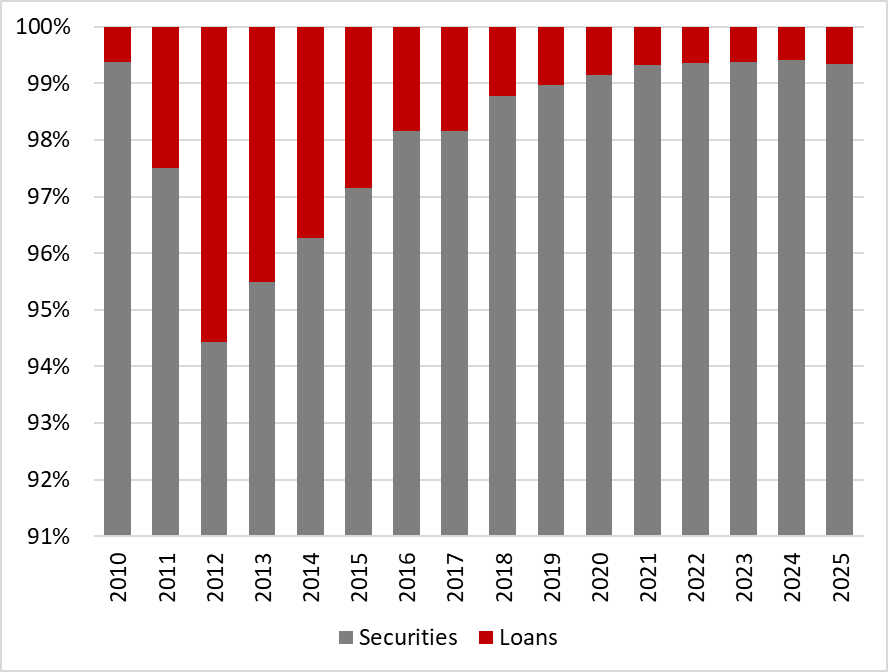}
    \end{minipage}
    \hfill
    \begin{minipage}{0.47\textwidth}
        \centering
        {\footnotesize\textbf{e)} Foreign debt by instrument}\par\vspace{2pt}
        \includegraphics[width=\textwidth]{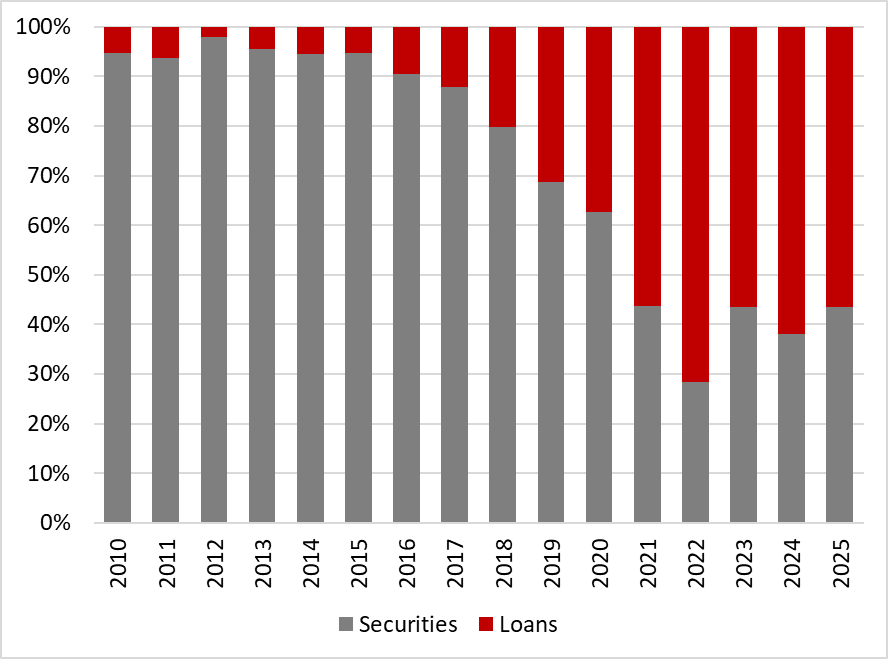}
    \end{minipage}

    \caption*{\footnotesize \textit{Sources: Eurostat, Slovenian Statistical Office and Slovenian Ministry of Finance.}}
    \label{fig:debt-structure-SLO}
\end{figure}

Slovenia's debt structure is broadly similar to Croatia's in terms of financial instruments. Long-term securities dominate, with long-term loans a distant second. Short-term loans and securities account for a small share of the total, as shown in detail in Fig.~\ref{fig:debt-structure-SLO}, panel (a). The main difference between the two countries lies in the domestic-foreign composition. Slovenian public debt has been predominantly domestic for most of the observed period, staying above 90 per cent except for a brief rise in foreign debt in the mid-2010s, as shown in Fig.~\ref{fig:debt-structure-SLO}, panel (b). Croatia followed a similar path but started later, and the shift is still ongoing. On the domestic side, securities are the primary financing instrument throughout, with loans rising only briefly between 2012 and 2015 by 3 to 4 percentage points. On the external side, securities dominated foreign debt until 2017, after which loan-financed borrowing grew steadily and overtook securities in 2021, following the pandemic.

Croatia and Slovenia are both small open post-transition economies and EU members. Since Croatia adopted the euro in 2023, both operate within the euro area and share the same constraint of having only fiscal policy available to absorb large shocks. Both economies have shown vulnerability to external shocks, from the financial crisis to the COVID-19 pandemic, as well as to severe natural disasters, the 2020 earthquakes in Croatia and the 2023 floods in Slovenia. Their broadly similar debt compositions and debt trajectories make their fiscal reactions to such events directly comparable.

\subsection{Natural disasters}

Although Croatia and Slovenia are not among the most disaster-prone countries globally, both have experienced significant natural disasters in recent decades. Table \ref{tab:disasters_cro} summarises major events with measurable economic and human impacts on Croatia. The most severe were the two earthquakes of 2020. The Zagreb earthquake in March caused damage estimated at around 11.7 per cent of GDP and directly affected roughly 2 per cent of the population. Nine months later, the Petrinja earthquake caused damage equivalent to about 10.7 per cent of GDP, affecting nearly 3.7 per cent of the population. Other disasters include floods, storms, droughts, and extreme temperatures. A major storm in the Zadar region in 2017 caused damage equal to 0.3 per cent of GDP, while an extreme heat wave in 2000 cost an estimated 1.1 per cent of GDP. Floods, such as those of 2010 and 2014, and periodic droughts have had more modest nationwide effects, though locally they can be highly destructive. The 2014 river flooding in Vukovar--Srijem county is a case in point.

\begin{table}[tbp]
\caption{Disasters in Croatia with measurable GDP and population impact.}
\label{tab:disasters_cro}
\resizebox{\textwidth}{!}{%
\begin{tabular}{ccp{6.5cm}cc}
\toprule
\textbf{Year} & \textbf{Disaster Type} & \textbf{Location} & \textbf{Damage to GDP (\%)} & \textbf{Affected Population (\%)} \\
\midrule
2023 & Storm & Brod-Posavina and Vukovar-Srijem Counties; Vinkovci & 0.00 & 0.16 \\
2020 & Earthquake & Zagreb & 11.69 & 1.95 \\
2020 & Earthquake & Sisak, Petrinja, Glina, Hrvatska Kostajnica, Zagreb county, Karlovac county & 10.70 & 3.69 \\
2017 & Storm & Zadar region, Bibinje, Sukošan, Biograd na Moru, Nin & 0.29 & 0.08 \\
2010 & Flood & Slavonski Brod, Vinkovci & 0.14 & 0.01 \\
2014 & Flood & Vukovar-Srijem County & 0.00 & 0.17 \\
2005 & Wildfire & Dubrovnik-Neretva, Lika-Senj & 0.06 & -- \\
2003 & Drought & Nationwide (20 counties) & 0.94 & -- \\
2000 & Extreme temperature & Zagreb, Split, Osijek, Rijeka & 1.08 & 0.01 \\
\bottomrule
\end{tabular}
} \scriptsize{Source: EM-DAT, The International Disaster Database (2025).}
\end{table}

Recent natural disasters in Slovenia are summarised in Table \ref{tab:disasters_slo}. The most severe were the 2023 floods, which affected almost 80 per cent of the population and caused damage equivalent to 0.72 per cent of GDP. Other events are comparatively smaller, ranging from extreme weather to earthquakes. Slovenia appears more prone to floods and storms than Croatia, and its seismic record is limited to a single earthquake in 2004 whose impact was negligible compared to the Croatian events of 2020.

\begin{table}[tbp]
\caption{Disasters in Slovenia with measurable GDP and population impact.}
\label{tab:disasters_slo}
\resizebox{\textwidth}{!}{%
\begin{tabular}{ccp{6.5cm}cc}
\toprule
\textbf{Year} & \textbf{Disaster Type} & \textbf{Location} & \textbf{Damage to GDP (\%)} & \textbf{Affected Population (\%)} \\
\midrule
2023 & Flood & Gorenjska, Goriska, Jugovzodna Slovenija, Koroska, Osrednjeslovenska, Podravska, Pomurska, Savinjska, Spodnjeposavska, Zasavska & 0.72 & 70.86 \\
2023 & Storm & Bled, Crnomelj, Savinja region & -- & 0.14 \\
2021 & Flood & Ljubljana & -- & 0.07 \\
2014 & Flood & Vranjsko municipality & -- & 0.12 \\
2014 & Extreme temperature & Notranjsko-kraska province & -- & 2.43 \\
2012 & Flood & Duplek, Dravograd, Bohinj, Kranj districts & 0.57 & 0.58 \\
2007 & Storm & Zelezniki, Skofja Loka, Cerklje na Gorenjskem, Cerkno, Velenje, Celje districts & 0.61 & 0.05 \\
2007 & Storm & Nationwide & 0.21 & -- \\
2004 & Earthquake & Kobarid, Bovec districts & 0.03 & 0.03 \\
2003 & Extreme temperature & Nationwide & 0.27 & 0.01 \\
\bottomrule
\end{tabular}
} \scriptsize{Source: EM-DAT, The International Disaster Database (2025).}
\end{table}


Croatia's experience in 2020 showed that a major disaster can coincide with other crises, in this case, the COVID-19 emergency, compounding fiscal stress. This is part of the reason why incorporating disaster scenarios into fiscal planning matters. In what follows, the 2020 earthquakes serve as the reference point for the shock magnitude applied in the Croatian stress test, and the 2023 floods serve the same role for Slovenia. For Croatia, we calibrate a scenario to mirror the fiscal and macroeconomic fallout of a disaster comparable to the 2020 earthquakes; for Slovenia, we construct a shock akin to the 2023 floods. We then examine what these shocks imply for fiscal sustainability.

\section{Methodology}

This study employs a debt-dynamics framework based on the IMF's Natural Disaster Debt Dynamic Tool (ND-DDT) approach \parencite{Acosta2025}. The ND-DDT framework evaluates a country's capacity to service debt under a baseline scenario and alternative stress scenarios. It projects the trajectory of the public debt-to-GDP ratio using projections of key macro-fiscal variables and policy targets. In our application to Croatia and Slovenia, we construct two scenarios: (a) baseline scenario reflecting expected macroeconomic conditions and current policies, and (b) natural-disaster shock scenario to assess how a severe adverse event would alter debt dynamics. The disaster scenario combines four complementary approaches, two empirical-distribution methods and two econometric methods, described in turn below.

The empirical-distribution component consists of two configurations. The first, called the \textit{distribution for period~$t$}, introduces a one-off shock in the initial year, capturing the immediate macro-fiscal effects of a single severe event. The second, referred to as the \textit{distribution for each period}, allows shocks to occur in every projection year, with magnitudes drawn from the empirical distribution of historical disaster impacts. Both are calibrated at the 5th percentile, corresponding to the worst 5 per cent of observed disaster outcomes. This makes the simulated scenario a rare but plausible tail-risk event rather than a central estimate. Together, the two configurations trace how disaster-induced shocks to real GDP growth and the primary balance propagate through debt dynamics.

The econometric component extends this framework to capture structural and cross-country dimensions of disaster impacts. The first method is the \textit{local projection} approach of \textcite{Jorda2005}, which estimates the dynamic multi-period response of macro-fiscal variables to a disaster event, evaluated at country-specific values of disaster severity, pre-disaster fiscal balance, adaptive capacity, income group, and disaster type. The second, \textit{quantile regression} \parencite{KoenkerBassett1978}, captures heterogeneous effects across the conditional distribution of outcomes. Focusing on the upper conditional quantile $\tau = 0.95$, it identifies how severe shocks affect countries in high-debt or high-vulnerability states, complementing the average responses estimated by local projections.

\subsection{Baseline debt dynamics}

The evolution of the debt-to-GDP ratio $d_t$ is governed by the standard debt accounting equation \parencite{IMF2013, Acosta2021}:

\begin{equation}
\Delta d_t = d_t - d_{t-1}
= \frac{i_t - (1+g_t)\pi_t}{(1+g_t)(1+\pi_t)} d_{t-1}
- \frac{g_t}{(1+g_t)(1+\pi_t)} d_{t-1}
- pb_t + of_t,
\label{eq:debt}
\end{equation}
where $i_t$ is the average nominal interest rate on debt, $g_t$ is real GDP growth, $\pi_t$ is the GDP deflator inflation rate, $pb_t$ is the primary budget balance, $of_t$ represents other debt-creating flows such as stock-flow adjustments or statistical discrepancies, and $\Delta$ is the difference operator. Notice that $pb_t>0$ denotes a surplus while $pb_t<0$ denotes a deficit. Since Croatia has no public debt denominated in a currency other than the euro, exchange-rate terms do not appear in the equation. Equation~\eqref{eq:debt} states that changes in the debt ratio are driven by four components: (a) interest payments on existing debt, (b) the dilution effect of real GDP growth and inflation, (c) the fiscal stance captured by the primary balance, and (d) other debt-creating flows. 

The baseline scenarios for Croatia and Slovenia use macro-fiscal projections from official and international sources. Real GDP growth, inflation, and interest rate assumptions are taken from the IMF's \textit{World Economic Outlook} and national forecasts, reflecting expectations of moderate growth and stable financing conditions. Fiscal projections (revenues, expenditures, and the primary balance) are based on each country's medium-term fiscal plan and IMF assessments, assuming gradual fiscal consolidation. The baseline therefore represents a continuation of current policies under normal macroeconomic conditions.

\subsection{Natural-disaster shock calibration}

To assess the impact of a natural disaster, we introduce exogenous shocks that affect real GDP growth ($g_t$) and the primary balance ($pb_t$) in the two-channel headline specification. The other macro-fiscal variables are held at their baseline levels; in Section~\ref{sec:robustness} and Appendix~\ref{AppendixA} we extend the analysis to a four-channel specification that also allows the GDP deflator and the effective interest rate to respond. The shock magnitudes and persistence are obtained from the cross-country panel coefficients of \textcite{NguyenFengGarciaEscribano2025}, which underpin the ND-DDT regression sheets, evaluated at country-specific values of disaster severity, pre-disaster overall fiscal balance, adaptive capacity, income group and disaster type. For Croatia, the calibration is anchored to the March 2020 Zagreb earthquake (EM-DAT event ID 12676; damage 11.74 per cent of GDP; population affected 2.02 per cent; pre-disaster overall fiscal balance $-1.10$ per cent of GDP; ND-GAIN adaptive ability 0.438). For Slovenia, the calibration is anchored to the August 2023 floods (EM-DAT event ID 14100; damage 0.72 per cent of GDP; population affected 70.86 per cent; pre-disaster overall fiscal balance $-2.35$ per cent of GDP; ND-GAIN adaptive ability 0.331). All country-specific calibration parameters are reported in Table~\ref{tab:calibration_cro_slo}. The disaster onset is set to $t = 2026$ in both cases.

Formally, the disaster shock is implemented as an exogenous deviation from the baseline projections starting in 2026, affecting only the real activity and fiscal balance channels in the headline specification. Let $s_h$ denote the deviation applied at horizon $h$ relative to the baseline projection. The shock-adjusted variables are thus expressed as:

\begin{equation}
\begin{aligned}
g_{t+h} &= g^{\text{base}}_{t+h} + s^{g}_h, \\
pb_{t+h} &= pb^{\text{base}}_{t+h} + s^{pb}_h,
\end{aligned}
\qquad h = 0,1,\ldots,5.
\label{eq:shock}
\end{equation}
For illustration, the vectors below present the output of the \textit{distribution for period~$t$} configuration for Croatia. The model generates a sequence of shocks to real GDP growth and the primary balance corresponding to the 5th percentile of the empirical distribution of earthquake-related impacts:

\begin{equation*}
\mathbf{s_h^{\,g}} = \{-3.0, -12.5, -5.2, -2.2, -1.2, -0.2\} \qquad
\mathbf{s_h^{\,pb}} = \{-2.5, -0.9, 0.1, 0.8, 0.1, 0.3\},
\label{eq:vectors}
\end{equation*}
where each element $s_h^i$ for $i \in \{g, pb\}$ represents the annual deviation from the baseline in percentage points of GDP at horizon $h$. These values describe the temporal propagation of the shock: an initial contraction in output and deterioration in the fiscal balance, followed by a gradual recovery.

\subsection{Empirical-distribution configuration of shocks}

In the empirical-distribution ND-DDT simulations, two configurations are used. The first, \textit{distribution for period~$t$}, applies a one-time shock:

\begin{equation}
Shock_{t+h} =
\begin{cases}
s_h, & h = 0, \\
0, & h > 0,
\end{cases}
\label{eq:onetimeshock}
\end{equation}
while the second, \textit{distribution for each period}, allows for shocks in every projection year:
\begin{equation}
Shock_{t+h} \sim F(\mu,\sigma^2), \quad \forall h \geq 0,
\label{eq:repeatedshock}
\end{equation}
where $F(\mu,\sigma^2)$ denotes the empirical distribution of disaster impacts. Both configurations are calibrated at the \textbf{5th percentile}, corresponding to the lower tail of the empirical distribution of historical macro-fiscal effects. The 5th percentile captures the worst 5 per cent of observed outcomes. They are rare but plausible events with significant macroeconomic and fiscal consequences. This tail calibration ensures that the disaster scenario reflects a severe yet empirically grounded stress event.

\subsection{Econometric analysis of disaster impacts}\label{sec:econometric}

While the empirical-distribution simulations provide a probabilistic range of debt outcomes under stochastic shocks, they do not explicitly capture structural heterogeneity across countries or the moderating role of adaptive capacity. To address this issue, we complement the empirical-distribution simulations with econometric approaches that estimate the propagation of disaster shocks and their dependence on country-specific characteristics.

First, the \textbf{local projection} method \parencite{Jorda2005} is used to estimate the dynamic response of key macro-fiscal variables to a disaster event:

\begin{equation}
y_{i,t+h} - y_{i,t-1} = \alpha_{i,h} + \mu_{t,h} + \beta_h \, \text{ND}_{i,t} +
\gamma_h \big( \text{ND}_{i,t} \times A_{i,t} \big)
+ \delta_h' X_{i,t-1} + \varepsilon_{i,t+h},
\label{eq:localproj}
\end{equation}
where $\text{ND}_{i,t}$ is a binary indicator equal to one if a large single-year non-overlapping natural disaster (damage exceeding 1 per cent of GDP) occurs in country $i$ in year $t$ and zero otherwise; $A_{i,t}$ is the ND-GAIN adaptive-capacity index; $X_{i,t-1}$ is a vector of lagged controls including the lagged dependent variable, the lagged fiscal balance, country-specific commodity-price shocks at $t-1$, $t$ and $t+1$, and additional disaster-characteristic interactions (damage, disaster-type and income-group dummies). The parameter $\alpha_{i,h}$ represents country fixed effects and $\mu_{t,h}$ year fixed effects. The coefficient $\beta_h$ traces the impulse response of variable $y$ at horizon $h$, $\gamma_h$ captures how the response is moderated by adaptive capacity, and $\delta_h'$ is the vector of coefficients on the controls. Finally, $\varepsilon_{i,t+h}$ is the error term. Eq.~\eqref{eq:localproj} corresponds to the heterogeneous-effects local-projection specification of \textcite{NguyenFengGarciaEscribano2025}, Tables 8 and 9.

Second, \textbf{quantile regression} analysis \parencite{KoenkerBassett1978} is used to examine heterogeneous effects across the conditional distribution of outcomes:
\begin{equation}
Q_{\tau}(y_{i,t+h} \mid X_{i,t}) = \alpha_{\tau, i,h} + \mu_{\tau, t,h} + \beta_\tau \, \text{ND}_{i,t} +
\gamma_\tau (\text{ND}_{i,t} \times A_{i,t}) +
\delta_\tau' X_{i,t-1},
\label{eq:quantile}
\end{equation}
where $Q_{\tau}(y \mid X)$ denotes the $\tau$-th conditional quantile of the dependent variable. Estimation at the upper quantile $\tau = 0.95$ captures high-risk states such as periods of elevated debt or weak fiscal positions. Coefficients $\alpha_{\tau, i,h}$, $\mu_{\tau, t,h}$, $\beta_\tau$, $\gamma_\tau$, and $\delta_\tau'$ have an interpretation analogous to the linear case.

\paragraph{Panel estimation versus country-specific evaluation.} It is important to distinguish the estimation step from the evaluation step. The coefficients $\beta_h, \gamma_h$ in Eq~\eqref{eq:localproj} and $\beta_\tau, \gamma_\tau$ in Eq.~\eqref{eq:quantile} are estimated by \textcite{NguyenFengGarciaEscribano2025} on a panel of 172 IMF member countries over 1980--2019, with country fixed effects, year fixed effects and country-specific commodity-price-shock controls. The Croatia- and Slovenia-specific impulse responses reported in Sections~\ref{sec:lp} and~\ref{sec:quantile} are then obtained by evaluating these panel coefficients at country-specific values of disaster damage, affected population, pre-disaster fiscal balance, ND-GAIN adaptive capacity, income-group dummies and disaster-type dummies, taken from Table~\ref{tab:calibration_cro_slo}. We do not re-estimate the panel; the country-specific dimension enters only through this evaluation step. Following the recent guidance of \textcite{OleaPlagborgQianWolf2025} on the use of local projections versus VARs, we use the LP estimator for impulse responses at horizons $h = 0, 1, 2$ and report quantile-regression results as a complementary distributional stress test.

\paragraph{Confidence bands and standard errors.} The local-projection coefficients in the ND-DDT engine reproduce the heterogeneous-effects regressions of \textcite{NguyenFengGarciaEscribano2025}, Tables 8 and 9. The standard errors are heteroskedasticity-robust and clustered at the country level, estimated using the Stata command \texttt{xtreg ..., fe vce(cluster country)} (confirmed by H.~M.~Nguyen in personal communication, 2026). Confidence bands in our figures are reported at the 95 per cent level (two-sided, $z = 1.96$), reflecting the active confidence-interval selector in the ND-DDT scenario files (cell \texttt{Shocks!L112} set to \texttt{95}). Whereas \textcite{NguyenFengGarciaEscribano2025} display 90 per cent bands in their working-paper figures, we set the level to 95 per cent in all our scenarios for a more conservative inference on the headline GDP response. For the primary balance, we report point estimates, consistent with the active configuration in the ND-DDT scenario files used for the quantile-regression and empirical-distribution simulations.

\subsection{Stochastic debt simulations}

Independently of the ND-DDT empirical-distribution scenarios described above, we additionally evaluate the uncertainty around debt outcomes through Monte Carlo simulations. The vector of shocks to key macro-fiscal variables $Z_t = (g_t, i_t, \pi_t, pb_t)$ is drawn from a multivariate normal distribution:
\[
Z_t \sim \mathcal{N}(\mu, \Sigma),
\]
where $\mu$ denotes historical means and $\Sigma$ the covariance matrix estimated from past data. Following the approach of \textcite{Celasun2007}, Monte Carlo simulations (10{,}000 iterations) produce a fan chart of debt paths, illustrating the range of possible debt outcomes under macro-fiscal uncertainty and situating the disaster scenarios within that probabilistic context. This stochastic fan chart is conceptually distinct from the empirical-distribution disaster scenarios of Section~3.3. It is based on a Gaussian approximation of historical macro-fiscal variation, whereas the empirical-distribution scenarios draw directly from the historical distribution of disaster impacts.

\subsection{Data}

The analysis covers 2015 to 2030, combining historical data (2015--2024) with projections (2025--2030). All debt indicators refer to general government gross debt as a percentage of GDP. Croatian historical debt figures are sourced from the Croatian National Bank and Ministry of Finance, aligned with Eurostat definitions. GDP and fiscal variables, including real GDP growth, the GDP deflator, and budget balances, come from the Croatian National Bank (CNB), the Croatian Bureau of Statistics, and the IMF databases. Slovenian data come from the Statistical Office of the Republic of Slovenia, the same international sources already listed, and the Federal Reserve Economic Data.

Table~\ref{tab:variables} lists the key variables and their sources. The Croatian nominal effective interest rate on public debt is computed from Ministry of Finance data on government securities, using a weighted average of interest rates on outstanding debt instruments. All historical interest rates and debt figures originally in Croatian kuna have been converted to euros at the fixed exchange rate established upon Croatia's entry into the euro area in 2023. For Slovenia, the nominal effective interest rate on public debt is proxied using long-term government bond yields for Slovenia, the Euro Area (19), and the United States.

Croatian real GDP growth rates come from the CNB and Eurostat, with inflation measured by the GDP deflator from the World Development Indicators. The primary balance, expressed as a percentage of GDP, uses IMF and Ministry of Finance data; for the projection years, we incorporate targets from Croatia's Medium-Term Fiscal Plan, extended to 2029. Slovenian macroeconomic data come from the IMF and the European Central Bank.

For other debt-creating flows, we use Eurostat's reported SFA for past years, covering items such as the net accumulation of financial assets, payments on called guarantees, and other transactions not reflected in the deficit. For the projection years, we assume these flows continue at roughly their historical average, adding a small positive increment to debt each year. This is a simplifying assumption given uncertainty about one-off operations. It rules out major privatisations, one-off debt reductions, and large extra-budgetary borrowing beyond what recent trends suggest.

All projections from 2025 onward assume moderate growth and low interest rates. Real GDP is expected to grow at around 3 to 4 per cent in 2024--2025 as the post-pandemic rebound continues, stabilising at around 2.5 per cent annually for Croatia and 2.1 per cent for Slovenia in the late 2020s. The GDP deflator is projected to ease to around 2 per cent for both countries, consistent with euro area price stability. Interest costs are expected to rise gradually as financial conditions normalise, though Croatia's euro adoption and improved credit ratings help contain borrowing costs, and Slovenia already held investment-grade ratings entering the period. The primary balance for Croatia is roughly in equilibrium in the baseline, with small deficits below 1 per cent of GDP in the mid-2020s approaching zero by 2028. Deviations from these assumptions under stress are outlined in the next section.

\begin{table}[tbp]
   \caption{List of Variables, Definitions, and Sources}
   \centering
   \renewcommand{\arraystretch}{1.3}
   \scriptsize
   \begin{tabularx}{\textwidth}{l l c c >{\raggedright\arraybackslash}X >{\raggedright\arraybackslash}X}
        \hline
        \textbf{Variable} & \textbf{Description} & \textbf{Period} & \textbf{Unit} & \textbf{Croatia} & \textbf{Slovenia}\\
        \hline
        \textbf{Historical data} \\
        \hline
        $d_t$ & Gross public debt & 2015--2024 & \% GDP & CNB, General government debt & SiStat\\
        $i^d_t$ & Nominal effective interest rate & 2015--2024 & \% & Ministry of Finance, bond issue data & Federal Reserve Economic Data \& SiStat\\
        $\pi_t$ & GDP deflator inflation & 2015--2024 & \% & World Bank, WDI & IMF database\\
        $g_t$ & Real GDP growth & 2015--2024 & \% & CNB, macroeconomic indicators & IMF database\\
        $pb_t$ & Primary balance & 2015--2024 & \% GDP & IMF database & IMF database\\
        $of_t$ & Other net debt-creating flows & 2015--2024 & \% GDP & Eurostat SFA & Eurostat SFA\\
        \hline
        \textbf{Projection data} \\
        \hline
        $i^d_t$ & Nominal effective interest rate & 2025--2029 & \% & IMF database & IMF database\\
        $\pi_t$ & GDP deflator inflation & 2025--2029 & \% & IMF database & IMF database\\
        $g_t$ & Real GDP growth & 2025--2029 & \% & IMF database & IMF database\\
        $pb_t$ & Primary balance & 2025--2028 & \% GDP & National Medium-term Fiscal Plan 2025--2028 & IMF database\\
        $of_t$ & Other net debt-creating flows & 2025--2028 & \% GDP & Average of previous years & Average of previous years\\
        \hline
   \end{tabularx}
   \label{tab:variables}
\end{table}

\section{Results}\label{sec:results}

\subsection{Baseline Debt Dynamics}\label{sec:baseline}

The baseline projections reveal a clear divergence between Croatia and Slovenia before any natural disaster shock is introduced. Although both are small open euro-area economies with similar post-transition characteristics, their projected debt paths differ substantially. This matters because the fiscal impact of the same class of adverse shock depends heavily on the starting trajectory. A country entering the shock with declining debt has more room to absorb temporary deterioration, while one entering from a rising path faces a higher risk of debt persistence. 

For Croatia, the debt-to-GDP ratio declines steadily over the projection horizon, falling from 63.04 per cent in 2024 to 58.76 per cent in 2025 and 52.70 per cent by 2034. The baseline rests on moderate real GDP growth, projected at 2.6 per cent from 2030 onwards, GDP deflator inflation of 2.22 per cent, an effective interest rate of 2.86 per cent, and a primary balance close to zero. Under these conditions, the interest-growth differential stays favourable, and the primary balance does not lead to persistent debt accumulation. The euro-denominated debt stock and the high share of fixed-rate public debt further limit financing risk. 
In the case of Slovenia, the baseline points in the opposite direction. Public debt stands at 66.6 per cent of GDP in 2024, dips slightly to 65.9 per cent in 2025, and then rises steadily to 74.71 per cent by 2034. The increase reflects a projected primary deficit averaging around -1.79 per cent of GDP over 2026--2034. Real GDP growth is at around 2.1 per cent, lower than Croatia, while the effective interest rate is broadly comparable at 2.8 per cent in the later projection years. The interest-growth differential is therefore less supportive, and the persistent primary deficit pushes the debt ratio upward.

\begin{figure}[tbp]
\centering
\caption{Individual Contributions to Public Debt, 2017--2031}
\label{fig:dcf-baseline}

\begin{subfigure}[t]{0.9\textwidth}
\centering
\caption{Croatia}
\label{fig:debt_contr_cro}
\includegraphics[width=\textwidth]{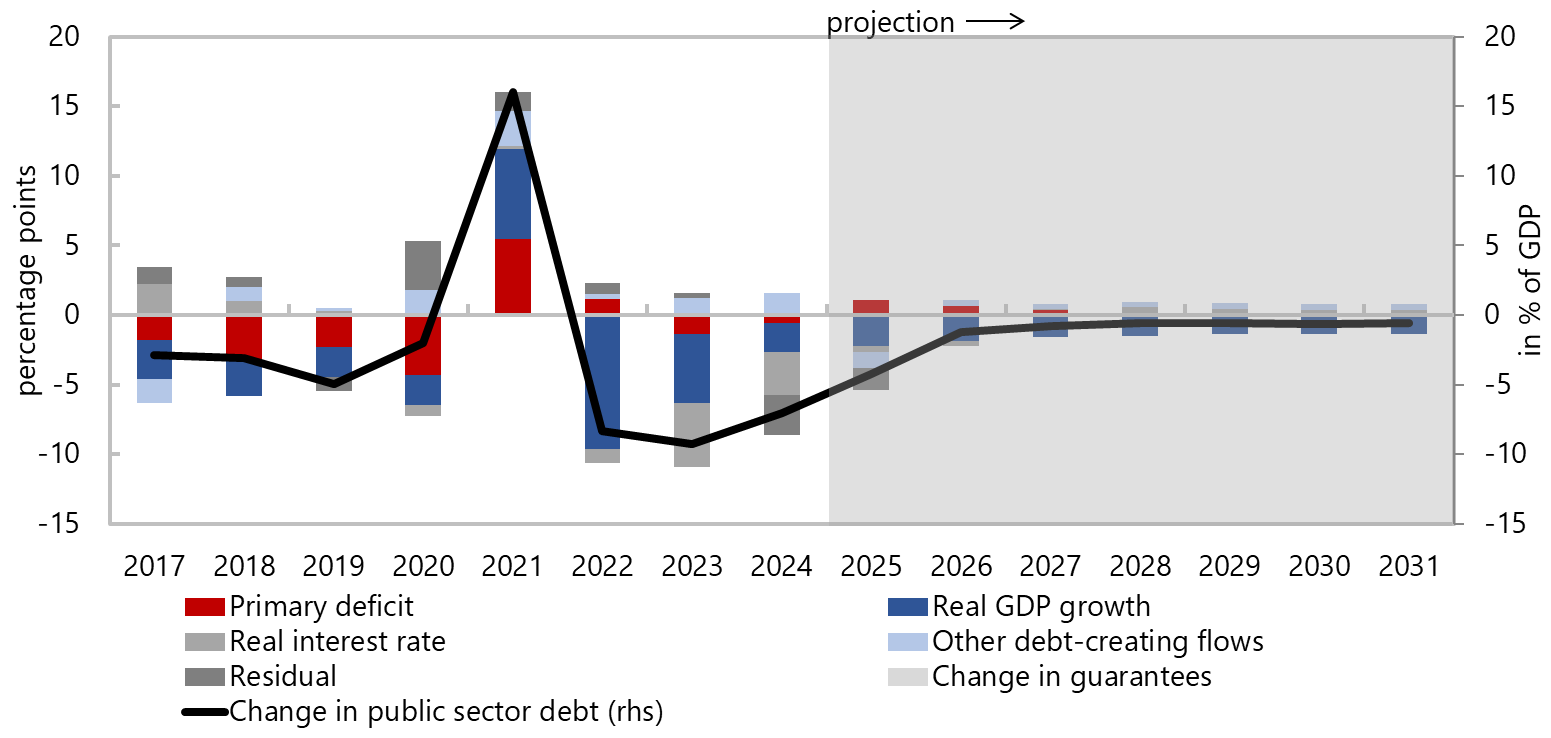}
\end{subfigure}

\vspace{0.5cm}

\begin{subfigure}[t]{0.9\textwidth}
\centering
\caption{Slovenia}
\label{fig:debt_contr_slo}
\includegraphics[width=\textwidth]{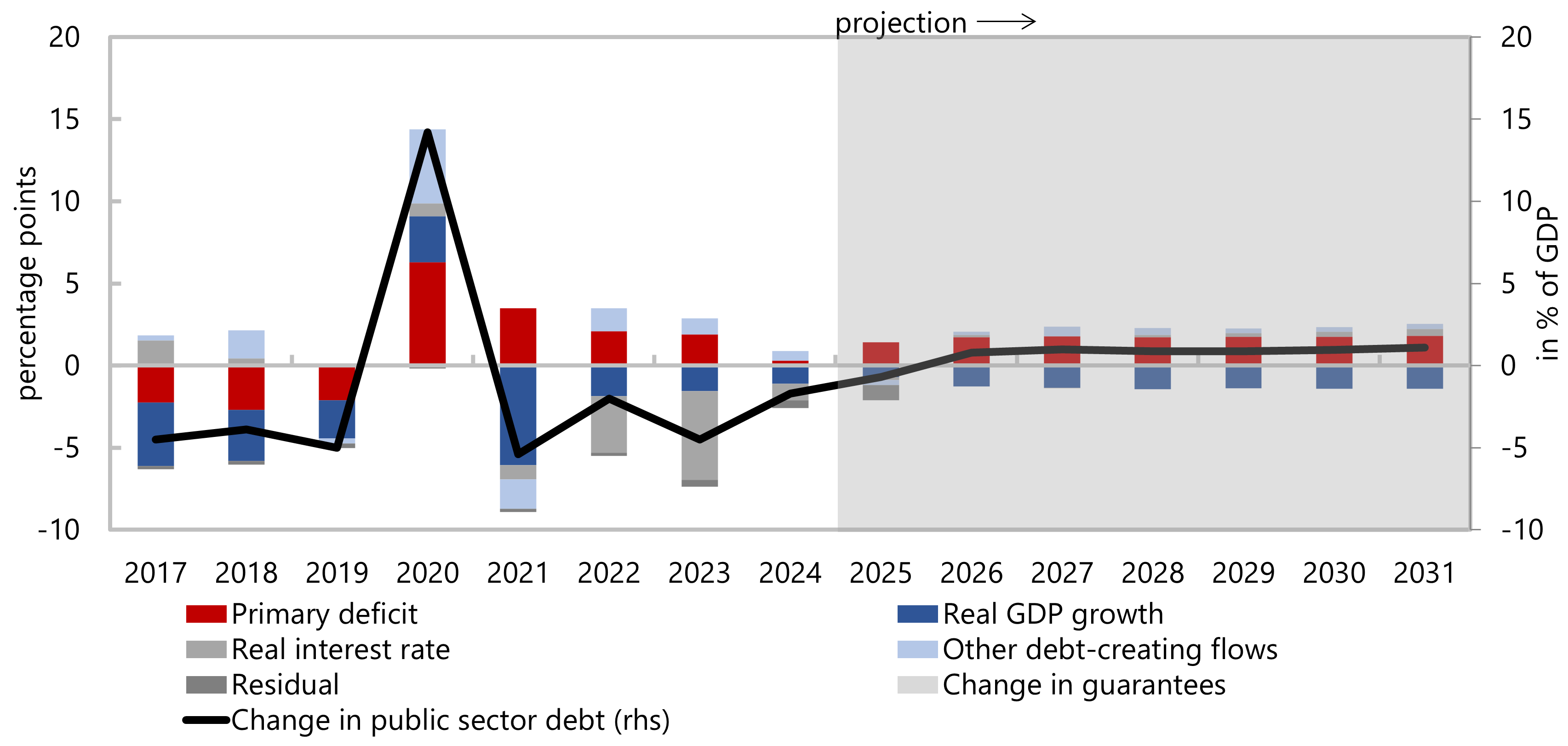}
\end{subfigure}

\caption*{\footnotesize \textit{Source:} Authors' calculation.}
\end{figure}

\paragraph{Debt decomposition.} Fig.~\ref{fig:dcf-baseline} decomposes the annual change in the debt ratio into its main components. In Croatia, real GDP growth provides a sizeable and persistent debt-reducing contribution, ranging between roughly  $-1.4$ and $-1.9$ percentage points per year over 2026--2031. It more than offsets the modest positive contribution of the real interest rate and other debt-creating flows. In Slovenia, growth also reduces the debt ratio, but by a smaller margin. The primary deficit is the dominant source of upward pressure, contributing around $+1.7$ to $+1.8$ percentage points per year. Other identified flows are more variable in the early projection years, though they do not change the main conclusion. Slovenia's rising baseline reflects a persistent fiscal gap.

\begin{figure}[tbp]
\centering
\caption{Public Sector Debt Dynamics, Baseline Scenario, 2024--2040}
\label{fig:fanchart}

\begin{subfigure}[t]{0.9\textwidth}
\centering
\caption{Croatia}
\label{fig:baseline_cro}
\includegraphics[width=\textwidth]{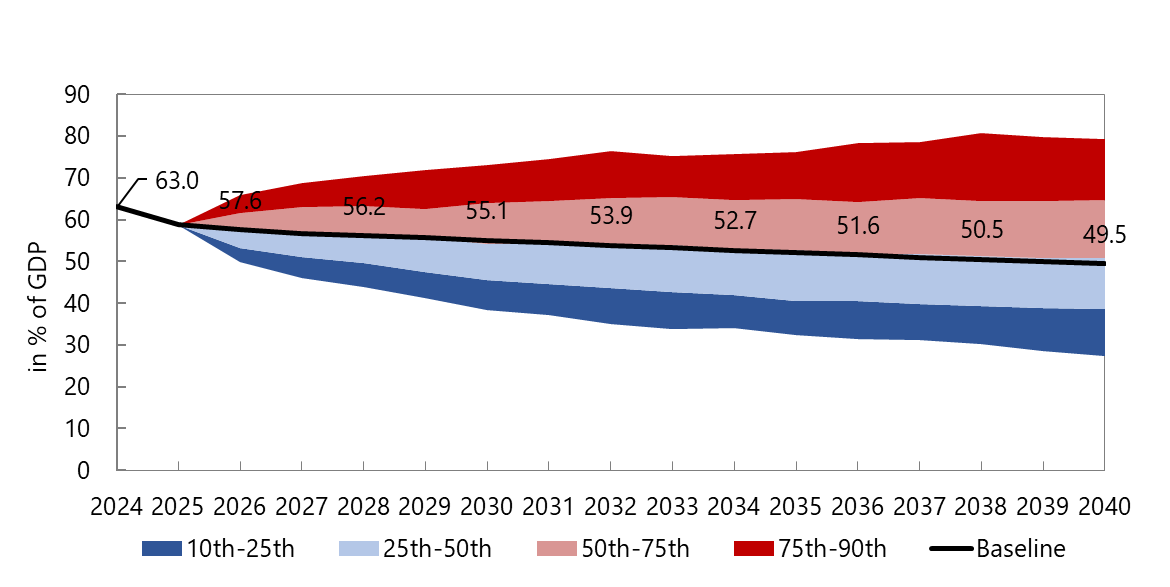}
\end{subfigure}

\vspace{0.5cm}

\begin{subfigure}[t]{0.9\textwidth}
\centering
\caption{Slovenia}
\label{fig:baseline_slo}
\includegraphics[width=\textwidth]{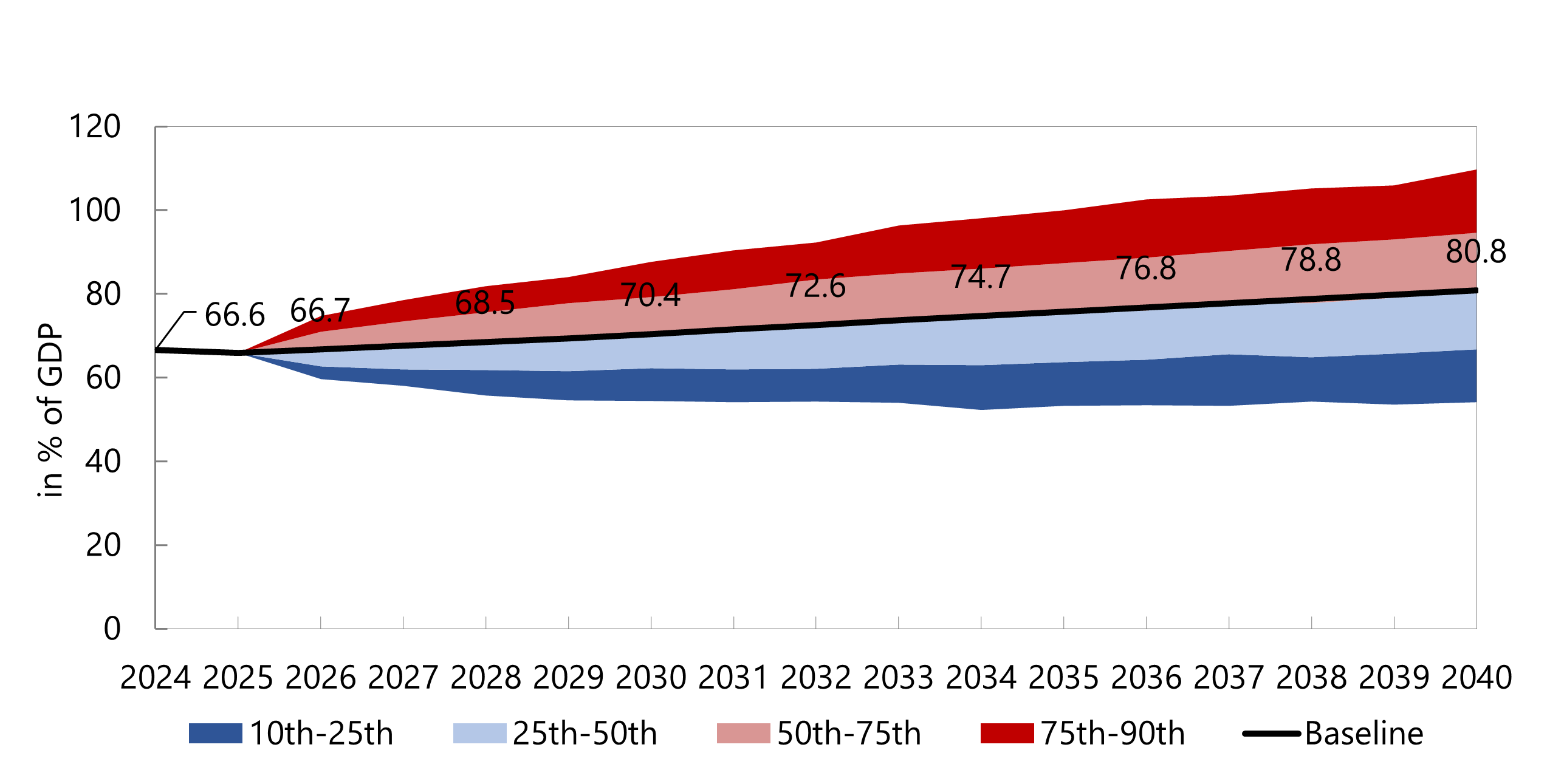}
\end{subfigure}

\caption*{\footnotesize \textit{Source:} Authors' calculation.}
\end{figure}

\paragraph{Probabilistic uncertainty.} The Monte Carlo fan charts in Fig.~\ref{fig:fanchart}, based on 10{,}000 simulations, highlight the difference between the two countries. The central simulated path places Croatian public debt at 55.1 per cent of GDP in 2030 and 49.5 per cent in 2040. For Slovenia, the corresponding figures are 70.4 per cent in 2030 and 80.8 per cent in 2040. Slovenia has a negligible probability of staying below the 60 per cent threshold over the projection horizon, while for Croatia, a path below that reference value is the central outcome.


\subsection{Debt Effects of Disaster Shocks under the Local Projection Scenario}\label{sec:lp}

We proceed by introducing natural disaster shocks. We assess their effect on public debt, beginning with the local-projection scenario. For Croatia, the scenario is calibrated to the March 2020 Zagreb earthquake (EM-DAT event ID 12676), associated with damage of 11.741 per cent of GDP, 2.017 per cent of the population affected, a pre-disaster fiscal balance of $-1.1$ per cent of GDP, and an ND-GAIN adaptive ability score of 0.438.\footnote{The figure of 2.017 per cent of the population affected by the earthquake reflects the EM-DAT methodology, which defines total affected as the sum of the number injured, the number affected, and the number homeless. The number of homeless and affected individuals is derived from reported figures multiplied by average household size where only family or housing counts are available. As a result, the measure captures those who lost their home or sustained injuries, not the broader population that experienced the earthquake. Thus, the 2.017 per cent figure is a lower bound on actual exposure.} For Slovenia, the scenario is calibrated to the August 2023 floods (EM-DAT event ID 14100). The Slovenian event has much lower direct damage at 0.723 per cent of GDP, but far wider population exposure at 70.862 per cent. Slovenia also enters the shock with a weaker pre-disaster fiscal balance of $-2.354$ per cent of GDP and a lower adaptive ability of 0.331.

\begin{table}[tbp]
\centering
\small
\caption{Country-specific calibration of the natural-disaster scenarios}
\label{tab:calibration_cro_slo}
\begin{tabularx}{\textwidth}{lcc}
\toprule
\textbf{Parameter} & \textbf{Croatia} & \textbf{Slovenia} \\
\midrule
Disaster type & Earthquake & Flood \\
Event year & 2020 & 2023 \\
EM-DAT event ID & 12676 & 14100 \\
Damage, \% of GDP & 11.741 & 0.723 \\
Affected population, \% & 2.017 & 70.862 \\
Pre-disaster overall fiscal balance, \% of GDP & $-1.100$ & $-2.354$ \\
ND-GAIN adaptive ability & 0.438 & 0.331 \\
Income group & Advanced economy & Advanced economy \\
Small island & No & No \\
\bottomrule
\end{tabularx}
\caption*{\footnotesize \textit{Source:} Authors' calculation based on EM-DAT, ND-GAIN and ND-DDT calibration files. The pre-disaster overall fiscal balance corresponds to the ND-DDT input cell \texttt{Options!C28}, defined as the pre-ND fiscal overall balance in per cent of GDP.}
\end{table}

The two events differ in type and exposure profile. Croatia's scenario is driven by large direct physical damage, while Slovenia's is characterised by exceptionally broad population exposure. The framework treats these cases as comparable because the local projection coefficients are evaluated at country-specific values of disaster damage, affected population, pre-disaster fiscal balance, adaptive ability, income group, and disaster type. Each disaster is therefore mapped through the same empirical structure without imposing a common shock size. The implied shocks are expressed as deviations from the no-disaster baseline.

For Croatia, the shock to real GDP growth is
\[
\mathbf{s}^{g}_{CRO} = \{-1.73,\,-1.68,\,-1.26,\,0,\,0,\,0\},
\]
while the shock to the primary balance is
\[
\mathbf{s}^{pb}_{CRO} = \{-2.21,\,-4.20,\,-4.49,\,0,\,0,\,0\}.
\]

For Slovenia, the corresponding vectors are
\[
\mathbf{s}^{g}_{SLO} = \{-2.28,\,-1.01,\,-0.86,\,0,\,0,\,0\},
\]
and
\[
\mathbf{s}^{pb}_{SLO} = \{-2.72,\,-3.04,\,-3.58,\,0,\,0,\,0\}.
\]
These shock vectors are the country-specific evaluations of Eq.~\eqref{eq:localproj} at the covariates reported in Table~\ref{tab:calibration_cro_slo}; the underlying panel coefficients are not re-estimated. The Slovenian GDP shock is larger in the first year because population exposure is much higher, while the Croatian primary-balance shock is more severe at the second and third horizons, consistent with a longer reconstruction-related fiscal deterioration after a major earthquake.

Resulting debt trajectories are reported in Table~\ref{tab:local_projection_debt}. In Croatia, the local projection scenario raises public debt from 58.76 per cent of GDP in 2025 to 69.64 per cent in 2028, after which it declines gradually to 64.67 per cent in 2034. Relative to the baseline, the debt effect peaks at 13.42 percentage points in 2028 and stands 11.97 percentage points above the no-disaster path by 2034. In Slovenia, debt rises from 65.90 per cent in 2025 to 80.45 per cent in 2028 and continues to 85.60 per cent in 2034. The gap relative to the baseline peaks at 11.91 percentage points in 2028 and remains at 10.89 percentage points in 2034.

\begin{table}[tbp]
\centering
\small
\caption{Baseline and local projection disaster debt trajectories, \% of GDP}
\label{tab:local_projection_debt}
\resizebox{\textwidth}{!}{%
\begin{tabular}{lcccccccccc}
\toprule
 & 2025 & 2026 & 2027 & 2028 & 2029 & 2030 & 2031 & 2032 & 2033 & 2034 \\
\midrule
Croatia baseline & 58.76 & 57.56 & 56.78 & 56.22 & 55.68 & 55.06 & 54.45 & 53.86 & 53.27 & 52.70 \\
Croatia local projection & 58.76 & 60.73 & 65.04 & 69.64 & 68.87 & 67.99 & 67.14 & 66.30 & 65.48 & 64.67 \\
\midrule
Slovenia baseline & 65.90 & 66.69 & 67.68 & 68.54 & 69.42 & 70.37 & 71.48 & 72.57 & 73.65 & 74.71 \\
Slovenia local projection & 65.90 & 70.90 & 75.53 & 80.45 & 81.12 & 81.89 & 82.84 & 83.77 & 84.69 & 85.60 \\
\bottomrule
\end{tabular}}
\caption*{\footnotesize \textit{Source:} Authors' calculation from ND-DDT local projection scenario files.}
\end{table}

The decomposition shows that the primary balance is the main channel through which the disaster shock affects public debt. In both countries, the primary deficit adds several percentage points of GDP to the debt ratio in the peak years 2026--2028. Real GDP growth partly offsets this once the immediate output loss fades, while the real interest rate contribution remains small. What happens after the peak differs between the two countries. Croatia's declining baseline allows debt to fall after 2028, whereas Slovenia's rising baseline means debt continues to increase even after the disaster shock itself has faded.

\subsection{Quantile-Regression Scenario}\label{sec:quantile}

Our quantile regression scenario focuses on the upper part of the conditional distribution, estimated at $\tau = 0.95$. This captures states in which countries are more vulnerable to debt accumulation following a disaster, offering a distributional perspective that complements the local projection results and is particularly relevant for fiscal risk analysis. For Croatia, it produces a smaller initial GDP shock than the local projection scenario, with a 2026 deviation of $-0.30$ percentage points, but a larger deterioration in the primary balance, at $-4.03$ percentage points. Public debt reaches 69.40 per cent of GDP in 2028 and then declines to 64.45 per cent in 2034. In Slovenia, the 2026 GDP shock is $-0.67$ percentage points and the primary-balance shock is $-2.51$ percentage points. Debt reaches 77.78 per cent of GDP in 2028 and 83.17 per cent by 2034.


\subsection{Empirical-Distribution Scenarios}\label{sec:dist}

The empirical distribution scenarios use the historical distribution of observed disaster impacts. In the \emph{distribution for period $t$} one, we apply a six-year shock profile once, starting in 2026. It should be read as a severe one-time disaster whose effects unfold over several years. Under this specification, Croatian debt reaches 68.91 per cent of GDP in 2034 and Slovenian debt reaches 88.80 per cent. The \emph{distribution for each period} scenario is more severe, as adverse shocks are repeatedly imposed over the projection horizon. Our estimates show that it raises Croatian debt to 70.44 per cent of GDP and Slovenian debt to 95.02 per cent by 2034.

On the other hand, the repeated-shock configuration is best read as a tail-risk envelope rather than a plausible expected path. It is useful precisely because it shows how debt dynamics respond when the fiscal system faces not one large shock but a sequence of adverse events or persistent reconstruction pressures. The effect is stronger in Slovenia, where repeated shocks compound with a rising baseline. In Croatia, the declining baseline partly absorbs the cumulative deterioration.

\begin{figure}[tbp]
\centering
\caption{Gross Nominal Public Debt in Alternative Disaster Scenarios}
\label{fig:debt_flows}

\begin{subfigure}[t]{0.9\textwidth}
\centering
\caption{Croatia}
\label{fig:debt_flows_cro}
\includegraphics[width=\textwidth]{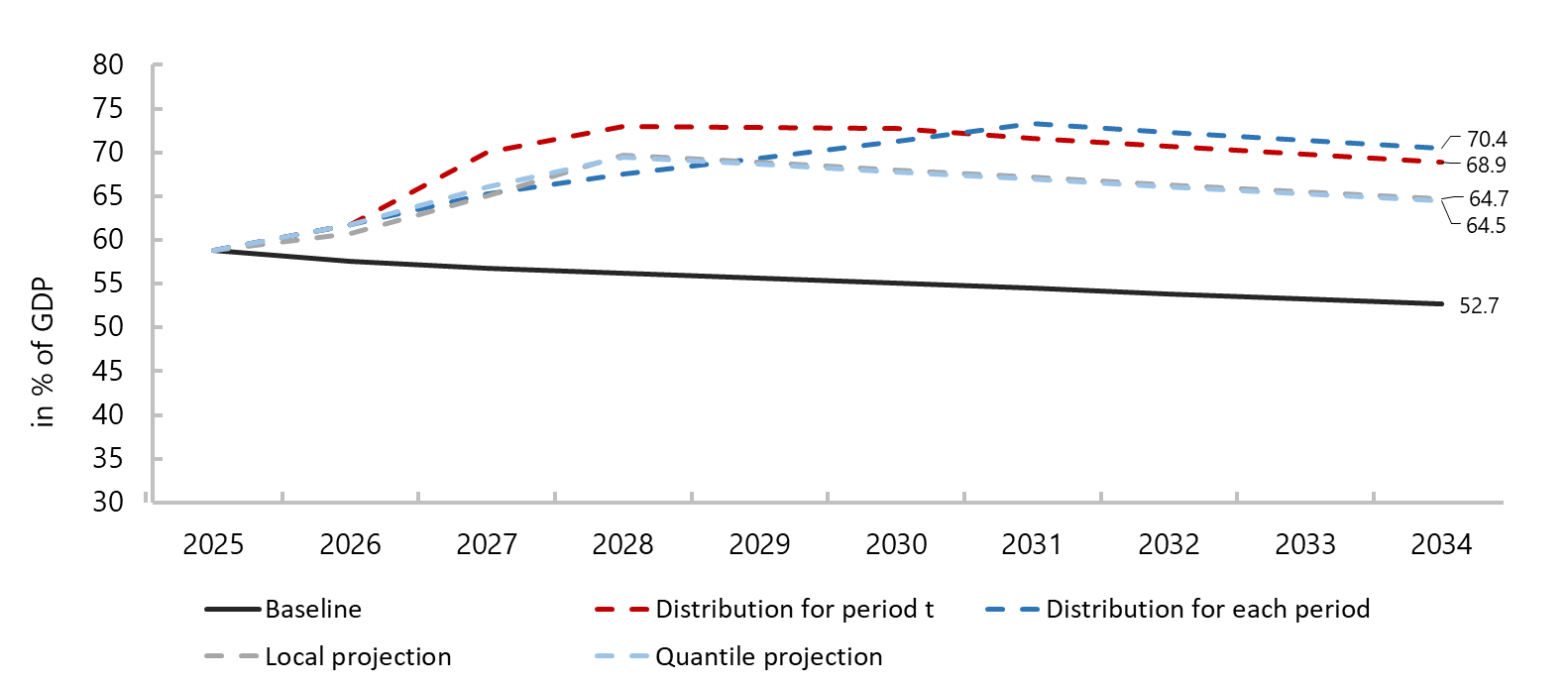}
\end{subfigure}

\vspace{0.5cm}

\begin{subfigure}[t]{0.9\textwidth}
\centering
\caption{Slovenia}
\label{fig:debt_flows_slo}
\includegraphics[width=\textwidth]{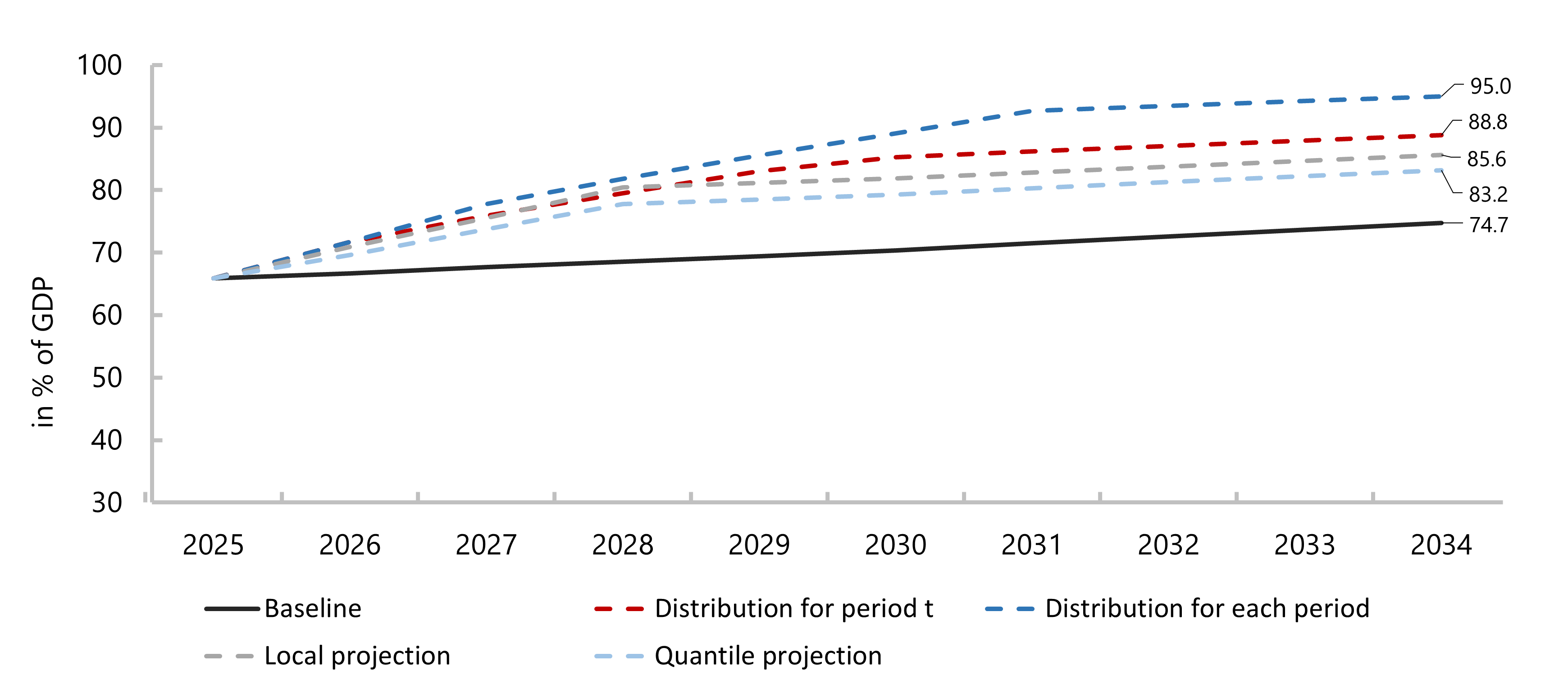}
\end{subfigure}

\caption*{\footnotesize \textit{Source:} Authors' calculation. Econometric assumptions: pre-disaster fiscal balance of $-1.1$ per cent of GDP and adaptive capacity of 0.44 for Croatia, and $-2.3$ per cent of GDP and 0.33 for Slovenia.}
\end{figure}

\begin{figure}[tbp]
\captionsetup{position=above}
\caption{Croatia Public Sector Debt Dynamics under Disaster Scenarios, 2026--2031}
\centering

{\footnotesize\textbf{a)} Distribution for period $t$} \\
\includegraphics[width=0.7\textwidth]{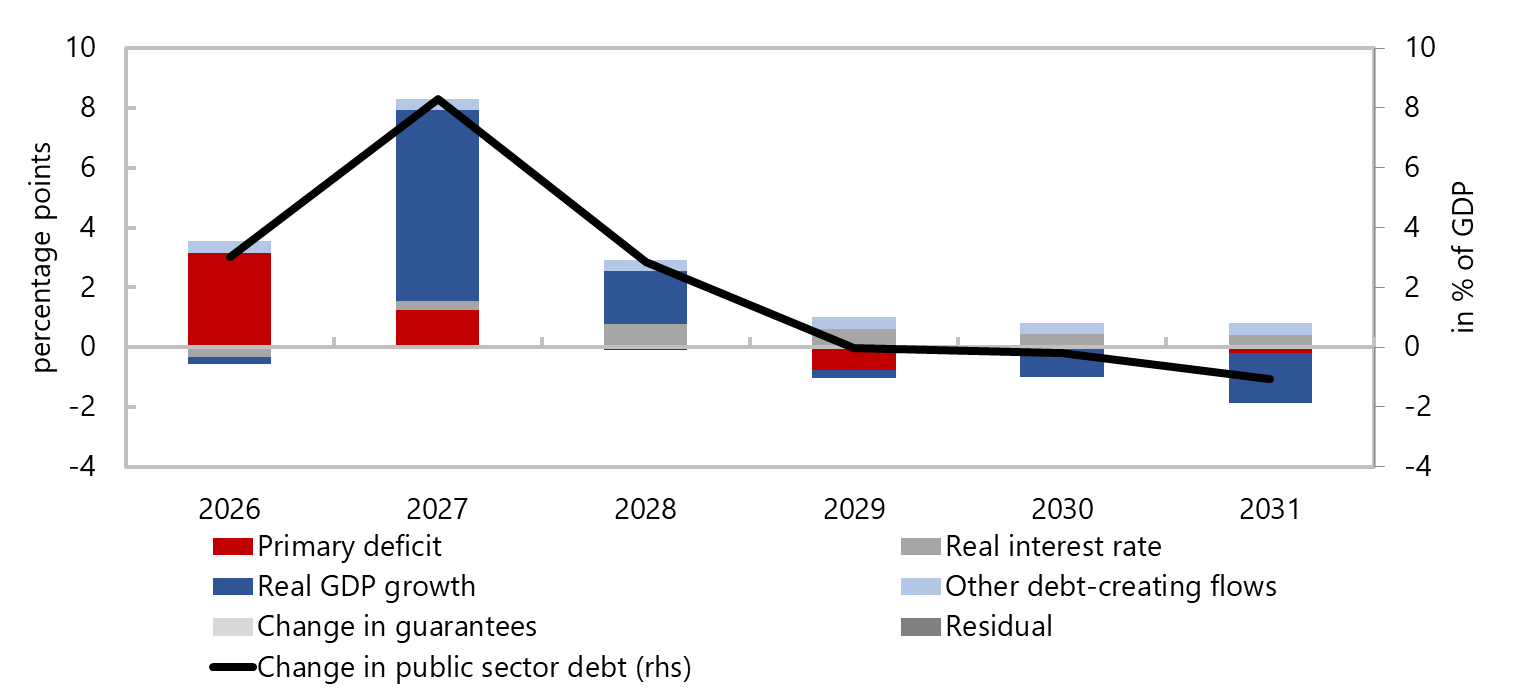} \\[6pt]

{\footnotesize\textbf{b)} Distribution for each period} \\
\includegraphics[width=0.7\textwidth]{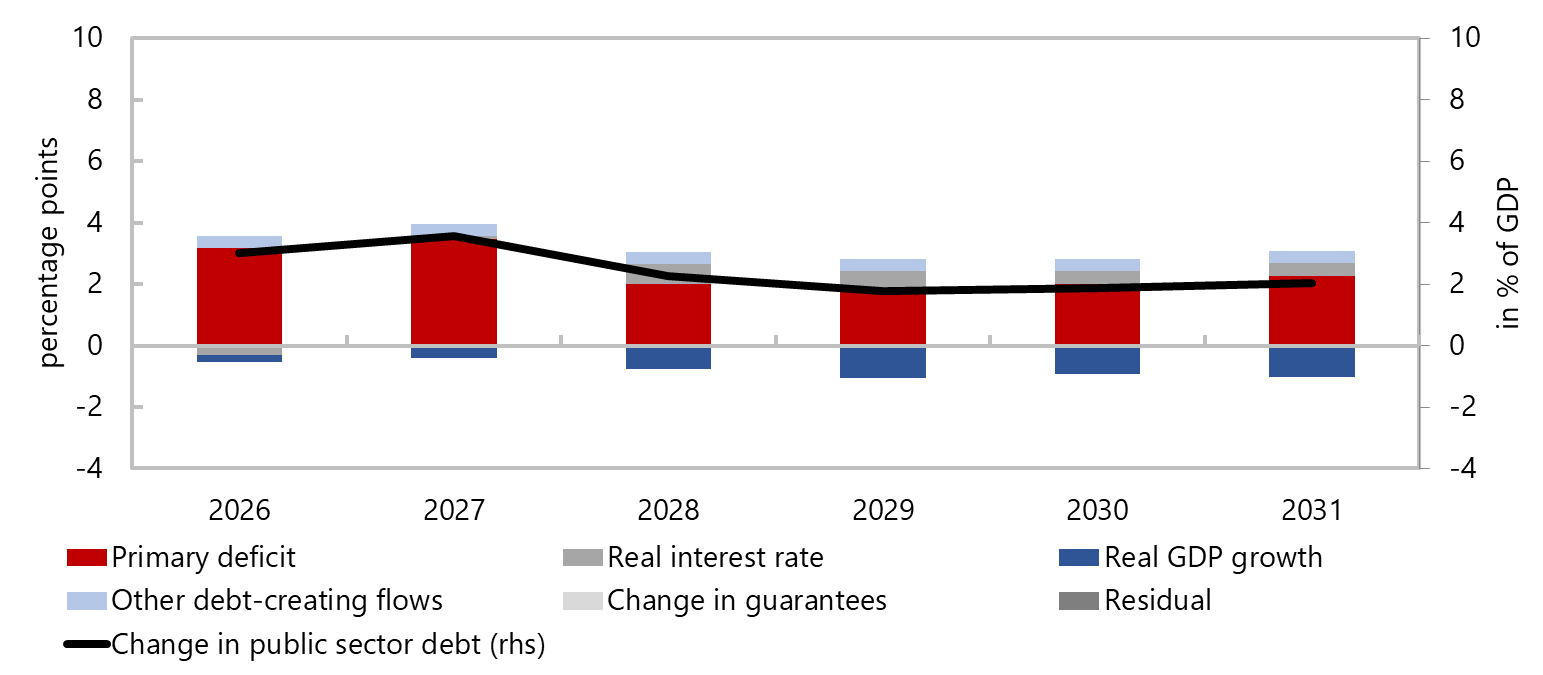} \\[6pt]

{\footnotesize\textbf{c)} Local projection} \\
\includegraphics[width=0.7\textwidth]{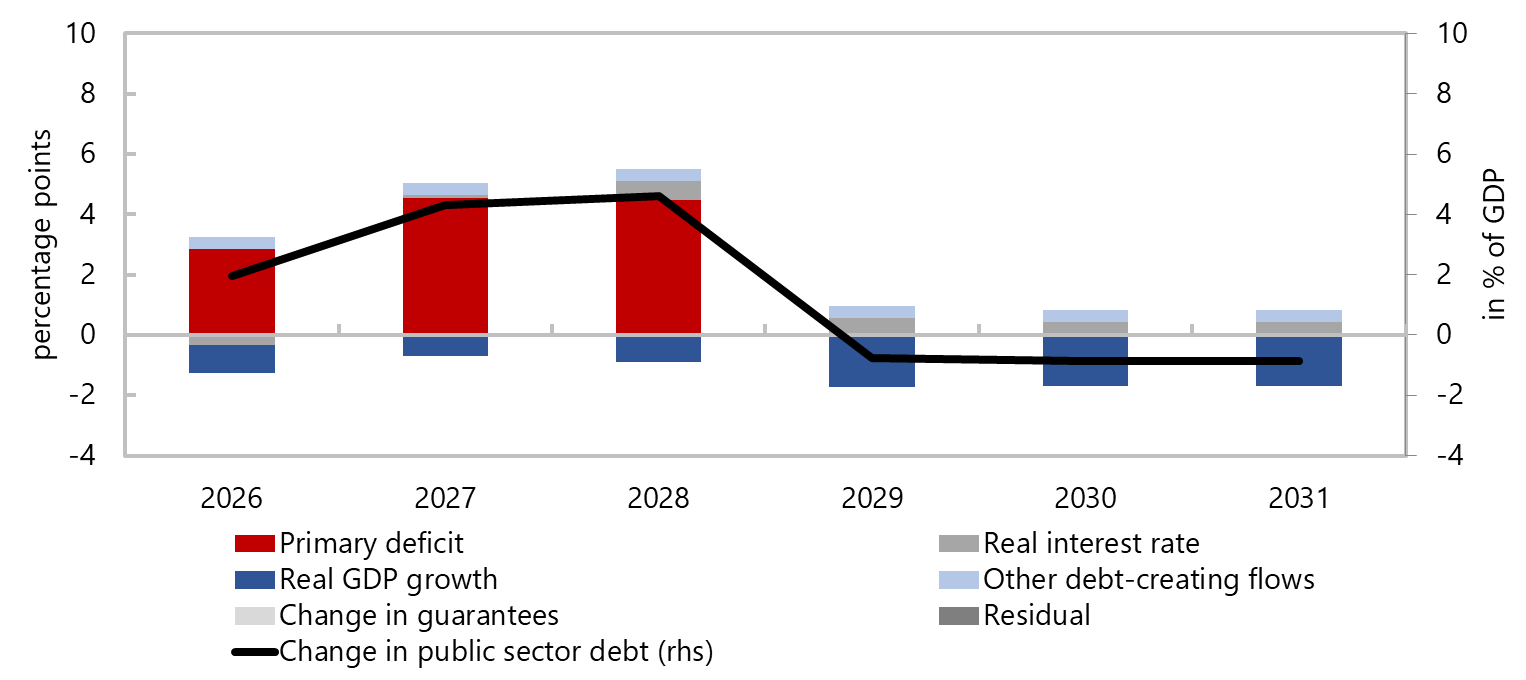} \\[6pt]

{\footnotesize\textbf{d)} Quantile projection} \\
\includegraphics[width=0.7\textwidth]{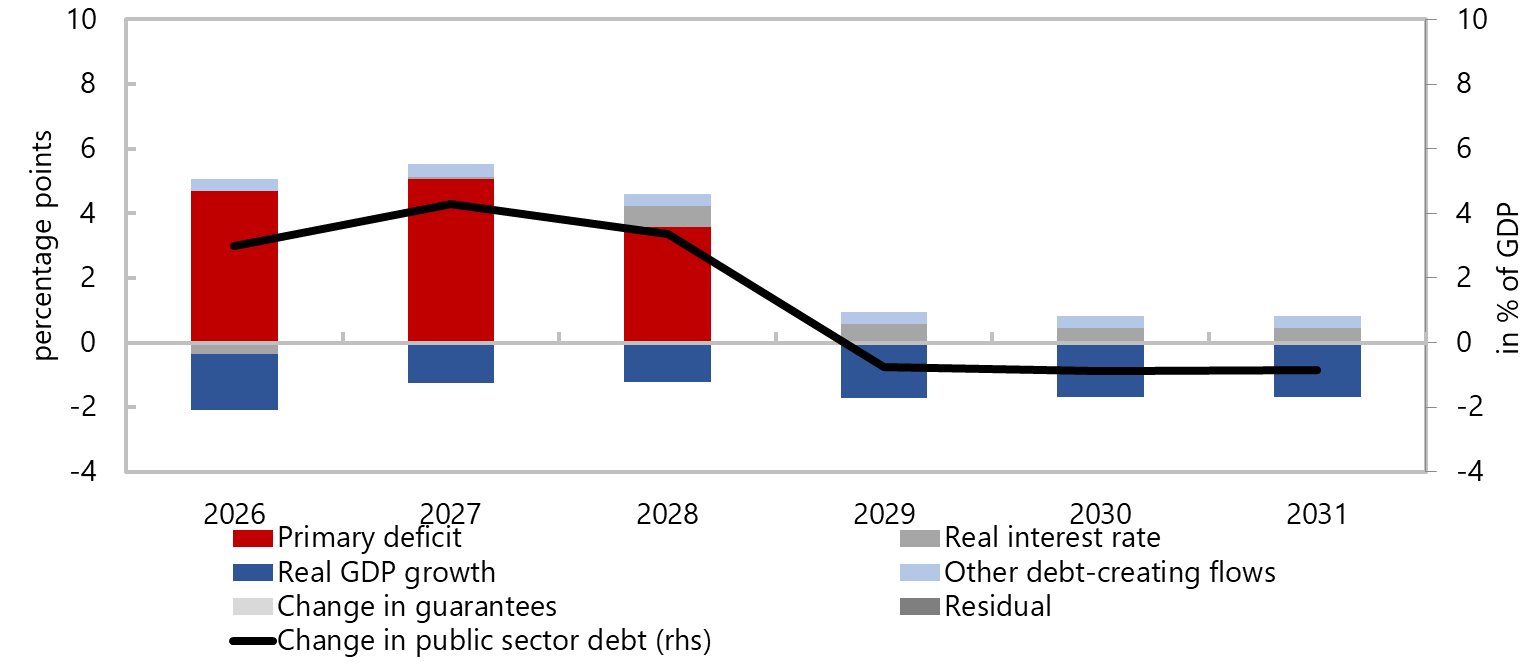}

\caption*{\footnotesize \textit{Source: Authors' calculation.}}
\label{fig:debt-structure-cro}
\end{figure}

\begin{figure}[tbp]
\captionsetup{position=above}
\caption{Slovenia Public Sector Debt Dynamics under Disaster Scenarios, 2026--2031}
\centering

{\footnotesize\textbf{a)} Distribution for period $t$} \\
\includegraphics[width=0.7\textwidth]{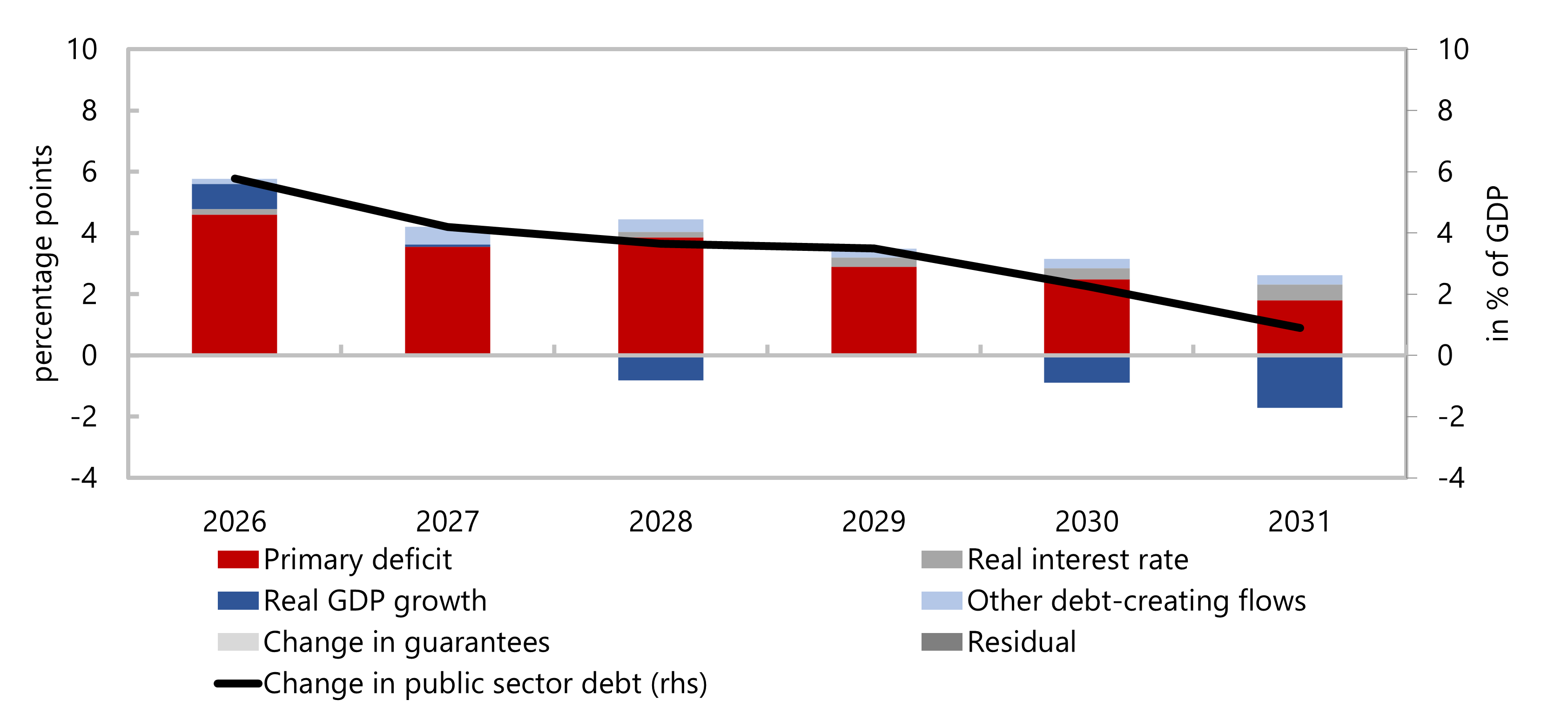} \\[6pt]

{\footnotesize\textbf{b)} Distribution for each period} \\
\includegraphics[width=0.7\textwidth]{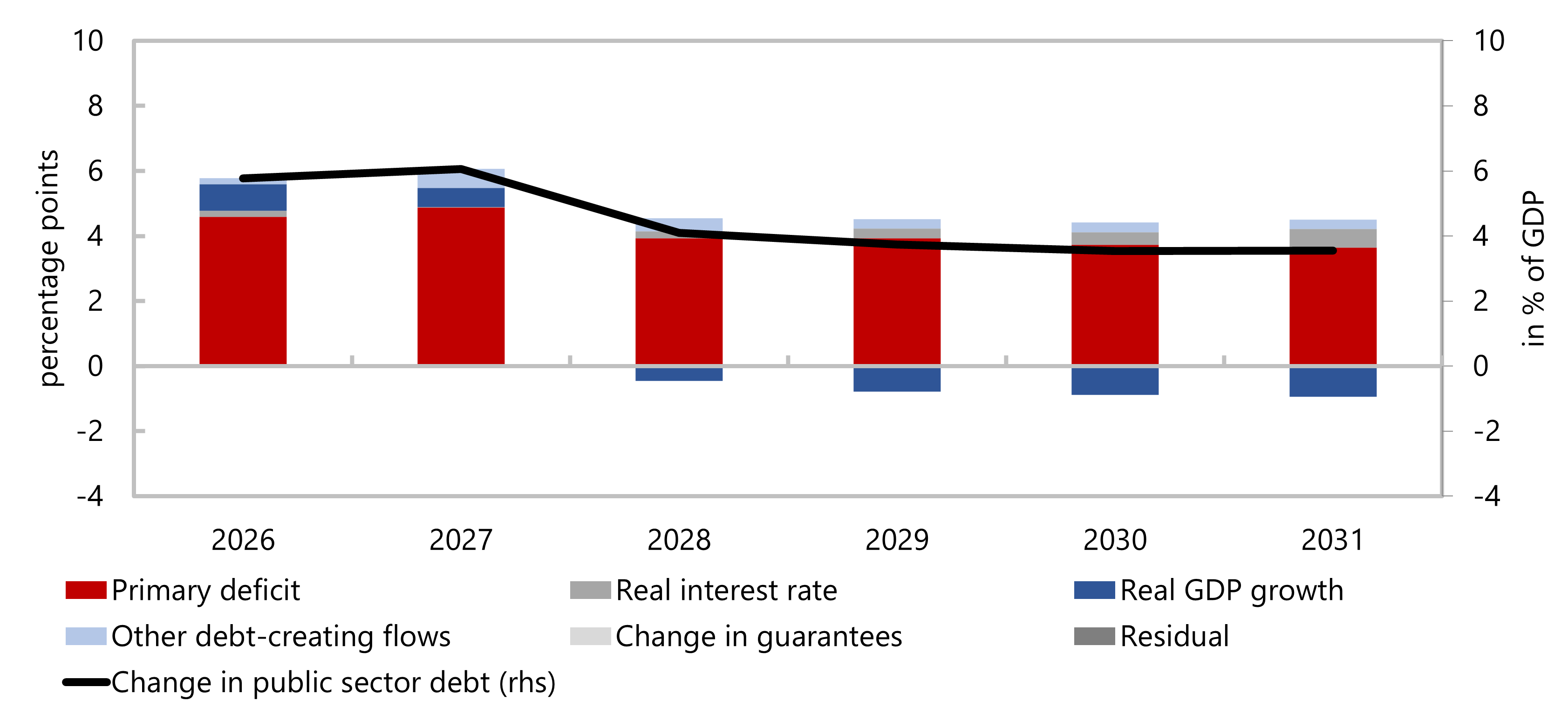} \\[6pt]

{\footnotesize\textbf{c)} Local projection} \\
\includegraphics[width=0.7\textwidth]{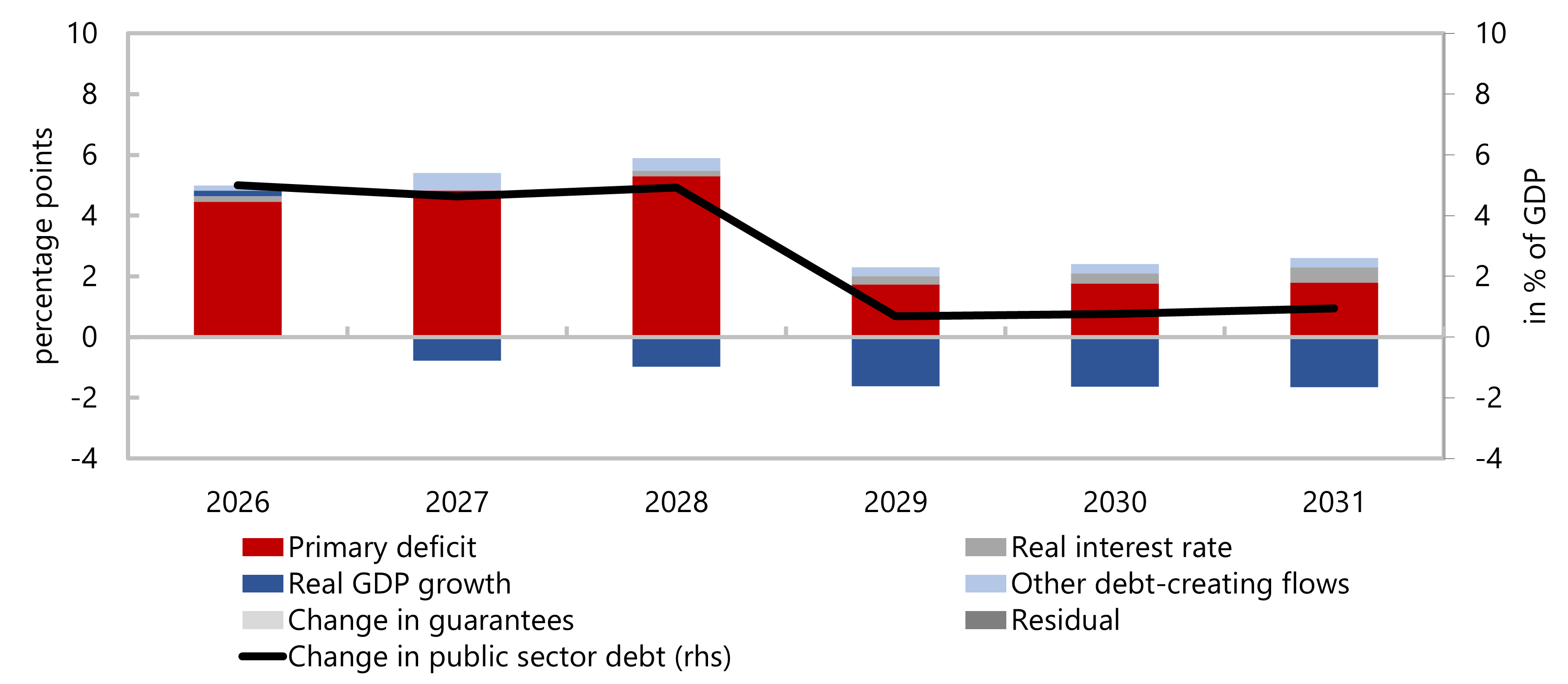} \\[6pt]

{\footnotesize\textbf{d)} Quantile projection} \\
\includegraphics[width=0.7\textwidth]{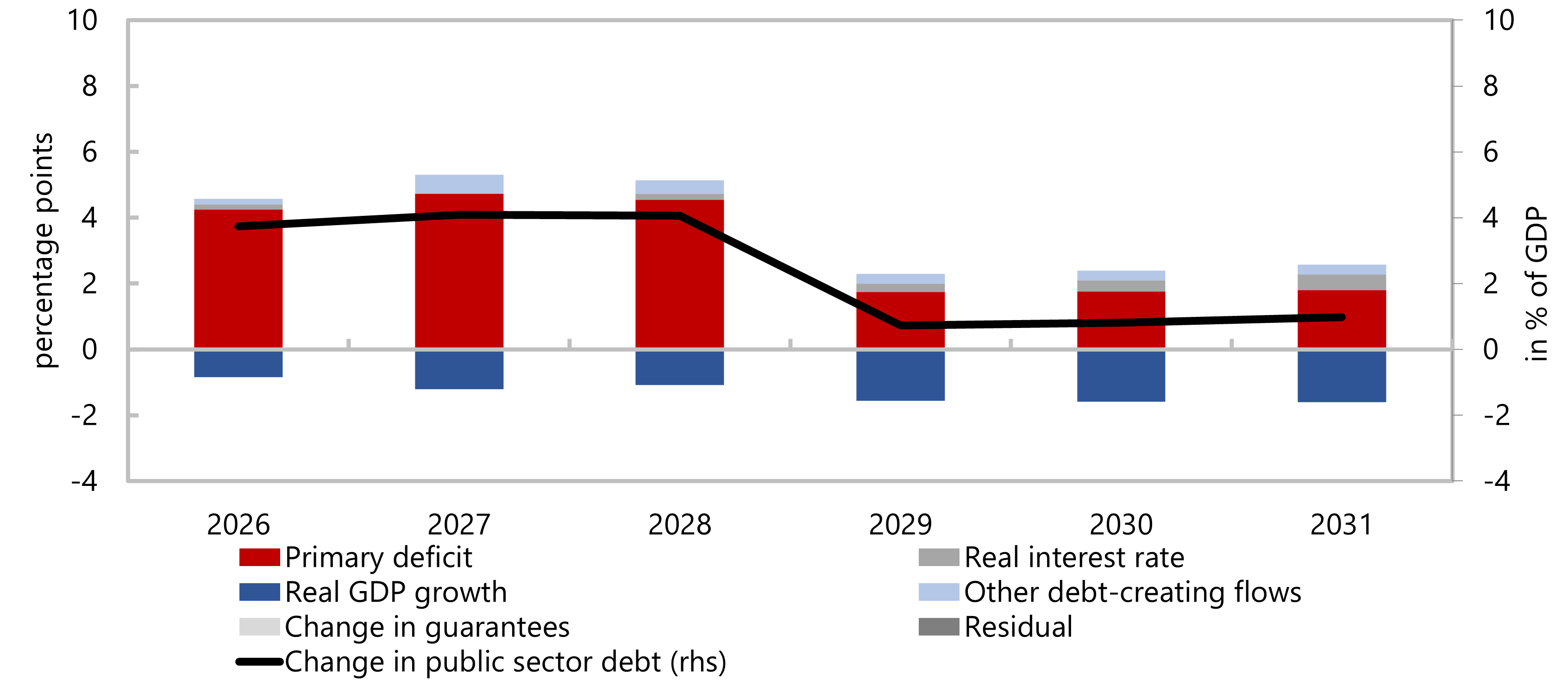}

\caption*{\footnotesize \textit{Source: Authors' calculation.}}
\label{fig:debt-structure-slo}
\end{figure}

\subsection{Fiscal Buffers and Adaptive Capacity}\label{sec:buffers}

So far, the analysis has taken the actual pre-disaster fiscal position and adaptive capacity as given. We now examine a counterfactual in which the pre-disaster primary balance is set to a surplus of +3.0 per cent of GDP, and adaptive capacity is equal to its reference value of zero. This specification is not a forecast but a policy counterfactual designed to quantify how much fiscal buffers and stronger adaptive capacity can attenuate debt accumulation after a severe disaster.\footnote{Achieving this counterfactual would require a sizeable ex-ante fiscal effort. Moving from the actual pre-disaster balance to a primary surplus of +3 per cent of GDP would require a cumulative adjustment of around 20.5 per cent of GDP in Croatia and 26.8 per cent of GDP in Slovenia over a five-year preparation window. A less ambitious target, such as a primary surplus of $+1\%$, would be more realistic but would also deliver a smaller reduction in post-disaster debt.} Improving the pre-disaster configuration reduces 2034 debt in both countries and under both econometric specifications. In the local projection scenario, the reduction amounts to 3.9 percentage points for Croatia and 4.64 percentage points for Slovenia. In the quantile regression scenario, the effect is larger, at 10.64 percentage points for Croatia and 8.25 percentage points for Slovenia. Pre-disaster fiscal buffers and adaptive capacity can therefore substantially attenuate how disaster shocks propagate to public debt.

\begin{table}[tbp]
\centering
\small
\caption{Debt-to-GDP ratio in 2034 under improved fiscal and adaptive conditions}
\label{tab:improved_summary}
\begin{tabularx}{\textwidth}{lcccccc}
\toprule
& \multicolumn{3}{c}{\textbf{Croatia}} & \multicolumn{3}{c}{\textbf{Slovenia}} \\
\cmidrule(lr){2-4}\cmidrule(lr){5-7}
\textbf{Scenario} & Default & Improved & Difference & Default & Improved & Difference \\
\midrule
Local projection & 64.67 & 60.77 & $-3.90$ & 85.60 & 80.96 & $-4.64$ \\
Quantile projection & 64.45 & 53.81 & $-10.64$ & 83.17 & 74.92 & $-8.25$ \\
\bottomrule
\end{tabularx}
\caption*{\footnotesize \textit{Source:} Authors' calculation. Improved scenarios assume a pre-disaster fiscal balance of $+3$ per cent of GDP and adaptive capacity equal to zero.}
\end{table}

Croatian quantile-regression counterfactuals should, however, be interpreted carefully. In that scenario, 2034 debt falls to 53.81 per cent of GDP, close to the no-disaster baseline of 52.70 per cent. This is an algebraic implication of evaluating cross-country elasticities at a stronger fiscal position and higher adaptive capacity, not a literal claim that a disaster-resilient economy would be better off than in the absence of a disaster. The same caveat applies to the four-channel results discussed in Section~\ref{sec:robustness}, where the Croatian and Slovenian improved-quantile outcomes fall below their respective no-disaster baselines.

Risk-transfer instruments such as EU Solidarity Fund grants, parametric catastrophe insurance, and post-disaster fiscal rules are not numerically simulated here and are treated as proposed extensions rather than reported results. Within the ND-DDT framework, such instruments could be represented by modifying the primary balance path $pb_t$ or other identified debt-creating flows $of_t$. 
Numerical simulation including them is left for future work because the size and timing of these disbursements depend on case-by-case European Commission assessments, parametric insurance contracts are not yet in place for either country, and post-disaster fiscal rules would require institutional design choices that go beyond the scope of this paper.

\begin{figure}[tbp]
\centering
\caption{Gross Nominal Public Debt under Improved Fiscal and Adaptive Conditions}
\label{fig:debt_flows_2}

\begin{subfigure}[t]{0.9\textwidth}
\centering
\caption{Croatia}
\label{fig:debt_flows_2_cro}
\includegraphics[width=\textwidth]{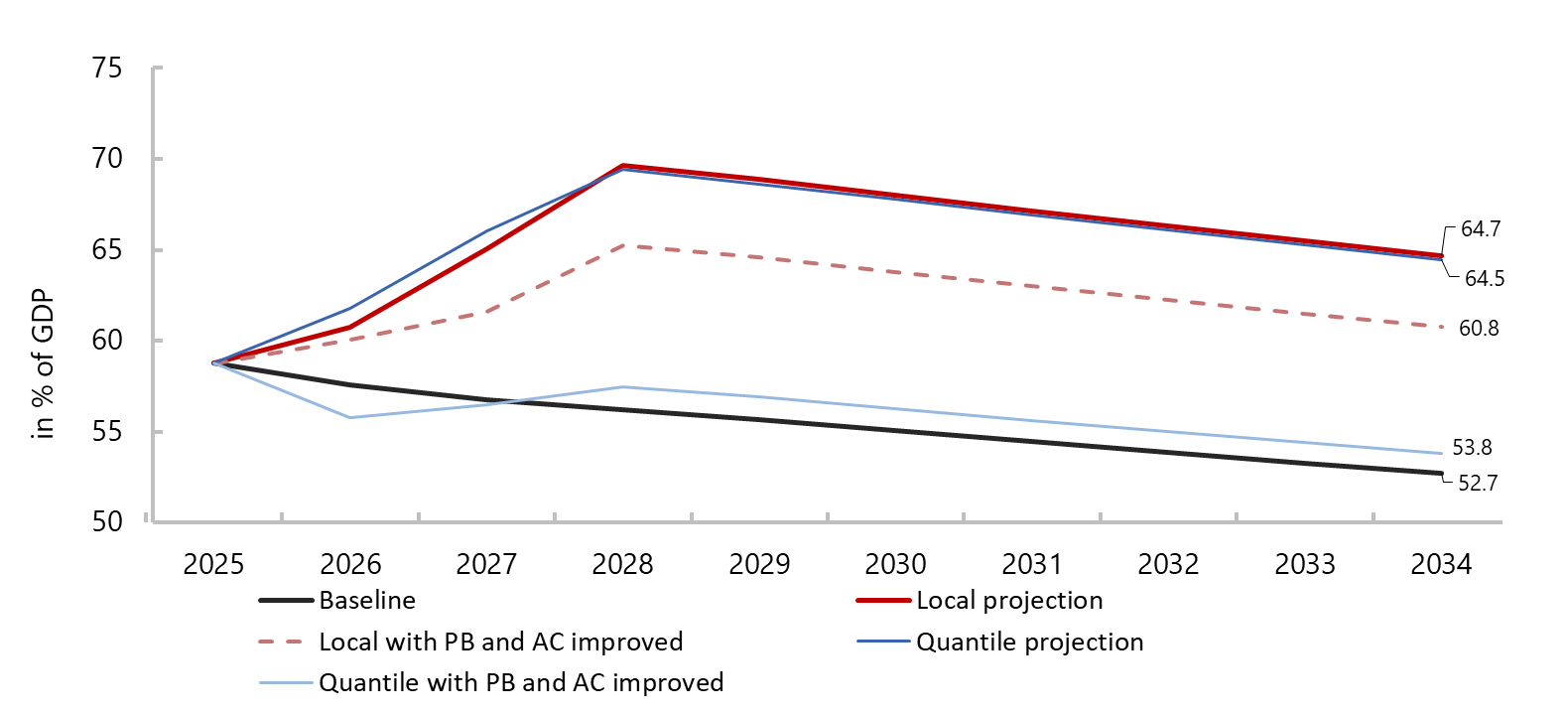}
\end{subfigure}

\vspace{0.5cm}

\begin{subfigure}[t]{0.9\textwidth}
\centering
\caption{Slovenia}
\label{fig:debt_flows_2_slo}
\includegraphics[width=\textwidth]{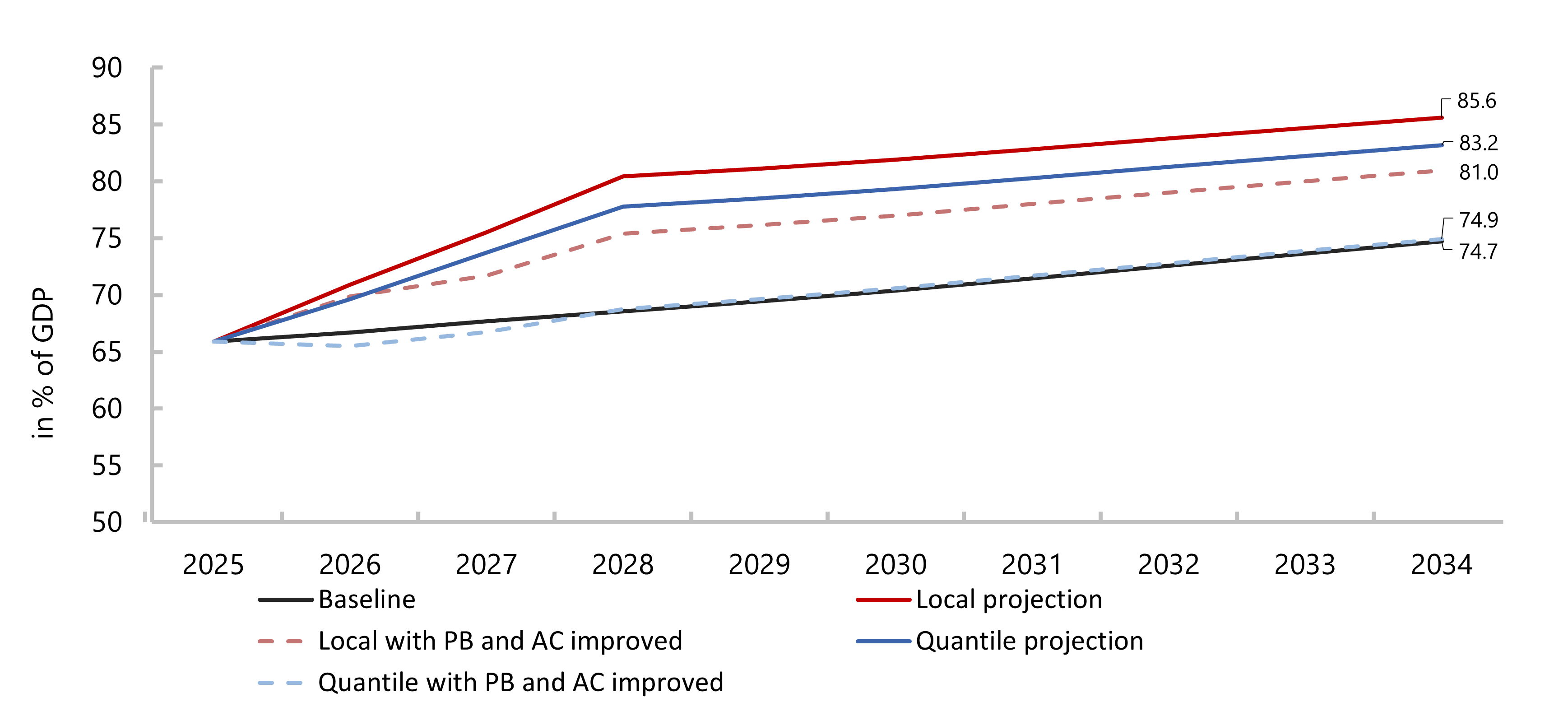}
\end{subfigure}

\caption*{\footnotesize \textit{Source:} Authors' calculation. Improved scenarios assume a pre-disaster fiscal balance of $+3$ per cent of GDP and adaptive capacity equal to zero.}
\end{figure}

\begin{figure}[tbp]
\captionsetup{position=above}
\caption{Croatia Public Sector Debt Dynamics under Improved Fiscal and Adaptive Conditions, 2026--2031}
\centering

{\footnotesize\textbf{a)} Local projection} \\
\includegraphics[width=0.7\textwidth]{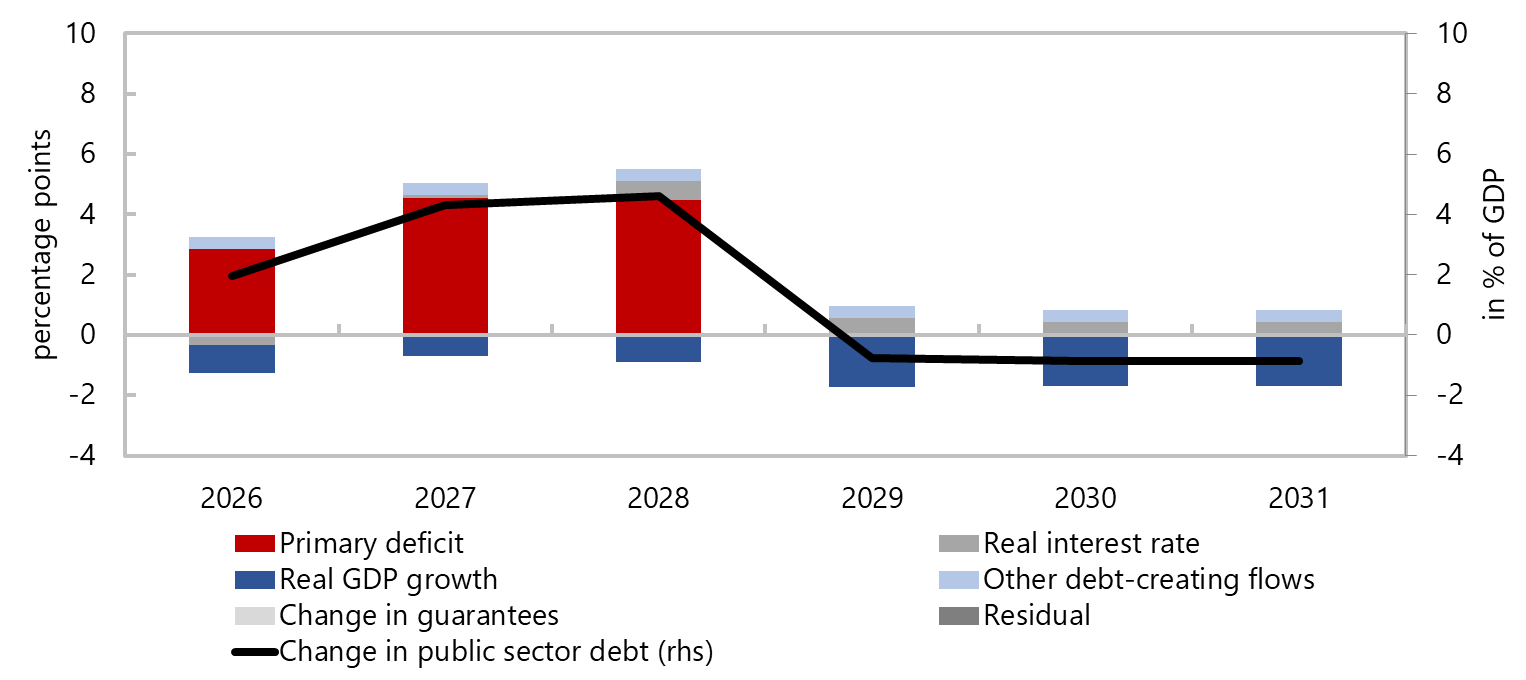} \\[6pt]

{\footnotesize\textbf{b)} Local projection with PB and AC improved} \\
\includegraphics[width=0.7\textwidth]{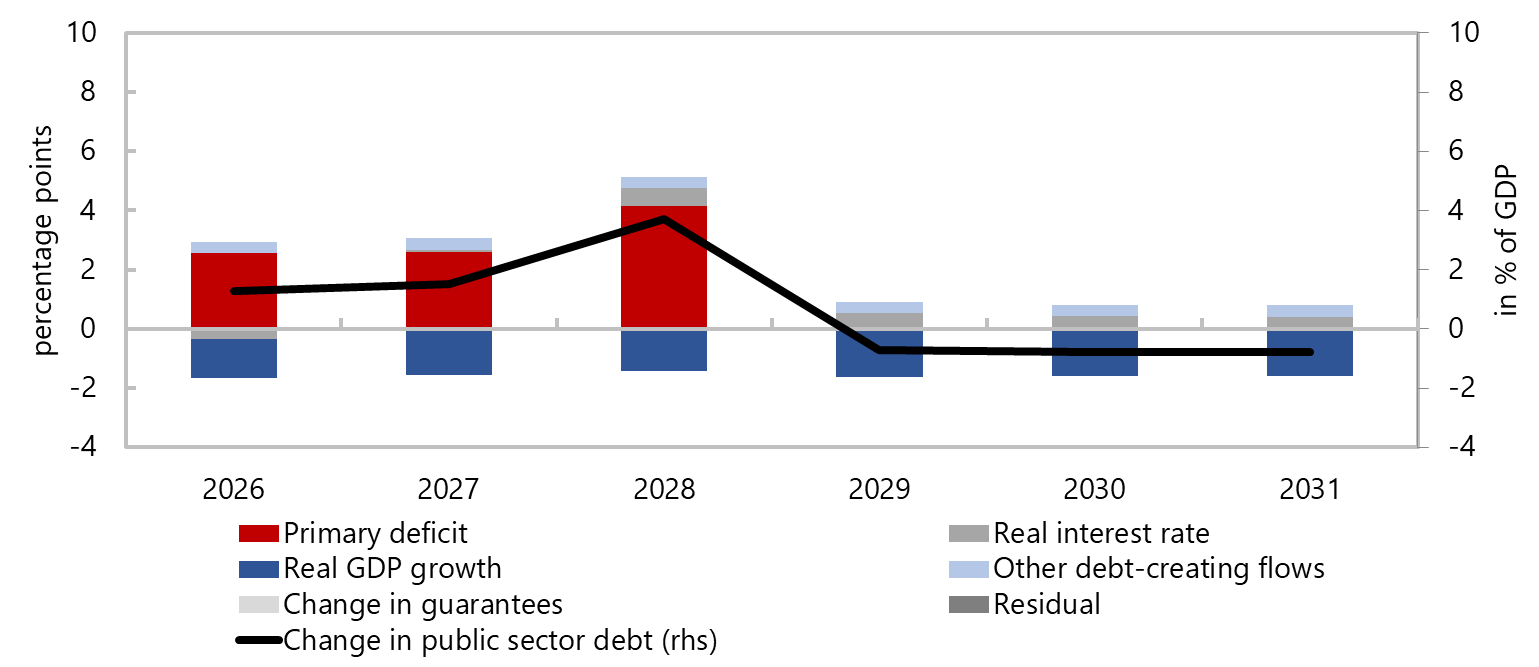} \\[6pt]

{\footnotesize\textbf{c)} Quantile projection} \\
\includegraphics[width=0.7\textwidth]{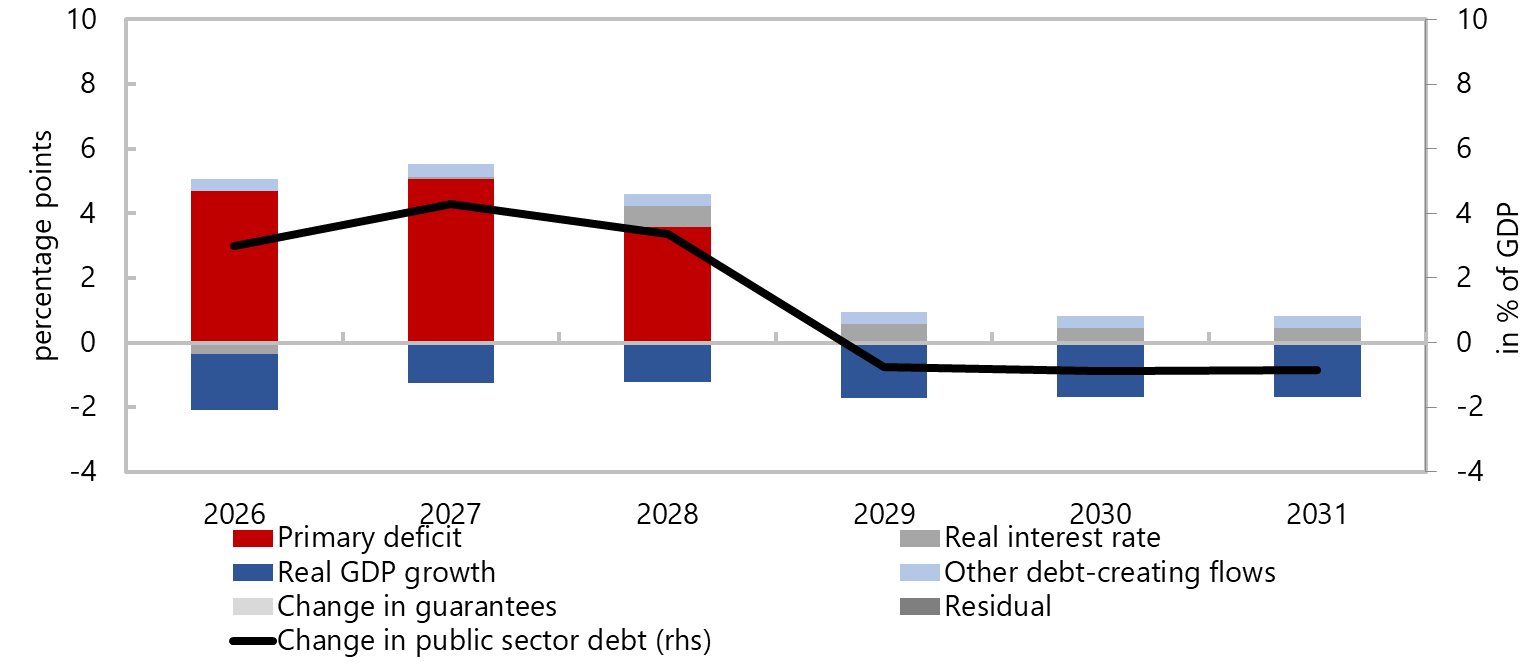} \\[6pt]

{\footnotesize\textbf{d)} Quantile projection with PB and AC improved} \\
\includegraphics[width=0.7\textwidth]{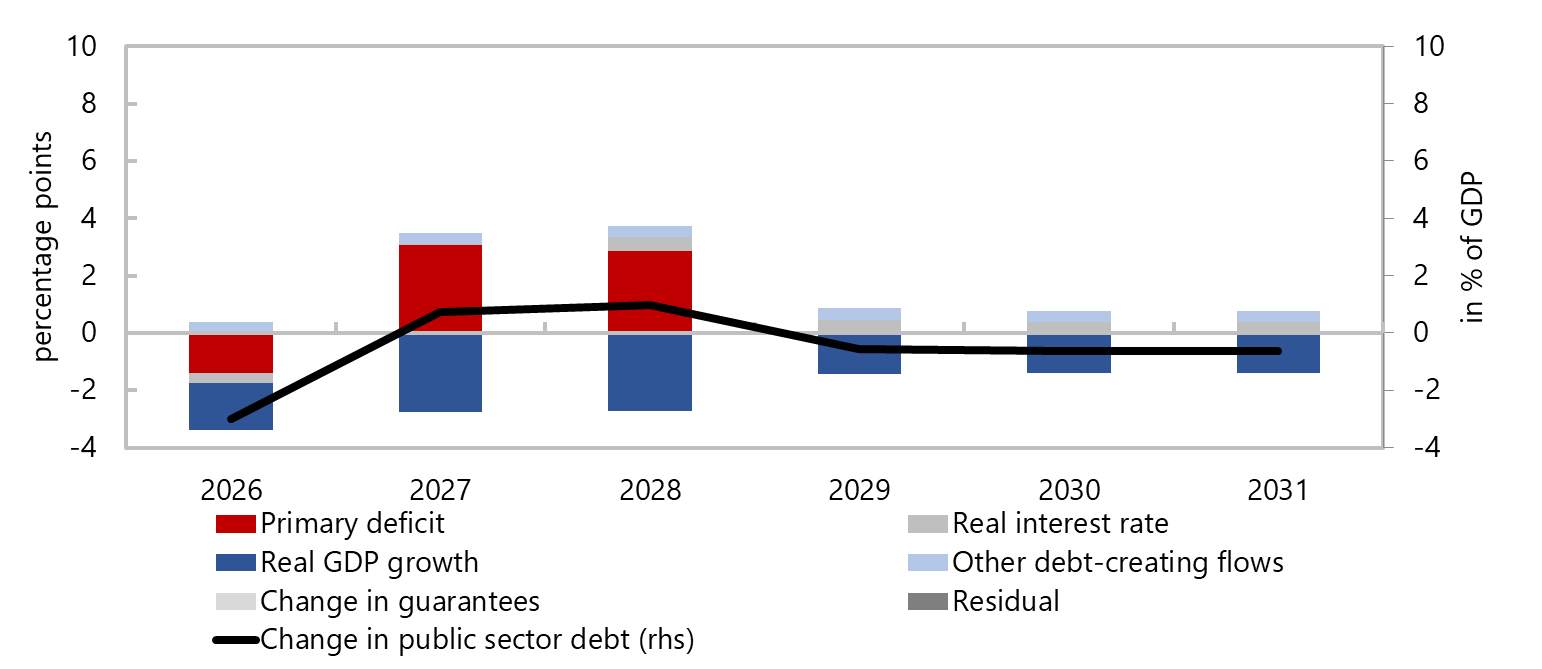}

\caption*{\footnotesize \textit{Source: Authors' calculation.}}
\label{fig:debt-structure-AC-and-PB-cro}
\end{figure}

\begin{figure}[tbp]
\captionsetup{position=above}
\caption{Slovenia Public Sector Debt Dynamics under Improved Fiscal and Adaptive Conditions, 2026--2031}
\centering

{\footnotesize\textbf{a)} Local projection} \\
\includegraphics[width=0.7\textwidth]{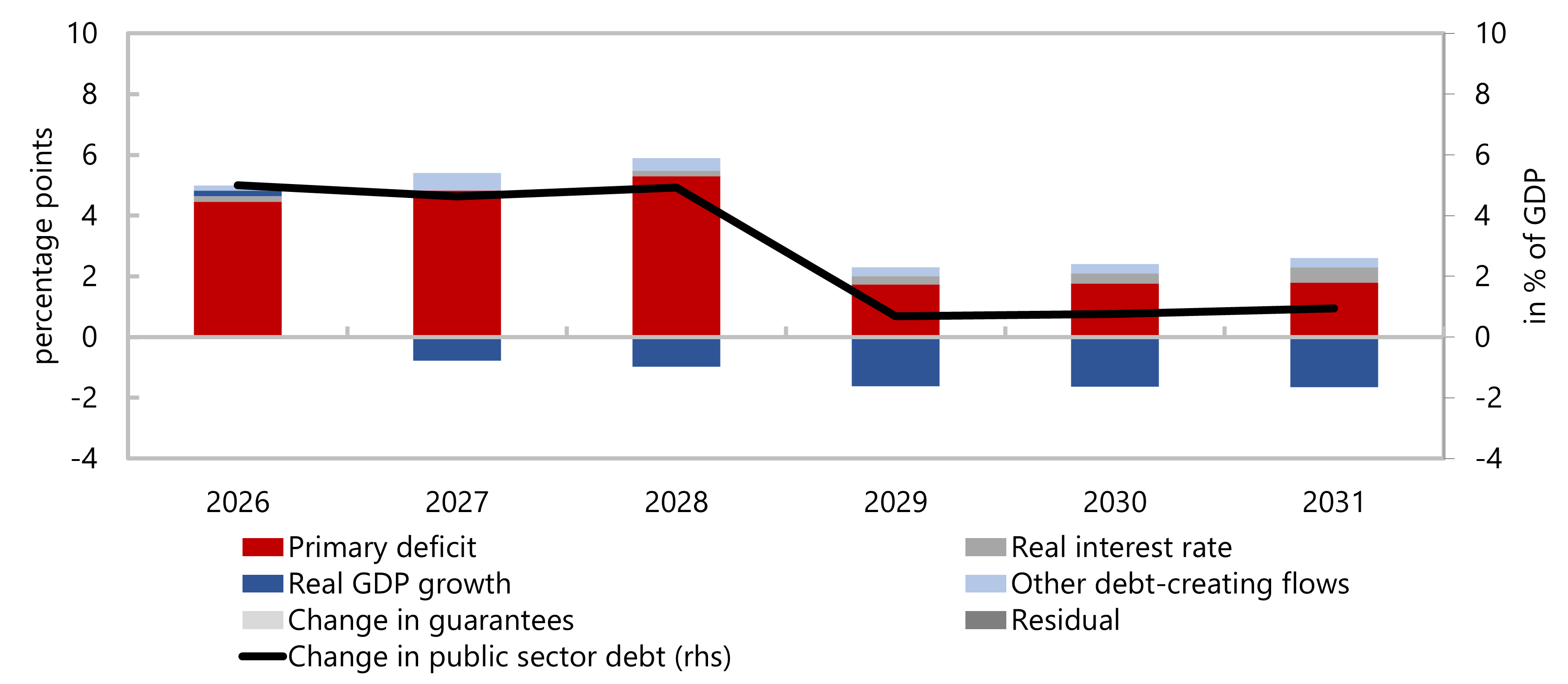} \\[6pt]

{\footnotesize\textbf{b)} Local projection with PB and AC improved} \\
\includegraphics[width=0.7\textwidth]{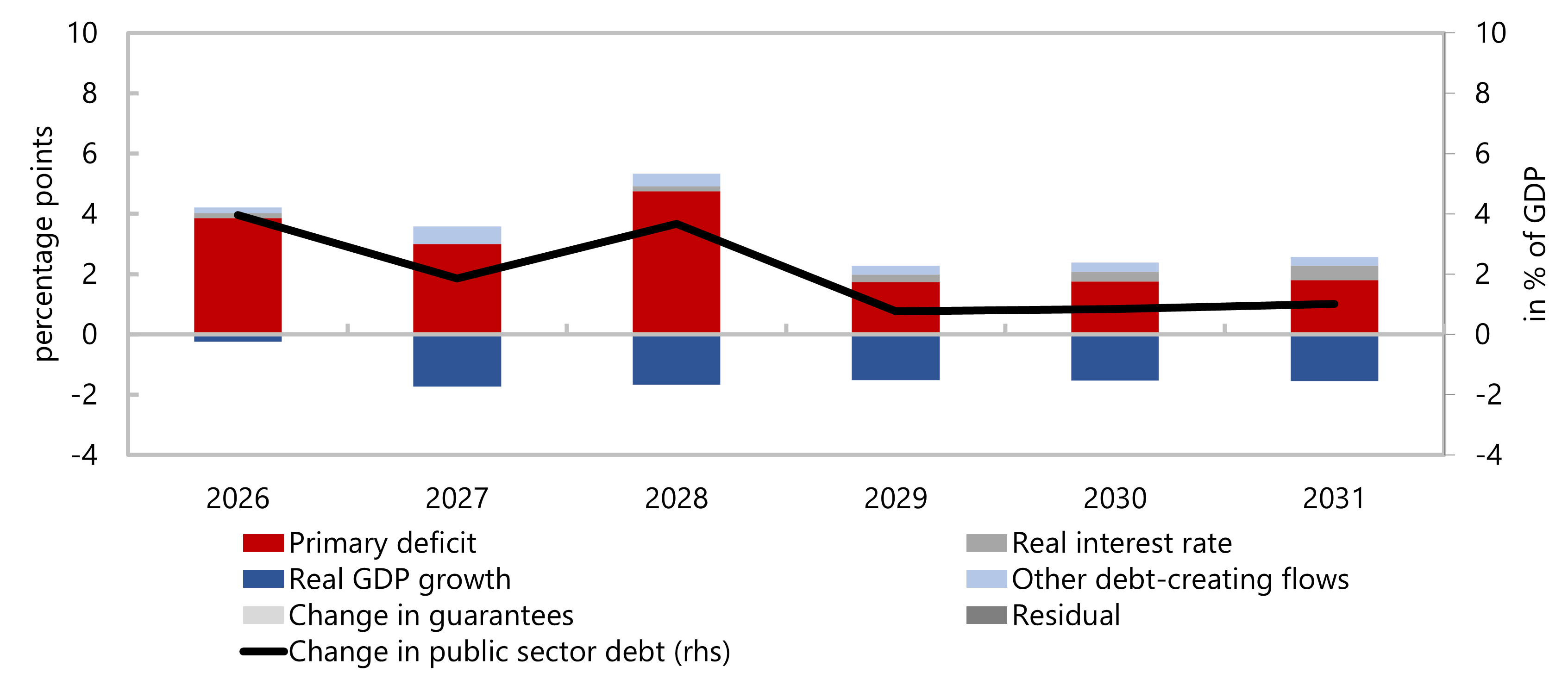} \\[6pt]

{\footnotesize\textbf{c)} Quantile projection} \\
\includegraphics[width=0.7\textwidth]{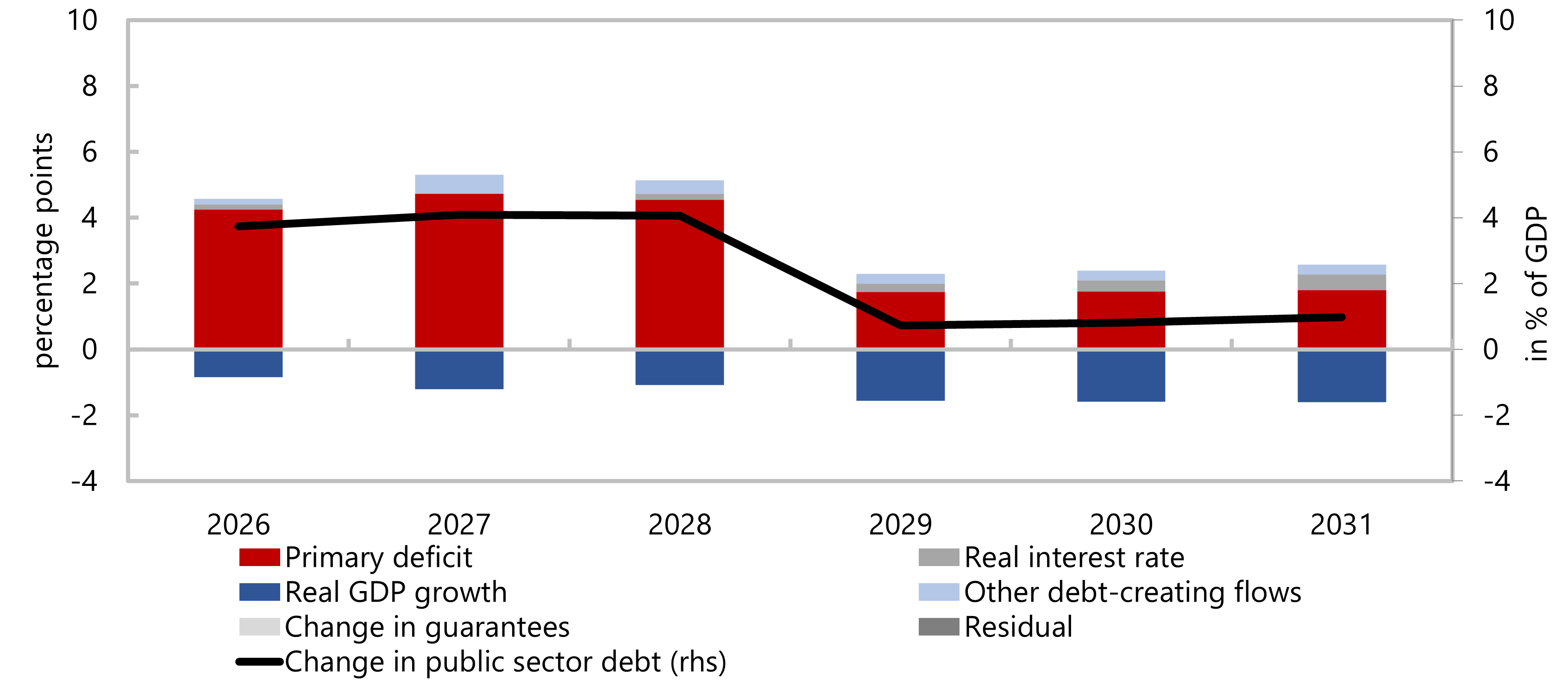} \\[6pt]

{\footnotesize\textbf{d)} Quantile projection with PB and AC improved} \\
\includegraphics[width=0.7\textwidth]{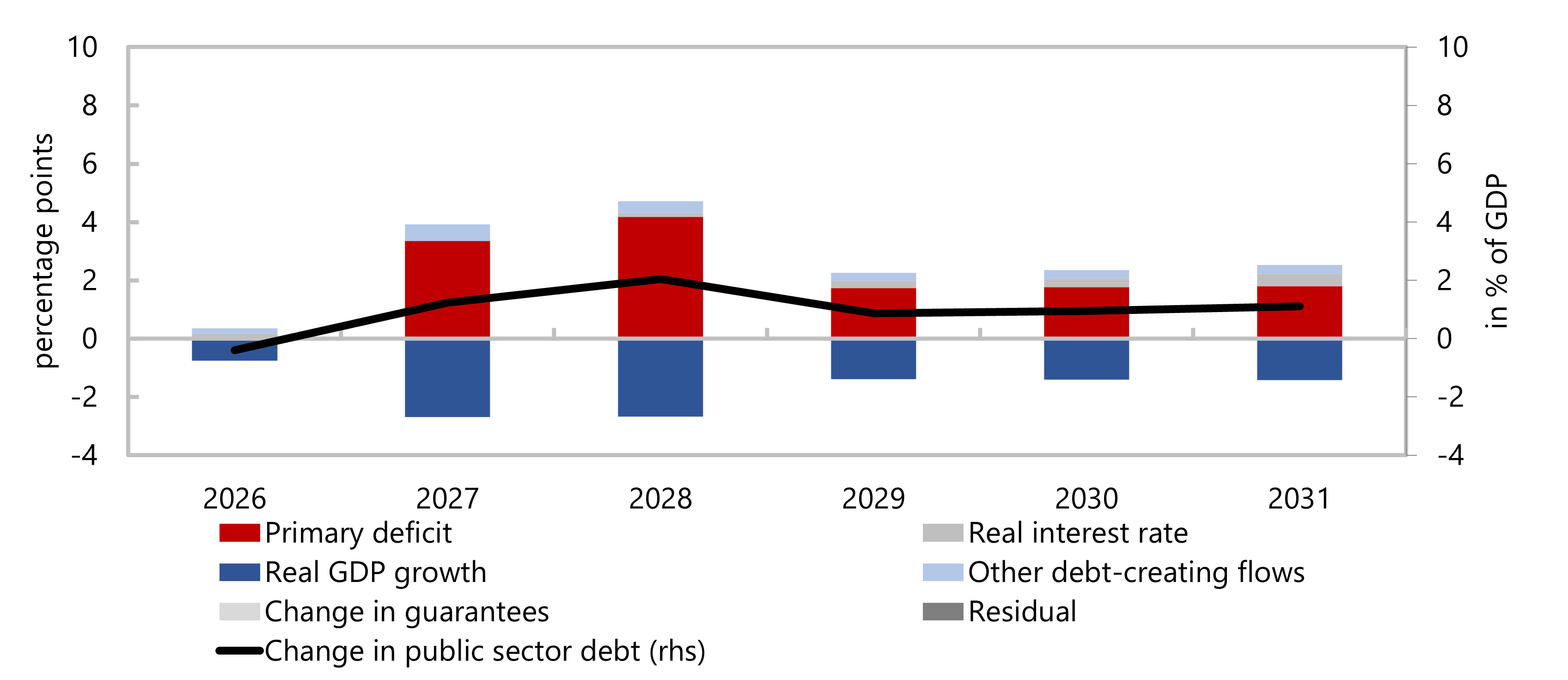}

\caption*{\footnotesize \textit{Source: Authors' calculation.}}
\label{fig:debt-structure-AC-and-PB-slo}
\end{figure}

\subsection{Robustness: Interest-Rate and Inflation Channels}\label{sec:robustness}

The baseline disaster scenarios allow natural disasters to affect real GDP growth and the primary balance, while holding the GDP deflator and the effective interest rate at their no-disaster paths. This assumption is appropriate for the headline analysis because both countries have euro-area monetary conditions and debt portfolios that attenuate short-run interest-rate pass-through. Croatia's public debt is fully euro-denominated and largely fixed rate, while Slovenia has a comparable euro-area debt-management environment. We proceed by estimating a four-channel specification in which the disaster shock affects real GDP growth, the primary balance, the GDP deflator and the effective interest rate. These results are reported as a robustness exercise rather than as the central scenario. The reason is that cross-country elasticities for inflation and interest rates are less directly transferable to Croatia and Slovenia than those for real activity and the fiscal balance.\footnote{Caution is warranted in interpreting the four-channel results. The effective interest rate on government debt is a slow-moving average cost, shaped by the maturity structure of outstanding debt and past borrowing conditions. Euro-area monetary conditions also limit the direct link between a domestic disaster and sovereign financing costs. Inflation responses to natural disasters are heterogeneous across countries, depending on the relative strength of supply-side disruptions, demand contraction, and policy support.}

\begin{table}[tbp]
\centering
\small
\caption{Debt-to-GDP ratio in 2034 when interest-rate and inflation channels are activated}
\label{tab:four_channel_summary}
\begin{tabularx}{\textwidth}{lcccc}
\toprule
& \multicolumn{2}{c}{\textbf{Croatia}} & \multicolumn{2}{c}{\textbf{Slovenia}} \\
\cmidrule(lr){2-3}\cmidrule(lr){4-5}
\textbf{Scenario} & 2 channels & 4 channels & 2 channels & 4 channels \\
\midrule
Local projection & 64.67 & 68.20 & 85.60 & 86.87 \\
Quantile projection & 64.45 & 63.35 & 83.17 & 77.42 \\
Local projection, improved PB and AC & 60.77 & 62.42 & 80.96 & 80.30 \\
Quantile projection, improved PB and AC & 53.81 & 49.15 & 74.92 & 65.76 \\
\bottomrule
\end{tabularx}
\caption*{\footnotesize \textit{Source:} Authors' calculation. The two-channel specification allows shocks to affect real GDP growth and the primary balance. The four-channel specification additionally allows shocks to affect the GDP deflator and the effective interest rate.}
\end{table}

For Croatia, activating the inflation and interest-rate channels raises the local projection outcome from 64.67 to 68.20 per cent of GDP in 2034, and the improved local projection estimate from 60.77 to 62.42 per cent. The quantile regression outcomes move in the opposite direction: the standard result falls from 64.45 to 63.35 per cent and the improved result from 53.81 to 49.15 per cent, reflecting the non-linear interaction between the estimated inflation response, the effective interest-rate path, and the debt stock. For Slovenia, our local projection estimate increases from 85.60 to 86.87 per cent when the additional channels are included, while the quantile regression result falls from 83.17 to 77.42 per cent. The improved quantile specification with four channels yields 65.76 per cent, below the Slovenian no-disaster baseline of 74.71 per cent. This is a corner case of the cross-country elasticity rather than a literal policy conclusion.

\begin{figure}[tbp]
\centering
\caption{Debt Trajectories when Interest-Rate and Inflation Channels are Activated}
\label{fig:robustness_ks}

\begin{subfigure}[t]{0.9\textwidth}
\centering
\caption{Croatia}
\label{fig:robustness_ks_cro}
\includegraphics[width=\textwidth]{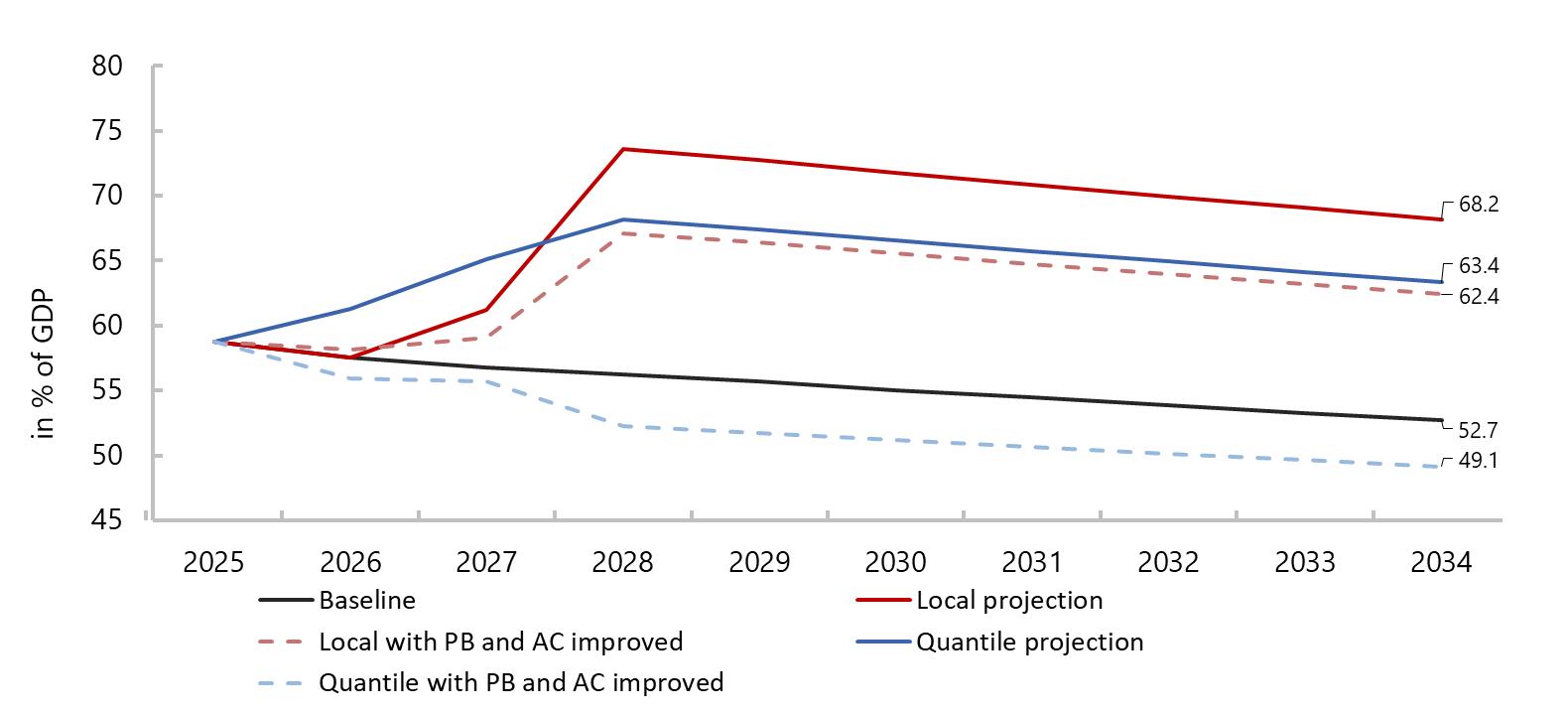}
\end{subfigure}

\vspace{0.5cm}

\begin{subfigure}[t]{0.9\textwidth}
\centering
\caption{Slovenia}
\label{fig:robustness_ks_slo}
\includegraphics[width=\textwidth]{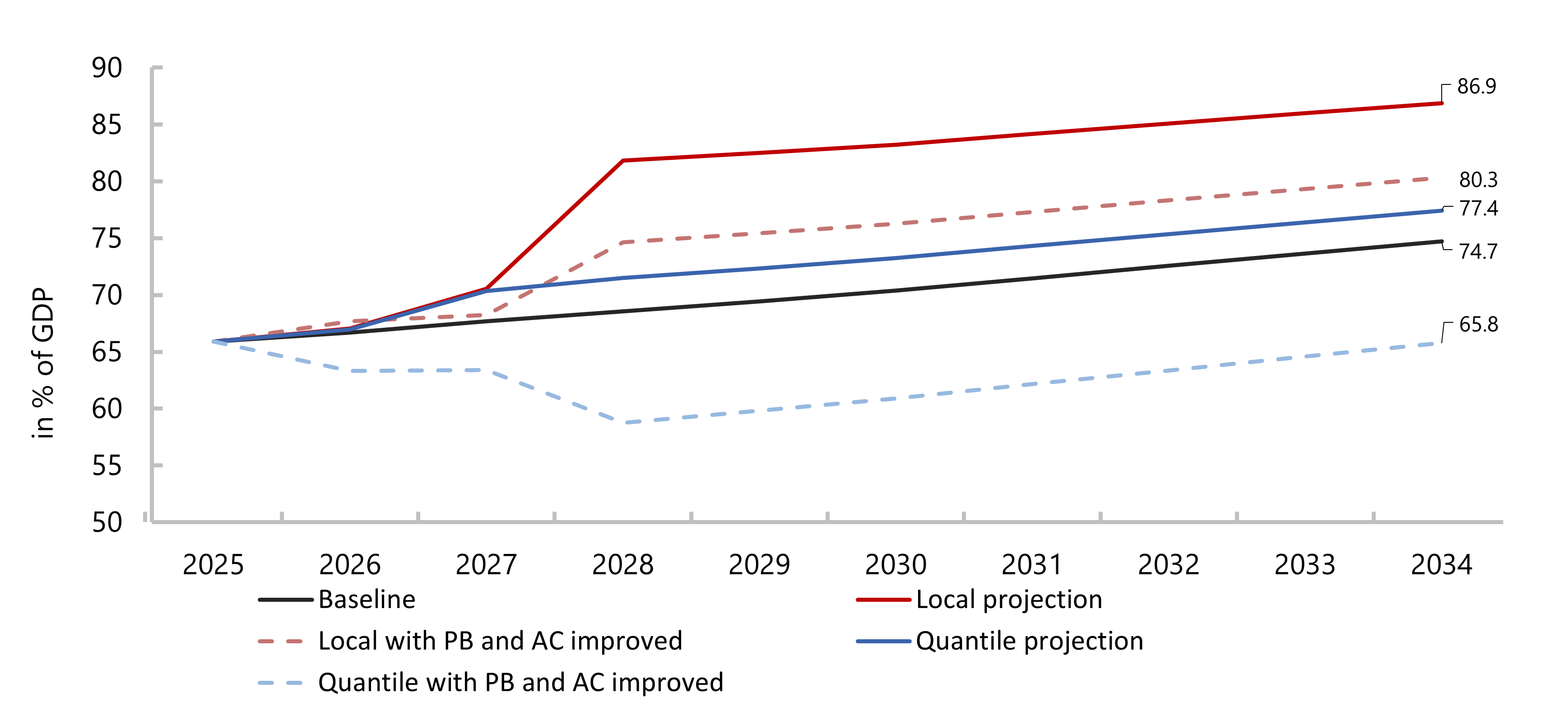}
\end{subfigure}

\caption*{\footnotesize \textit{Source:} Authors' calculation. The figure reports the four-channel specification in which real GDP growth, the primary balance, the GDP deflator and the effective interest rate are allowed to respond to the disaster shock. Improved scenarios assume a pre-disaster fiscal balance of $+3$ per cent of GDP and adaptive capacity equal to zero.}
\end{figure}


\subsection{Summary of Implications for Debt Sustainability}\label{sec:summary}

Table~\ref{tab:scenario_summary_2034} summarises the 2034 debt outcomes across the main scenarios. Three findings are worth highlighting. First, the baseline trajectories of Croatia and Slovenia diverge strongly. Croatia moves from 63.04 per cent of GDP in 2024 to 52.70 per cent in 2034; Slovenia moves from 66.60 to 74.71 per cent. This baseline difference shapes our interpretation of all disaster scenarios. Second, disaster shocks raise debt materially in both countries, but the absolute debt level reached by Slovenia is substantially higher. In Croatia, the 2034 debt ratio ranges from 64.45 per cent in the quantile-regression scenario to 70.44 per cent in the recurring empirical-distribution scenario. In Slovenia, the corresponding range is 83.17 to 95.02 per cent. The impact of disasters is therefore broadly comparable in percentage-point terms, but it happens against very different baseline trajectories.

Third, fiscal buffers and adaptive capacity reduce the post-disaster debt burden but do not eliminate disaster-related fiscal risk. The improved local projection scenario reduces 2034 debt by 3.90 percentage points in Croatia and 4.64 percentage points in Slovenia, with larger effects in the quantile regression configuration. As discussed in Section~\ref{sec:buffers}, the Croatian improved-quantile outcome should be read as an algebraic implication of evaluating cross-country elasticities at a stronger fiscal position rather than as a literal forecast. Pre-disaster fiscal space is therefore the most important policy lever in the model, with risk-transfer arrangements serving as complementary instruments.

\begin{table}[tbp]
\centering
\small
\caption{Debt-to-GDP ratio in 2034 across scenarios}
\label{tab:scenario_summary_2034}
\begin{tabularx}{\textwidth}{lcc}
\toprule
\textbf{Scenario} & \textbf{Croatia} & \textbf{Slovenia} \\
\midrule
Baseline & 52.70 & 74.71 \\
Distribution for period $t$ & 68.91 & 88.80 \\
Distribution for each period & 70.44 & 95.02 \\
Local projection & 64.67 & 85.60 \\
Quantile projection & 64.45 & 83.17 \\
Local projection, improved PB and AC & 60.77 & 80.96 \\
Quantile projection, improved PB and AC & 53.81 & 74.92 \\
Local projection, four channels & 68.20 & 86.87 \\
Quantile projection, four channels & 63.35 & 77.42 \\
Local projection, improved PB and AC, four channels & 62.42 & 80.30 \\
Quantile projection, improved PB and AC, four channels & 49.15 & 65.76 \\
\bottomrule
\end{tabularx}
\caption*{\footnotesize \textit{Source:} Authors' calculation from ND-DDT scenario files. Improved scenarios assume a pre-disaster fiscal balance of $+3$ per cent of GDP and adaptive capacity equal to zero. Four-channel scenarios allow disaster shocks to affect real GDP growth, the primary balance, inflation and the effective interest rate.}
\end{table}

Across all scenarios considered, natural disaster debt risk is not determined by the size of the event alone. It reflects the interaction between the shock, the baseline fiscal trajectory, the primary balance, and adaptive capacity. Croatia's declining debt path provides partial protection against persistent debt increases, while Slovenia's rising baseline makes disaster shocks harder to absorb. The results point to the importance of building fiscal resilience before disasters occur.

\section{Final considerations}

This paper examines how severe natural disasters affect public debt sustainability in Croatia and Slovenia, two small open euro-area economies exposed to different types of natural hazards. Using the IMF's Natural Disaster Debt Dynamics Tool, we combine baseline debt projections, empirical distribution scenarios, local projections, quantile regressions, and stochastic simulations to assess how disaster shocks propagate through public debt. Fiscal vulnerability depends not only on the size of the shock but also on the baseline debt path, the pre-disaster fiscal balance, and adaptive capacity.

Croatia and Slovenia enter disaster scenarios from markedly different baseline positions. Croatia's public debt is projected to decline from 63.04 per cent of GDP in 2024 to 52.70 per cent in 2034, while Slovenia's rises from 66.60 to 74.71 per cent over the same horizon. This reflects Croatia's near-balanced primary position and more favourable debt dynamics, against Slovenia's persistent projected primary deficit. The baseline divergence is not a secondary detail but a key determinant of fiscal resilience. Severe natural disasters produce persistent upward shifts in debt trajectories. Under the local projection scenario, Croatian debt reaches 64.67 per cent of GDP in 2034, almost 12 percentage points above baseline, while Slovenian debt reaches 85.6 per cent, around 11 percentage points above its own baseline. Under recurring empirical distribution shocks, the 2034 debt ratio rises to 70.44 per cent in Croatia and 95 per cent in Slovenia. Even when the percentage-point debt effect is comparable across countries, the one entering the shock with a weaker baseline ends up with substantially higher debt.

Fiscal buffers and adaptive capacity attenuate the debt effect of natural disasters. Assuming a pre-disaster primary balance of $+3$ per cent of GDP and stronger adaptive capacity, 2034 debt falls by 3.90 percentage points in Croatia and 4.64 percentage points in Slovenia relative to the default local projection. The reduction is larger under the quantile regression specification, although that case should be read with the caveat noted in Sections~\ref{sec:buffers} and~\ref{sec:summary}. Fiscal space accumulated before a disaster is the most effective protection against post-disaster debt persistence.


Our analysis points to the importance of integrating disaster risk into medium-term fiscal planning. Baseline projections can appear benign, particularly in countries with declining debt ratios, but severe natural disasters can generate lasting fiscal costs. Maintaining adequate fiscal buffers, strengthening adaptive capacity, and developing risk-transfer instruments all help attenuate the propagation of disaster shocks to public debt. In that sense, EU Solidarity Fund support, parametric catastrophe insurance, and post-disaster fiscal rules are promising avenues for future numerical extensions of the framework.

Debt sustainability analysis should not treat natural disasters as exceptional add-ons to standard macro-fiscal projections. Climate change is expected to raise the frequency and severity of weather- and climate-related hazards, further strengthening the case for integrating the two. For economies exposed to earthquakes, floods, storms, and climate-related hazards, disaster risk is part of the fiscal environment. The ND-DDT framework offers a practical way to quantify this risk and to compare how different fiscal starting points shape the propagation of shocks. In the Croatian and Slovenian cases, the results are unambiguous: resilience is built before the shock occurs, not after it has already entered the debt dynamics.

\newpage

\appendix
\numberwithin{equation}{section}
\numberwithin{figure}{section}
\numberwithin{table}{section}

\section{Appendix}\label{AppendixA}

This Appendix reports the econometric specification and supporting regression tables used to construct the local projection and quantile-regression scenarios. The main text focuses on the preferred two-channel specification, in which natural disasters affect real GDP growth and the primary balance. As a robustness exercise, the main text also reports a four-channel specification in which the GDP deflator and the effective interest rate are allowed to respond to the shock.

The four-channel results should be interpreted cautiously. Inflation and effective interest-rate responses are more difficult to generalise across countries than output and fiscal-balance responses. In Croatia and Slovenia, the euro-area monetary environment and the structure of outstanding debt attenuate the direct pass-through from a domestic natural disaster to the average effective interest rate on public debt. For this reason, the four-channel results are reported in Section~\ref{sec:robustness} as a robustness check, while the detailed decomposition charts are retained below for transparency.

To understand the accounting role of the additional channels, consider the following approximation of Eq.~\eqref{eq:debt}:
\begin{equation}
\Delta d_t \approx (i_t-\pi_t-g_t)\,d_{t-1}-pb_t+of_t .
\end{equation}
The term $(i_t-\pi_t-g_t)\,d_{t-1}$ captures the debt-stock amplification effect. A decline in inflation or an increase in the effective interest rate can raise the real debt burden, especially when the initial debt stock is high. This explains why the four-channel specification can generate stronger debt effects in the local projection case. At the same time, the quantile-regression results show that these channels are sensitive to the estimated cross-country elasticity and should not be read as country-specific forecasts.

\begin{figure}[tbp]
\captionsetup{position=above}
\caption{Croatia Public Sector Debt Dynamics in the Four-Channel Specification, 2026--2031}
\centering

{\footnotesize\textbf{a)} Local projection} \\
\includegraphics[width=0.7\textwidth]{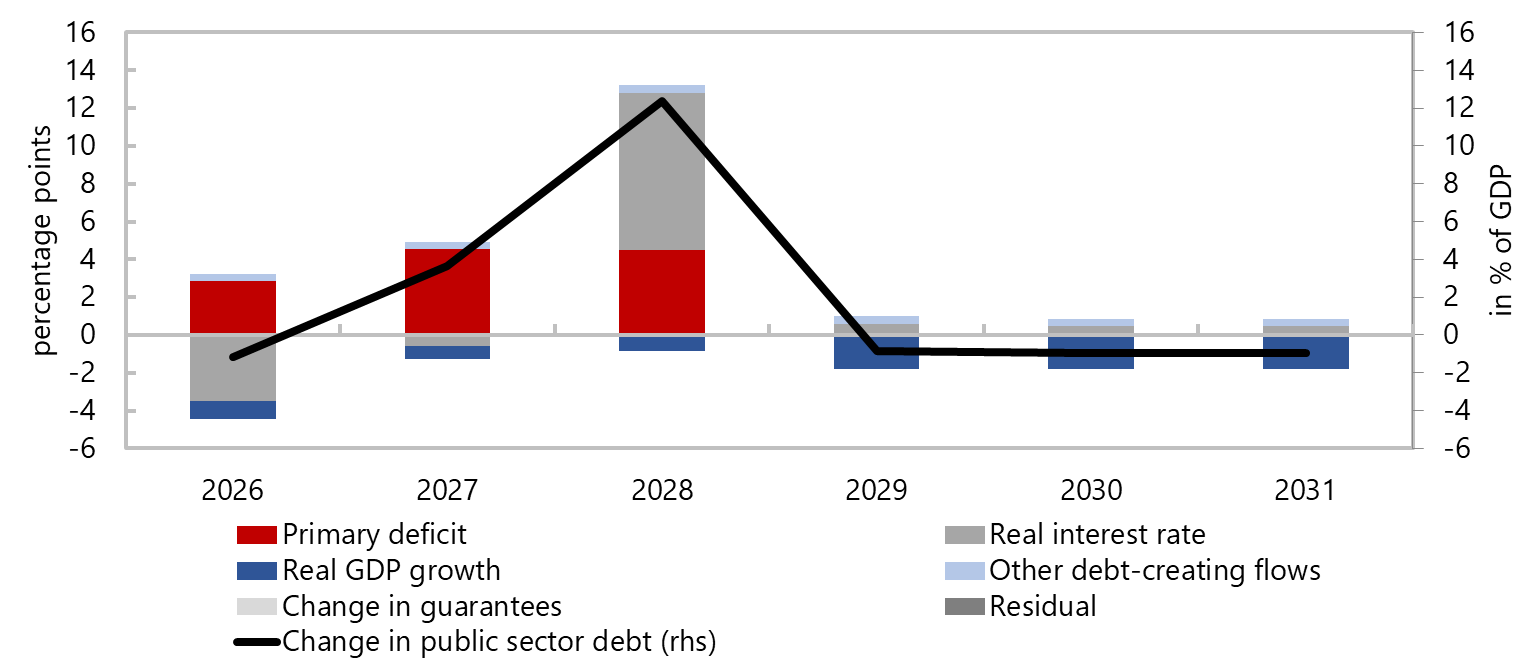} \\[6pt]

{\footnotesize\textbf{b)} Local projection with PB and AC improved} \\
\includegraphics[width=0.7\textwidth]{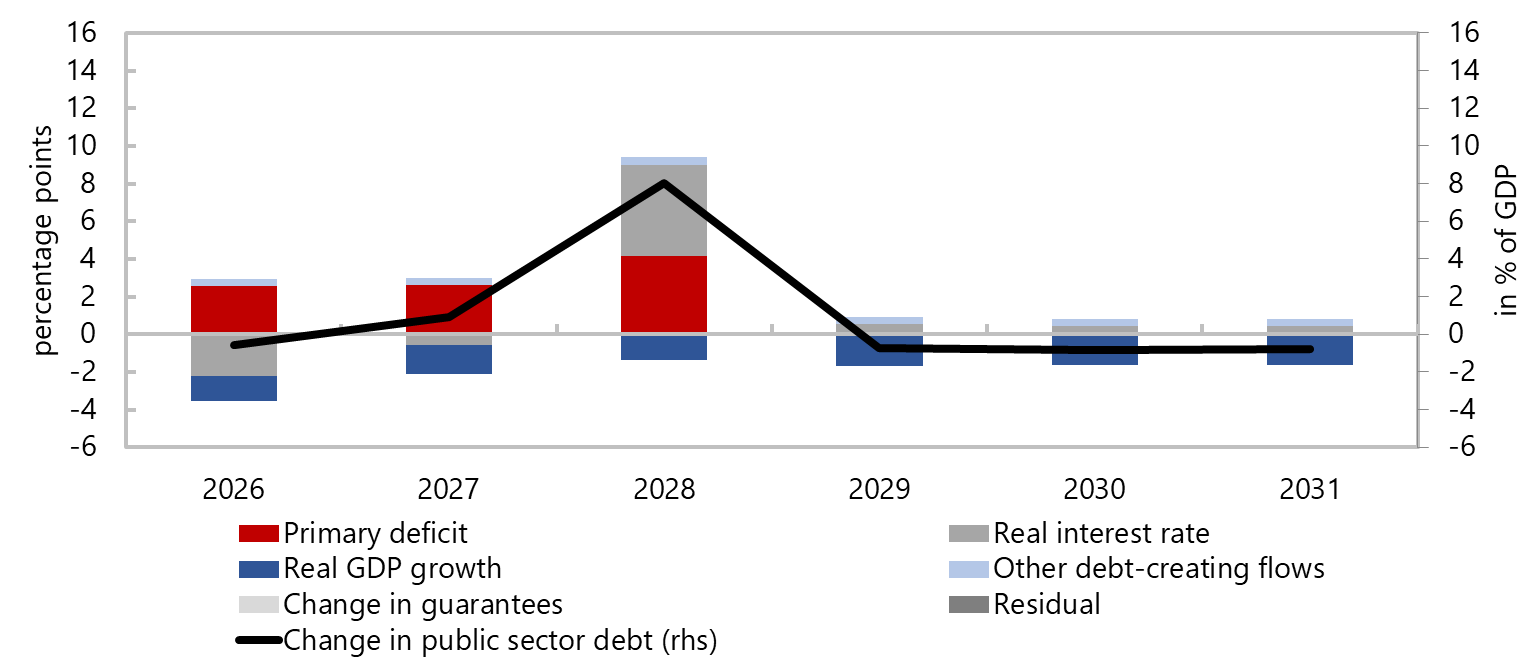} \\[6pt]

{\footnotesize\textbf{c)} Quantile projection} \\
\includegraphics[width=0.7\textwidth]{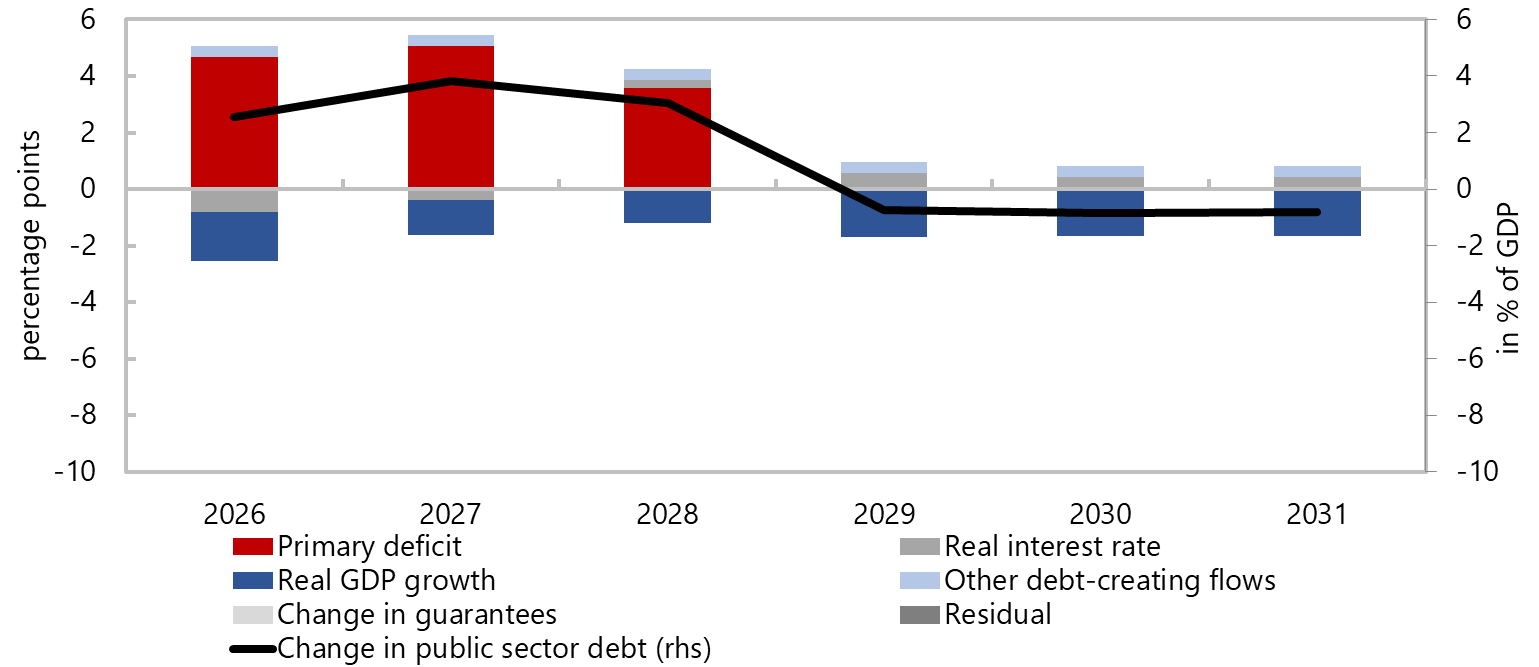} \\[6pt]

{\footnotesize\textbf{d)} Quantile projection with PB and AC improved} \\
\includegraphics[width=0.7\textwidth]{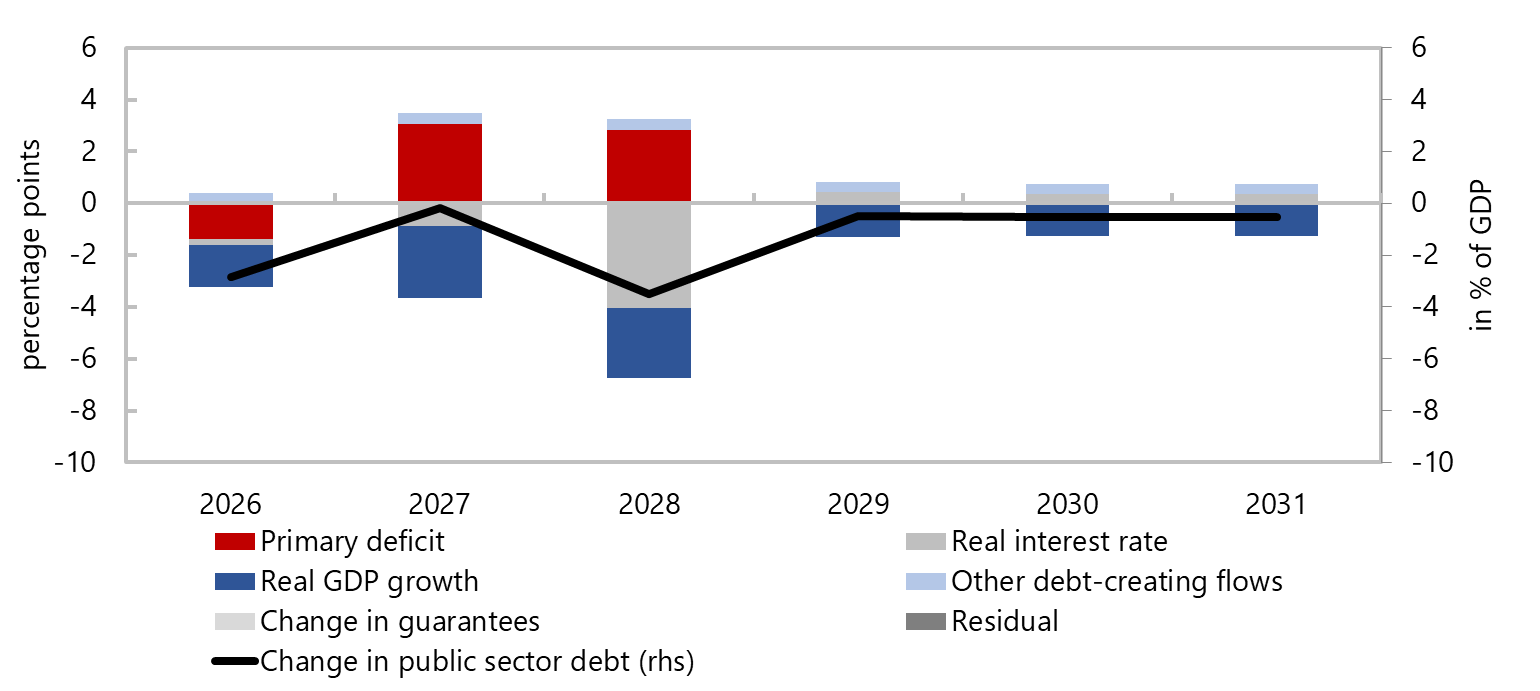}

\caption*{\footnotesize \textit{Source: Authors' calculation.}}
\label{fig:debt-structure-3-cro}
\end{figure}

\begin{figure}[tbp]
\captionsetup{position=above}
\caption{Slovenia Public Sector Debt Dynamics in the Four-Channel Specification, 2026--2031}
\centering

{\footnotesize\textbf{a)} Local projection} \\
\includegraphics[width=0.7\textwidth]{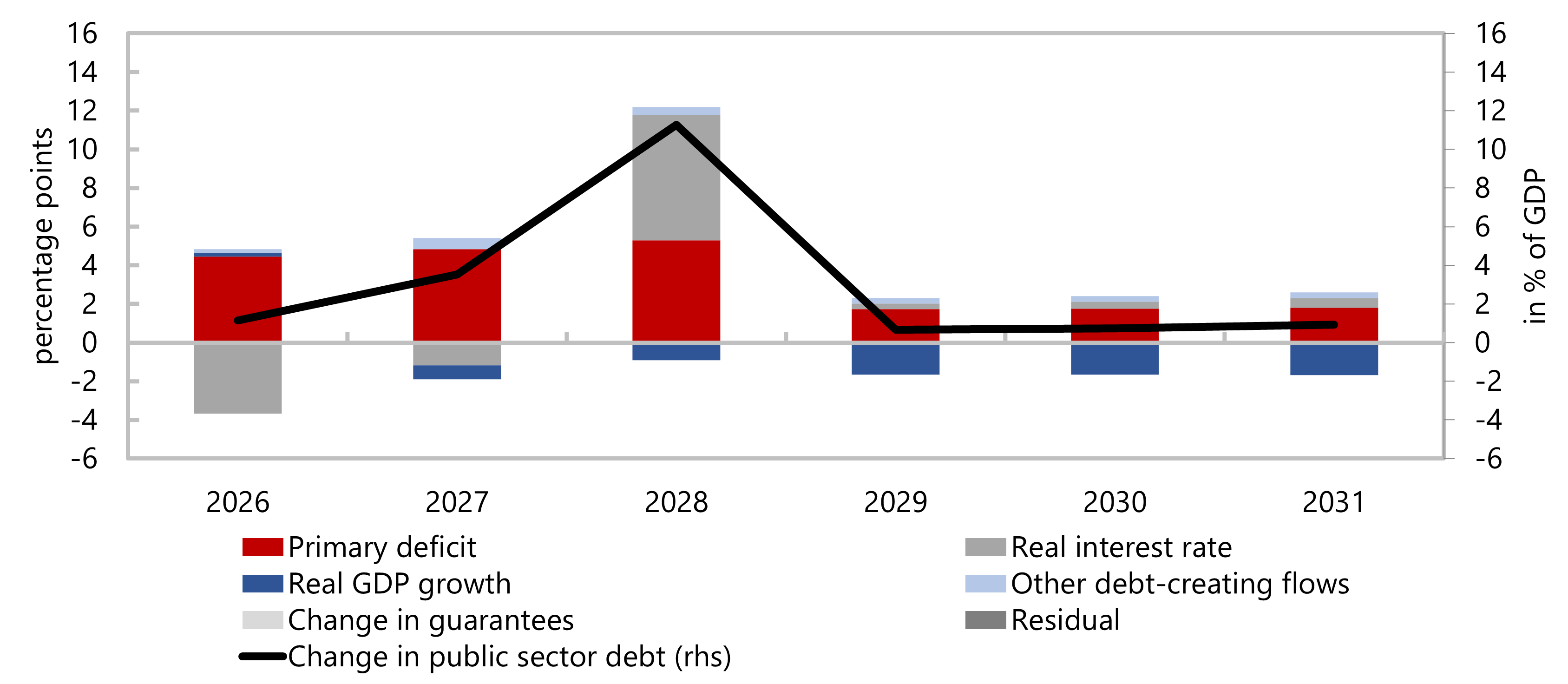} \\[6pt]

{\footnotesize\textbf{b)} Local projection with PB and AC improved} \\
\includegraphics[width=0.7\textwidth]{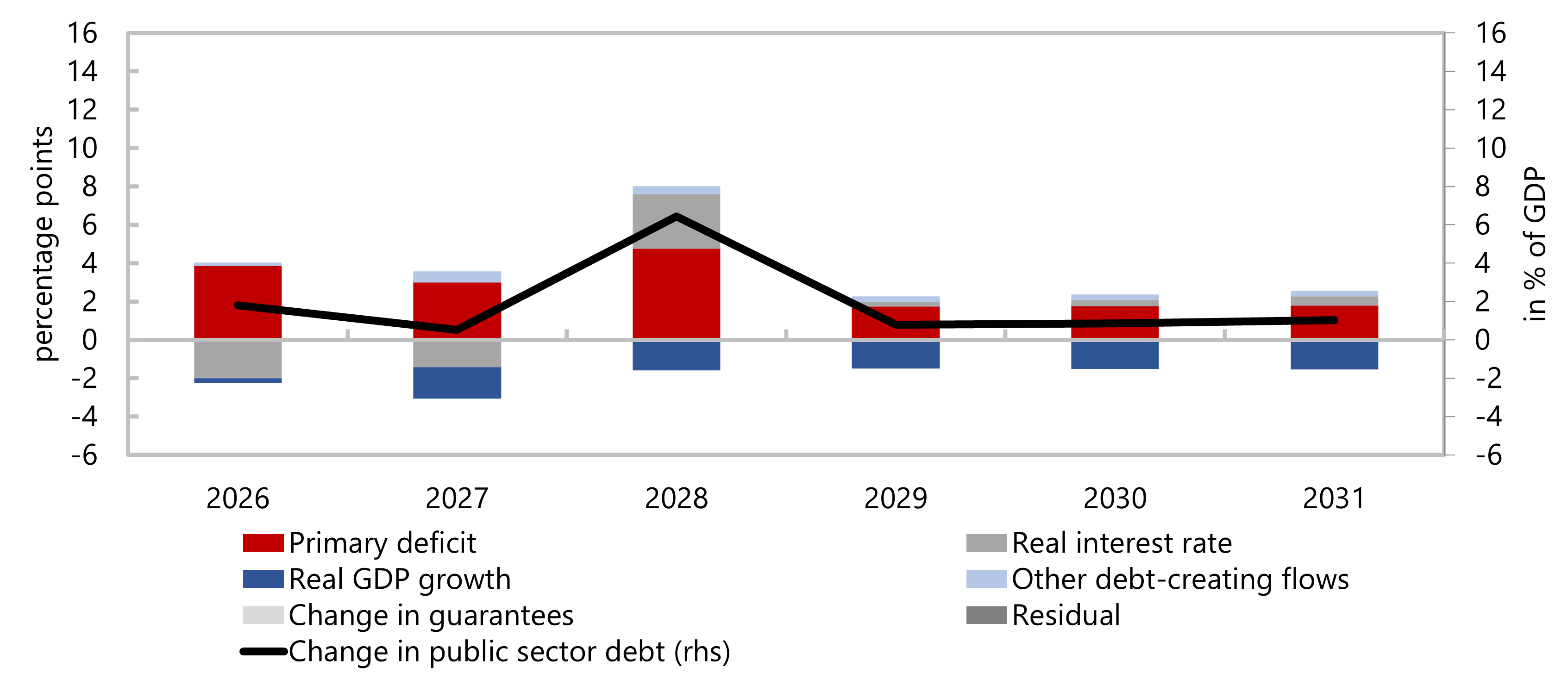} \\[6pt]

{\footnotesize\textbf{c)} Quantile projection} \\
\includegraphics[width=0.7\textwidth]{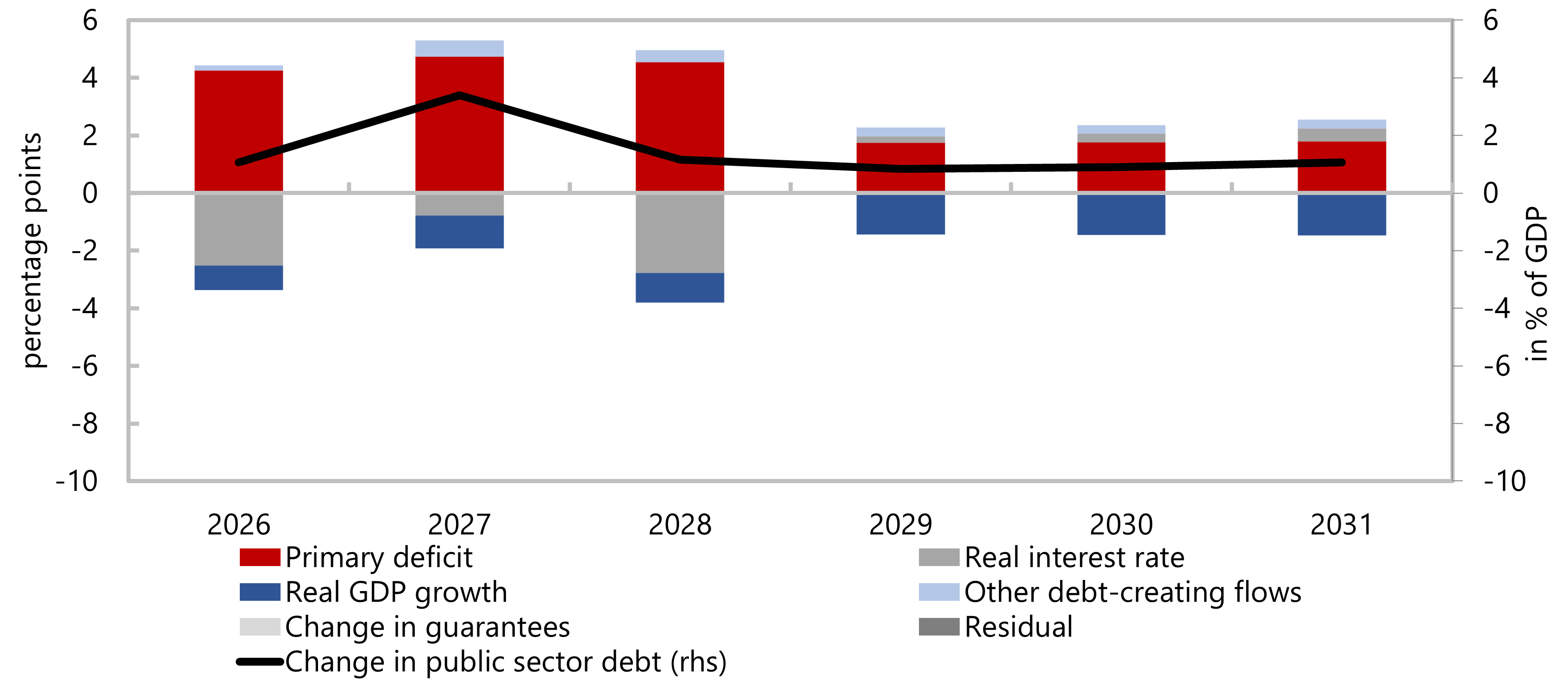} \\[6pt]

{\footnotesize\textbf{d)} Quantile projection with PB and AC improved} \\
\includegraphics[width=0.7\textwidth]{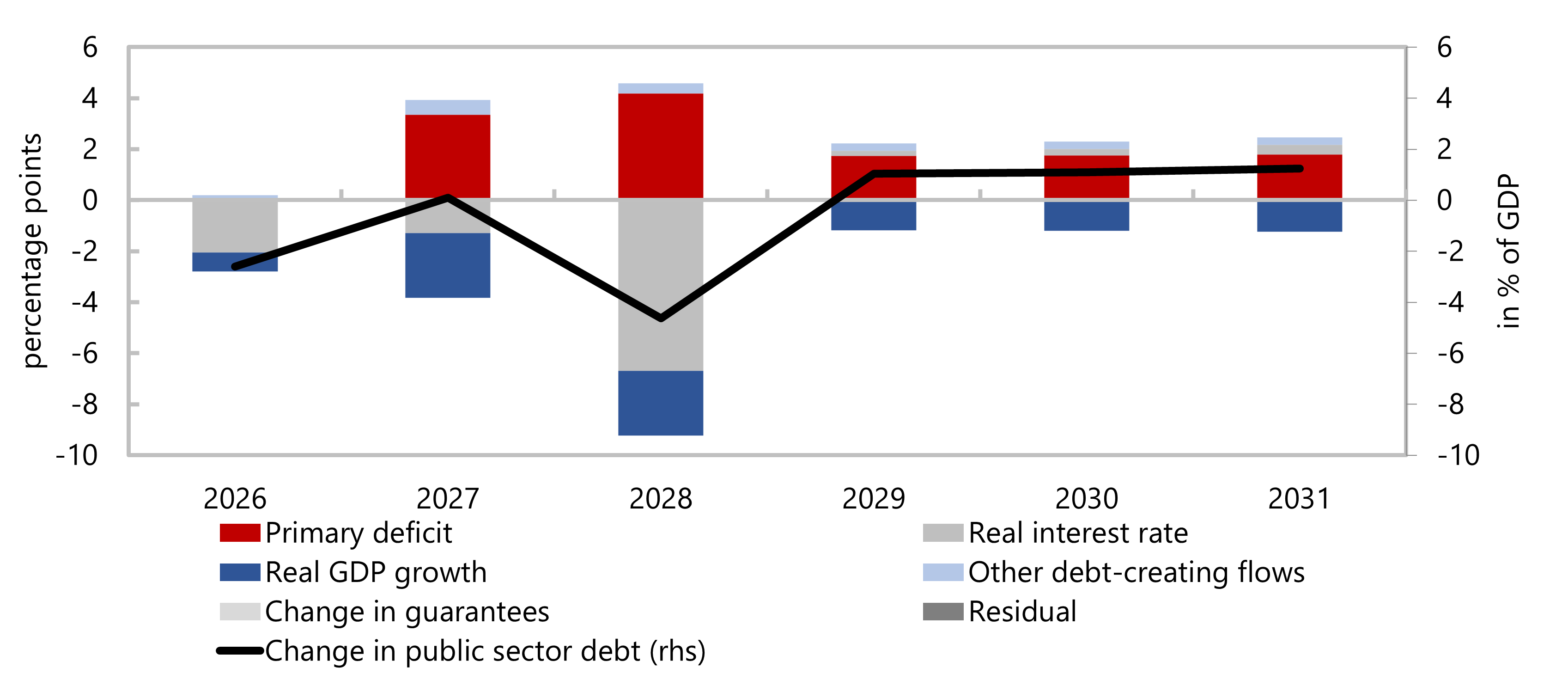}

\caption*{\footnotesize \textit{Source: Authors' calculation.}}
\label{fig:debt-structure-3-slo}
\end{figure}

\begin{framed}
\begin{adjustwidth}{-0.3cm}{-0.3cm}
\small
\noindent \textbf{Empirical LP specification}

\begin{equation}
\resizebox{0.95\textwidth}{!}{$
\begin{aligned}
y_{i,t+h} - y_{i,t-1} =\; & \alpha_{i,h} + \mu_{t,h}
+ \beta_h \, \text{ND}_{i,t}  \\
& + \gamma_{1h} (\text{ND}_{i,t} \times \text{Damage}_{i,t})
+ \gamma_{2h} (\text{ND}_{i,t} \times AE_{i,t})
+ \gamma_{3h} (\text{ND}_{i,t} \times \text{NDcapacity}_{i,t}) \\
& + \gamma_{4h} (\text{ND}_{i,t} \times fb_{i,t})
+ \delta_{1h} y_{i,t-1}
+ \delta_{2h} fb_{i,t-1}
+ \delta_{3h} \text{ExtraND1995}_{i,t-1}
+ \delta_{4h} \text{lcompshock}_{i,t+h-1} \\
& + \varepsilon_{i,t+h}.
\end{aligned}
$}
\end{equation}

\begin{minipage}{\textwidth}
\footnotesize
\textit{Note:} The equation is estimated separately for each horizon $h = 0, 1, 2$, corresponding to contemporaneous, one-year-ahead and two-year-ahead responses.
\end{minipage}

\medskip

\begin{minipage}{\textwidth}
\footnotesize\emph{Variable definitions:}
$y_{i,t+h}$ denotes the macroeconomic indicator of interest, namely real GDP growth, the GDP deflator, the effective interest rate or the primary balance. $\alpha_{i,h}$ represents country fixed effects and $\mu_{t,h}$ year fixed effects. $\text{ND}_{i,t}$ is a binary indicator equal to one in the year a large single-year non-overlapping natural disaster (damage exceeding 1 per cent of GDP) occurs and zero otherwise. $\text{Damage}_{i,t}$ measures the physical intensity of the disaster as a percentage of GDP. $AE_{i,t}$ is a dummy variable equal to one for advanced economies. $\text{NDcapacity}_{i,t}$ is the ND-GAIN climate adaptive-capacity index. $fb_{i,t}$ represents the overall fiscal balance as a share of GDP. $y_{i,t-1}$ is the lagged dependent variable, $fb_{i,t-1}$ is the lagged fiscal balance, $\text{ExtraND1995}_{i,t-1}$ indicates whether a country had an established national disaster-policy framework or institutional arrangement since 1995, and $\text{lcompshock}_{i,t+h-1}$ is the country-specific commodity-price shock from \textcite{GrussKebhaj2019}. Equation~(A.2) corresponds to the heterogeneous-effects local-projection specification of \textcite{NguyenFengGarciaEscribano2025}, Tables 8 and 9. Tables~\ref{tab:LP_outputG}--\ref{tab:LP_eff_int_fx} report the local-projection coefficients reproduced in the ND-DDT regression sheets.
\end{minipage}
\end{adjustwidth}
\end{framed}

\subsection*{Local-projection regression tables}

The eight tables below reproduce the local-projection coefficients used in the ND-DDT regression sheets. Each row reports the point estimate; the corresponding robust standard error is shown in brackets immediately below. Coefficients for the headline output and primary-balance regressions match \textcite{NguyenFengGarciaEscribano2025}, Tables 8 and 9, to three decimals.

\begin{table}[H]
\centering
\footnotesize
\caption{Real GDP growth (heterogeneous-effects local projection)}
\label{tab:LP_outputG}
\begin{tabular}{lccc}
\toprule
\textbf{VARIABLES} & GDP growth (t) & GDP growth (t+1) & GDP growth (t+2) \\
\midrule
lcompshock(t-1)        & 6.160***  &           &           \\
                       & [1.886]   &           &           \\
lcompshock(t)          &           & 8.101***  &           \\
                       &           & [2.201]   &           \\
lcompshock(t+1)        &           &           & 8.998***  \\
                       &           &           & [2.080]   \\
GDP growth(t-1)        & 0.299***  & 0.106***  & 0.070***  \\
                       & [0.027]   & [0.023]   & [0.022]   \\
ND (onset)             & 0.160     & 1.480     & 1.972     \\
                       & [2.493]   & [2.846]   & [2.604]   \\
ND $\times$ damage     & $-0.031$***  & 0.018     & $-0.032$     \\
                       & [0.007]   & [0.018]   & [0.021]   \\
ND $\times$ storm      & 0.040     & 1.602*    & 0.203     \\
                       & [0.705]   & [0.844]   & [0.815]   \\
ND $\times$ flood      & $-0.903$     & 0.802     & 0.159     \\
                       & [0.804]   & [0.882]   & [0.691]   \\
ND $\times$ drought    & $-0.489$     & 2.259**   & $-0.163$     \\
                       & [0.995]   & [1.038]   & [1.369]   \\
ND $\times$ AE         & 0.364     & $-0.569$     & $-0.570$     \\
                       & [1.465]   & [1.496]   & [1.386]   \\
ND $\times$ LIDC       & 1.503*    & $-0.571$     & $-0.432$     \\
                       & [0.797]   & [1.071]   & [1.018]   \\
ND $\times$ small island & $-1.324$**  & 0.309     & 0.232     \\
                       & [0.670]   & [0.738]   & [0.843]   \\
ND $\times$ adapt. cap. & $-0.879$     & $-2.202$     & $-0.788$     \\
                       & [3.994]   & [5.270]   & [4.649]   \\
ND $\times$ fb         & 0.070     & 0.131*    & 0.155**   \\
                       & [0.111]   & [0.079]   & [0.066]   \\
fb(t-1)                & 0.029**   & 0.032**   & 0.013     \\
                       & [0.012]   & [0.015]   & [0.013]   \\
ExtraND1995(t-1)       & 4.634     & 10.487**  & 9.040*    \\
                       & [3.922]   & [4.820]   & [5.082]   \\
Constant               & $-1.206$     & $-5.445$*    & $-3.244$     \\
                       & [2.312]   & [2.968]   & [2.980]   \\
\midrule
Observations           & 4{,}321  & 4{,}151  & 3{,}984  \\
R-squared              & 0.199     & 0.132     & 0.123     \\
Number of countries    & 172       & 172       & 172       \\
\bottomrule
\end{tabular}
\caption*{\footnotesize \textit{Notes:} Country and year fixed effects included. Heteroskedasticity-robust standard errors clustered at the country level in brackets (Stata command \texttt{xtreg ..., fe vce(cluster country)}; confirmed by H.~M.~Nguyen, personal communication, 2026). *, **, *** denote statistical significance at the 10, 5 and 1 per cent levels respectively. Coefficients reproduce \textcite{NguyenFengGarciaEscribano2025}, Table 8.}
\end{table}

\begin{table}[H]
\centering
\small
\caption{Primary balance}
\label{tab:LP_pb}
\begin{tabular}{lccc}
\toprule
\textbf{VARIABLES} & PB (t) & PB (t+1) & PB (t+2) \\
\midrule
lcompshock(t-1)        & 13.566*** &           &           \\
                       & [4.163]   &           &           \\
lcompshock(t)          &           & 35.889*** &           \\
                       &           & [5.211]   &           \\
lcompshock(t+1)        &           &           & 34.052*** \\
                       &           &           & [4.661]   \\
PB(t-1)                & 0.985***  & 0.698***  & 0.613***  \\
                       & [0.183]   & [0.111]   & [0.137]   \\
ND (onset)             & 0.527     & 2.785     & 0.242     \\
                       & [2.447]   & [2.398]   & [2.396]   \\
ND $\times$ damage     & $-0.001$     & 0.010     & $-0.020$     \\
                       & [0.014]   & [0.010]   & [0.015]   \\
ND $\times$ storm      & $-0.304$     & 1.805**   & 0.607     \\
                       & [0.707]   & [0.795]   & [0.948]   \\
ND $\times$ flood      & $-0.233$     & 1.162     & 0.920     \\
                       & [0.762]   & [0.728]   & [0.798]   \\
ND $\times$ drought    & $-1.856$**   & 1.969     & 2.121     \\
                       & [0.780]   & [2.130]   & [2.655]   \\
ND $\times$ AE         & $-0.444$     & $-2.744$*    & $-0.975$     \\
                       & [1.326]   & [1.413]   & [1.434]   \\
ND $\times$ LIDC       & $-0.019$     & $-2.060$*    & $-1.636$*    \\
                       & [0.720]   & [1.215]   & [0.909]   \\
ND $\times$ small island & $-0.569$     & $-3.580$***  & $-2.980$**   \\
                       & [1.213]   & [1.309]   & [1.222]   \\
ND $\times$ adapt. cap. & 0.859     & $-2.943$     & 0.542     \\
                       & [3.961]   & [4.116]   & [4.094]   \\
ND $\times$ fb         & 0.165*    & 0.159     & 0.135     \\
                       & [0.091]   & [0.125]   & [0.135]   \\
fb(t-1)                & $-0.534$***  & $-0.438$***  & $-0.490$***  \\
                       & [0.182]   & [0.118]   & [0.156]   \\
ExtraND1995(t-1)       & 2.255     & 1.813     & $-1.871$     \\
                       & [6.506]   & [8.312]   & [10.530]  \\
Constant               & $-3.592$     & $-3.035$     & $-2.158$     \\
                       & [3.815]   & [4.791]   & [6.054]   \\
\midrule
Observations           & 4{,}390  & 4{,}223  & 4{,}054  \\
R-squared              & 0.324     & 0.237     & 0.166     \\
Number of countries    & 170       & 170       & 170       \\
\bottomrule
\end{tabular}
\caption*{\footnotesize \textit{Notes:} Country and year fixed effects included. Heteroskedasticity-robust standard errors clustered at the country level in brackets (see notes to Table~\ref{tab:LP_outputG}). *, **, *** denote statistical significance at the 10, 5 and 1 per cent levels respectively. Source: ND-DDT \texttt{LP primarybalance} sheet.}
\end{table}

\begin{table}[H]
\centering
\small
\caption{Real exports growth}
\label{tab:LP_exportG}
\begin{tabular}{lccc}
\toprule
\textbf{VARIABLES} & Exports (t) & Exports (t+1) & Exports (t+2) \\
\midrule
lcompshock(t-1)        & $-10.136$    &           &           \\
                       & [28.667]  &           &           \\
lcompshock(t)          &           & $-40.287$    &           \\
                       &           & [29.836]  &           \\
lcompshock(t+1)        &           &           & 26.012    \\
                       &           &           & [24.202]  \\
Exports(t-1)           & 0.007***  & $-0.010$**   & 0.136***  \\
                       & [0.002]   & [0.004]   & [0.005]   \\
ND (onset)             & $-19.155$*   & $-3.843$     & $-6.987$     \\
                       & [9.838]   & [12.369]  & [8.553]   \\
ND $\times$ damage     & $-0.041$     & 0.079     & $-0.087$     \\
                       & [0.052]   & [0.135]   & [0.055]   \\
ND $\times$ storm      & $-5.232$*    & $-0.386$     & $-7.692$     \\
                       & [3.113]   & [3.505]   & [5.143]   \\
ND $\times$ flood      & $-1.739$     & 7.108**   & $-4.313$     \\
                       & [3.724]   & [3.151]   & [3.854]   \\
ND $\times$ drought    & 2.208     & 4.578*    & $-2.905$     \\
                       & [4.236]   & [2.781]   & [4.287]   \\
ND $\times$ AE         & 15.469*** & 2.580     & 2.736     \\
                       & [5.477]   & [5.939]   & [4.900]   \\
ND $\times$ LIDC       & 0.110     & 0.846     & $-0.339$     \\
                       & [3.838]   & [5.239]   & [4.194]   \\
ND $\times$ small island & 0.064     & $-2.208$     & 4.289     \\
                       & [2.623]   & [5.757]   & [3.764]   \\
ND $\times$ adapt. cap. & 39.720**  & 5.737     & 19.110    \\
                       & [18.008]  & [23.685]  & [17.014]  \\
ND $\times$ fb         & 1.309***  & 1.020***  & $-0.010$     \\
                       & [0.484]   & [0.324]   & [0.373]   \\
fb(t-1)                & $-0.841$     & $-0.374$**   & $-0.317$     \\
                       & [0.528]   & [0.182]   & [0.304]   \\
ExtraND1995(t-1)       & $-75.244$    & $-101.727$   & $-89.068$    \\
                       & [70.719]  & [79.422]  & [83.644]  \\
Constant               & 41.798    & 56.719    & 61.588    \\
                       & [40.071]  & [44.465]  & [46.872]  \\
\midrule
Observations           & 4{,}022  & 3{,}864  & 3{,}706  \\
R-squared              & 0.049     & 0.032     & 0.278     \\
Number of countries    & 158       & 158       & 158       \\
\bottomrule
\end{tabular}
\caption*{\footnotesize \textit{Notes:} As in Table~\ref{tab:LP_outputG}. Source: ND-DDT \texttt{LP exportG} sheet.}
\end{table}

\begin{table}[H]
\centering
\small
\caption{Domestic-currency debt change ($\Delta$ enda)}
\label{tab:LP_enda}
\begin{tabular}{lccc}
\toprule
\textbf{VARIABLES} & enda (t) & enda (t+1) & enda (t+2) \\
\midrule
lcompshock(t-1)        & $-26.322$*   &           &           \\
                       & [15.340]  &           &           \\
lcompshock(t)          &           & $-17.939$    &           \\
                       &           & [16.995]  &           \\
lcompshock(t+1)        &           &           & $-14.377$    \\
                       &           &           & [19.149]  \\
enda(t-1)              & 0.300***  & 0.071*    & 0.047*    \\
                       & [0.052]   & [0.041]   & [0.027]   \\
ND (onset)             & $-22.945$*   & 0.062     & $-13.933$    \\
                       & [12.643]  & [9.829]   & [33.210]  \\
ND $\times$ damage     & 0.043     & 0.056     & $-0.076$     \\
                       & [0.054]   & [0.037]   & [0.125]   \\
ND $\times$ storm      & $-3.477$     & $-8.937$**   & $-16.655$    \\
                       & [4.019]   & [4.490]   & [12.946]  \\
ND $\times$ flood      & $-2.531$     & $-6.141$     & $-11.734$    \\
                       & [4.126]   & [4.426]   & [10.582]  \\
ND $\times$ drought    & $-5.121$**   & $-5.123$     & $-13.850$    \\
                       & [2.563]   & [4.114]   & [10.482]  \\
ND $\times$ AE         & 8.549     & 4.442     & 8.425     \\
                       & [6.638]   & [5.161]   & [13.522]  \\
ND $\times$ LIDC       & $-6.173$*    & 0.495     & $-4.497$     \\
                       & [3.462]   & [3.805]   & [13.298]  \\
ND $\times$ small island & $-2.023$     & 1.545     & 1.103     \\
                       & [3.710]   & [2.890]   & [3.699]   \\
ND $\times$ adapt. cap. & 43.293**  & 7.541     & 39.363    \\
                       & [21.500]  & [17.137]  & [71.540]  \\
ND $\times$ fb         & $-0.563$     & 0.334     & $-0.856$     \\
                       & [0.551]   & [0.295]   & [1.003]   \\
fb(t-1)                & $-0.065$     & $-0.134$     & $-0.067$     \\
                       & [0.062]   & [0.096]   & [0.079]   \\
ExtraND1995(t-1)       & 14.958    & 20.589    & 19.265    \\
                       & [13.875]  & [20.177]  & [23.047]  \\
Constant               & 7.615     & 4.622     & 19.827    \\
                       & [9.144]   & [12.580]  & [16.715]  \\
\midrule
Observations           & 4{,}370  & 4{,}200  & 4{,}032  \\
R-squared              & 0.218     & 0.118     & 0.113     \\
Number of countries    & 170       & 170       & 170       \\
\bottomrule
\end{tabular}
\caption*{\footnotesize \textit{Notes:} As in Table~\ref{tab:LP_outputG}. Source: ND-DDT \texttt{LP enda\_pch} sheet.}
\end{table}

\begin{table}[H]
\centering
\small
\caption{Foreign-currency debt change ($\Delta$ ende)}
\label{tab:LP_ende}
\begin{tabular}{lccc}
\toprule
\textbf{VARIABLES} & ende (t) & ende (t+1) & ende (t+2) \\
\midrule
lcompshock(t-1)        & $-12.369$    &           &           \\
                       & [14.246]  &           &           \\
lcompshock(t)          &           & $-14.970$    &           \\
                       &           & [18.990]  &           \\
lcompshock(t+1)        &           &           & $-8.860$     \\
                       &           &           & [20.015]  \\
ende(t-1)              & 0.212***  & 0.037     & 0.009     \\
                       & [0.070]   & [0.050]   & [0.040]   \\
ND (onset)             & $-14.619$    & 25.027**  & 5.664     \\
                       & [11.049]  & [11.873]  & [10.779]  \\
ND $\times$ damage     & 0.134***  & $-0.025$     & 0.020     \\
                       & [0.038]   & [0.038]   & [0.037]   \\
ND $\times$ storm      & $-5.665$     & $-5.850$*    & $-1.095$     \\
                       & [5.195]   & [3.307]   & [2.557]   \\
ND $\times$ flood      & $-4.620$     & $-2.272$     & $-3.096$     \\
                       & [4.262]   & [5.595]   & [3.935]   \\
ND $\times$ drought    & $-4.255$     & $-3.306$     & 0.895     \\
                       & [3.161]   & [5.876]   & [3.206]   \\
ND $\times$ AE         & 7.445     & $-7.744$     & 5.709     \\
                       & [7.118]   & [7.191]   & [6.005]   \\
ND $\times$ LIDC       & $-2.029$     & 8.253**   & 3.352     \\
                       & [4.603]   & [3.766]   & [4.477]   \\
ND $\times$ small island & $-1.176$     & 4.247     & 1.147     \\
                       & [4.401]   & [2.710]   & [3.076]   \\
ND $\times$ adapt. cap. & 29.116    & $-41.045$**  & $-9.923$     \\
                       & [20.355]  & [20.172]  & [18.829]  \\
ND $\times$ fb         & $-0.391$     & 0.250     & 0.422*    \\
                       & [0.855]   & [0.389]   & [0.253]   \\
fb(t-1)                & $-0.120$*    & $-0.090$     & $-0.054$     \\
                       & [0.072]   & [0.079]   & [0.065]   \\
ExtraND1995(t-1)       & 12.339    & 19.570    & 3.582     \\
                       & [14.151]  & [18.141]  & [18.525]  \\
Constant               & 7.561     & 9.705     & 17.104    \\
                       & [8.721]   & [11.556]  & [11.811]  \\
\midrule
Observations           & 4{,}322  & 4{,}153  & 3{,}989  \\
R-squared              & 0.155     & 0.120     & 0.120     \\
Number of countries    & 170       & 170       & 170       \\
\bottomrule
\end{tabular}
\caption*{\footnotesize \textit{Notes:} As in Table~\ref{tab:LP_outputG}. Source: ND-DDT \texttt{LP ende\_pch} sheet.}
\end{table}

\begin{table}[H]
\centering
\small
\caption{GDP deflator inflation}
\label{tab:LP_GDPdef}
\begin{tabular}{lccc}
\toprule
\textbf{VARIABLES} & Deflator (t) & Deflator (t+1) & Deflator (t+2) \\
\midrule
lcompshock(t-1)        & $-13.764$    &           &           \\
                       & [11.370]  &           &           \\
lcompshock(t)          &           & $-10.856$    &           \\
                       &           & [10.797]  &           \\
lcompshock(t+1)        &           &           & $-1.603$     \\
                       &           &           & [15.248]  \\
GDP deflator(t-1)      & 0.512***  & 0.437***  & 0.203**   \\
                       & [0.108]   & [0.050]   & [0.088]   \\
ND (onset)             & $-17.113$    & $-11.527$    & 9.858     \\
                       & [16.026]  & [12.234]  & [7.451]   \\
ND $\times$ damage     & 0.064     & 0.040     & 0.025     \\
                       & [0.044]   & [0.025]   & [0.023]   \\
ND $\times$ storm      & 7.171*    & 1.968     & $-1.840$     \\
                       & [3.805]   & [2.478]   & [2.216]   \\
ND $\times$ flood      & 0.409     & $-0.472$     & $-0.022$     \\
                       & [2.656]   & [2.491]   & [2.439]   \\
ND $\times$ drought    & 1.154     & 1.038     & $-0.080$     \\
                       & [2.630]   & [1.911]   & [3.118]   \\
ND $\times$ AE         & 3.992     & 6.665     & $-0.527$     \\
                       & [6.416]   & [4.952]   & [3.481]   \\
ND $\times$ LIDC       & $-12.669$    & $-7.997$     & 6.800*    \\
                       & [7.700]   & [5.171]   & [3.576]   \\
ND $\times$ small island & $-8.446$     & $-3.088$     & 4.613*    \\
                       & [6.016]   & [3.881]   & [2.756]   \\
ND $\times$ adapt. cap. & 30.607    & 20.690    & $-22.103$    \\
                       & [30.274]  & [22.555]  & [13.425]  \\
ND $\times$ fb         & $-0.747$*    & 0.041     & 0.144     \\
                       & [0.446]   & [0.173]   & [0.244]   \\
fb(t-1)                & $-0.044$     & $-0.058$     & $-0.049$     \\
                       & [0.052]   & [0.062]   & [0.057]   \\
ExtraND1995(t-1)       & 0.746     & $-3.941$     & 4.915     \\
                       & [9.835]   & [11.793]  & [13.883]  \\
Constant               & 11.883    & 9.389     & 15.802    \\
                       & [7.234]   & [6.708]   & [9.806]   \\
\midrule
Observations           & 4{,}348  & 4{,}177  & 4{,}006  \\
R-squared              & 0.357     & 0.270     & 0.129     \\
Number of countries    & 170       & 170       & 170       \\
\bottomrule
\end{tabular}
\caption*{\footnotesize \textit{Notes:} As in Table~\ref{tab:LP_outputG}. Source: ND-DDT \texttt{LP GDPdeflator} sheet.}
\end{table}

\begin{table}[H]
\centering
\small
\caption{Effective interest rate, local currency}
\label{tab:LP_eff_int_lc}
\begin{tabular}{lccc}
\toprule
\textbf{VARIABLES} & $i^d_{lc}$ (t) & $i^d_{lc}$ (t+1) & $i^d_{lc}$ (t+2) \\
\midrule
lcompshock(t-1)        & 12.440    &           &           \\
                       & [13.358]  &           &           \\
lcompshock(t)          &           & 23.842    &           \\
                       &           & [20.239]  &           \\
lcompshock(t+1)        &           &           & 35.566    \\
                       &           &           & [28.851]  \\
$i^d_{lc}$(t-1)        & 0.592***  & 0.246***  & $-0.135$**   \\
                       & [0.044]   & [0.012]   & [0.056]   \\
ND (onset)             & $-1.119$     & 1.093     & 6.017     \\
                       & [0.950]   & [1.695]   & [3.682]   \\
ND $\times$ damage     & 0.419     & 2.839**   & 4.296*    \\
                       & [0.971]   & [1.325]   & [2.333]   \\
ND $\times$ storm      & 0.138     & $-1.861$     & $-5.133$     \\
                       & [0.560]   & [1.169]   & [3.847]   \\
ND $\times$ flood      & 0.051     & $-0.647$     & $-2.997$     \\
                       & [0.574]   & [1.348]   & [2.169]   \\
ND $\times$ drought    & $-3.027$*    & $-2.662$     & $-0.817$     \\
                       & [1.657]   & [2.147]   & [1.253]   \\
ND $\times$ AE         & $-1.119$     & 1.093     & 6.017     \\
                       & [0.950]   & [1.695]   & [3.682]   \\
ND $\times$ LIDC       & 0.419     & 2.839**   & 4.296*    \\
                       & [0.971]   & [1.325]   & [2.333]   \\
ND $\times$ small island & $-0.367$     & 2.261*    & 4.572     \\
                       & [0.611]   & [1.159]   & [3.214]   \\
ND $\times$ adapt. cap. & $-1.639$     & $-1.718$     & 8.585     \\
                       & [3.420]   & [4.388]   & [7.207]   \\
ND $\times$ fb         & 0.381**   & $-0.172$     & $-0.425$     \\
                       & [0.150]   & [0.416]   & [0.354]   \\
fb(t-1)                & $-0.195$     & $-0.304$     & $-0.281$     \\
                       & [0.156]   & [0.232]   & [0.186]   \\
ExtraND1995(t-1)       & 5.932     & 10.435    & 13.170    \\
                       & [4.935]   & [7.562]   & [9.974]   \\
Constant               & $-1.982$     & $-2.092$     & $-2.052$     \\
                       & [3.072]   & [4.630]   & [6.020]   \\
\midrule
Observations           & 2{,}216  & 2{,}099  & 1{,}983  \\
R-squared              & 0.404     & 0.129     & 0.091     \\
Number of countries    & 118       & 118       & 118       \\
\bottomrule
\end{tabular}
\caption*{\footnotesize \textit{Notes:} As in Table~\ref{tab:LP_outputG}. Source: ND-DDT \texttt{LP eff\_int\_lc} sheet.}
\end{table}

\begin{table}[H]
\centering
\small
\caption{Effective interest rate, foreign currency}
\label{tab:LP_eff_int_fx}
\begin{tabular}{lccc}
\toprule
\textbf{VARIABLES} & $i^d_{fx}$ (t) & $i^d_{fx}$ (t+1) & $i^d_{fx}$ (t+2) \\
\midrule
lcompshock(t-1)        & 0.383     &           &           \\
                       & [0.503]   &           &           \\
lcompshock(t)          &           & 0.630     &           \\
                       &           & [0.770]   &           \\
lcompshock(t+1)        &           &           & 0.683     \\
                       &           &           & [0.751]   \\
$i^d_{fx}$(t-1)        & 0.523***  & 0.357***  & 0.175***  \\
                       & [0.078]   & [0.071]   & [0.055]   \\
ND (onset)             & $-0.225$     & $-0.610$     & 0.501     \\
                       & [1.130]   & [1.389]   & [1.417]   \\
ND $\times$ damage     & 0.007*    & $-0.007$     & $-0.019$***  \\
                       & [0.004]   & [0.007]   & [0.004]   \\
ND $\times$ storm      & $-0.148$     & $-0.199$     & $-0.292$     \\
                       & [0.299]   & [0.406]   & [0.392]   \\
ND $\times$ flood      & 0.119     & 0.046     & 0.258     \\
                       & [0.355]   & [0.555]   & [0.423]   \\
ND $\times$ drought    & $-0.013$     & 0.308     & 0.188     \\
                       & [0.353]   & [0.567]   & [0.655]   \\
ND $\times$ AE         & 0.679     & 0.992     & $-0.061$     \\
                       & [0.699]   & [0.951]   & [0.856]   \\
ND $\times$ LIDC       & 0.168     & 0.213     & 0.094     \\
                       & [0.475]   & [0.618]   & [0.662]   \\
ND $\times$ small island & 0.119     & 0.246     & 0.612     \\
                       & [0.300]   & [0.330]   & [0.435]   \\
ND $\times$ adapt. cap. & 0.292     & 0.653     & $-0.816$     \\
                       & [2.024]   & [2.476]   & [2.521]   \\
ND $\times$ fb         & 0.091**   & 0.047     & 0.029     \\
                       & [0.038]   & [0.039]   & [0.042]   \\
fb(t-1)                & $-0.012$**   & $-0.013$**   & $-0.015$**   \\
                       & [0.006]   & [0.006]   & [0.007]   \\
ExtraND1995(t-1)       & 1.700     & 2.292     & 3.275     \\
                       & [2.713]   & [3.520]   & [4.387]   \\
Constant               & 0.113     & 0.170     & 0.104     \\
                       & [1.527]   & [1.976]   & [2.449]   \\
\midrule
Observations           & 3{,}738  & 3{,}571  & 3{,}410  \\
R-squared              & 0.395     & 0.215     & 0.102     \\
Number of countries    & 170       & 170       & 170       \\
\bottomrule
\end{tabular}
\caption*{\footnotesize \textit{Notes:} As in Table~\ref{tab:LP_outputG}. Source: ND-DDT \texttt{LP eff\_int\_fx} sheet.}
\end{table}

\newpage

\printbibliography
\end{document}